\documentclass[aps,prx,amsmath,floats,floatfix, twocolumn,
superscriptaddress,nofootinbib,showpacs]{revtex4-1}

\usepackage[T1]{fontenc}
\usepackage[utf8]{inputenc}
\usepackage{lmodern}
\usepackage{tensor}
\usepackage{verbatim}
\usepackage{siunitx}
\usepackage[caption=false]{subfig}
\usepackage{wasysym}
\usepackage[english]{babel}
\usepackage{booktabs}
\usepackage{mathrsfs}
\usepackage{multirow}

\usepackage[dvipsnames, usenames]{xcolor}
\definecolor{linkcolor}{rgb}{0.0,0.3,0.5}
\usepackage[hypertexnames=false, unicode, colorlinks=true, linkcolor=linkcolor,
citecolor=linkcolor, filecolor=linkcolor,urlcolor=linkcolor,
pdfusetitle]{hyperref}

\usepackage[all]{hypcap}
\usepackage{graphicx}
\usepackage{xspace}
\usepackage{amssymb}
\usepackage[normalem]{ulem} %for \sout
\usepackage{bm} % boldmath
\usepackage{enumitem,amssymb}
\usepackage{mathtools} 

\usepackage{microtype}

\usepackage[english]{babel}
\usepackage{blindtext}
\usepackage{tikz}
\usetikzlibrary{decorations.pathmorphing}
\usetikzlibrary{decorations.markings}
\usepackage{orcidlink}

\begin{document}
\title{Tidal Excitation of Relativistic Stars}

\newcommand{\Tsinghua}{\affiliation{Department of Astronomy, Tsinghua University, Beijing 100084, China}}

\author{Zhiqiang Miao\orcidlink{0000-0003-1197-3329}}
\affiliation{Tsung-Dao Lee Institute, Shanghai Jiao Tong University, Shanghai, 1 Lisuo Road, 201210, China}

\author{Junquan Su \orcidlink{0009-0008-1901-533X}}
\Tsinghua
\author{Neev Khera \orcidlink{0000-0003-3515-2859}}
\Tsinghua

\author{Huan Yang\orcidlink{0000-0002-9965-3030}}
\email{hyangdoa@tsinghua.edu.cn}
\Tsinghua

\begin{abstract}

How a relativistic star responds to a time-dependent tide has lacked a first-principles counterpart of the Newtonian normal-mode expansion. In general relativity, the tidal field is gauge sensitive, radiative modes are non-Hermitian and incomplete, and the direct and late-time responses are intrinsically nonmodal. In this work we present, to our knowledge, the first causal relativistic theory that treats all of these components on equal footing. We first develop the spectral decomposition of the retarded Green function of the matched stellar interior--Schwarzschild exterior, which classifies four separate regimes in time domain: acausal region, prompt wave, surface-reflected wave, and mode$+$tail. The prompt signal is identical to that of a Schwarzschild black hole with the same mass, which is combined with the surface-reflected wave to be referred as the "direct part". 
With the Green function and a complex bilinear form on the full metric--fluid phase space, we decompose the linear response of a nonrotating perfect-fluid star into direct, quasinormal-mode, and late-time-tail sectors. Crucially, every sector admits an explicit first-order driven evolution equation: discrete bilinear projections determine the dynamically evolving quasinormal-mode amplitudes, while continuum projections of the same tidal source onto branch-cut states determine the direct and tail amplitudes. The result is a complete modal--nonmodal decomposition of the spacetime and fluid perturbations, which is physically described by the dynamics of a discrete and a continuous-frequency set of oscillators. 
We verify the decomposition numerically under generic time-dependent tidal driving and discuss applications to black hole perturbations.

\end{abstract}

\maketitle

\section{Introduction}

Binary neutron-star mergers are among the principal targets of ground-based gravitational-wave astronomy and offer a unique opportunity to probe matter at supranuclear densities. The time-dependent tidal field of a companion can deform a neutron star, excite its internal oscillation modes, and thereby imprint its internal dynamics on the orbital evolution and gravitational waveform. These effects are conventionally described in terms of ``equilibrium tides'' and ``dynamical tides'', a framework inherited largely from Newtonian stellar perturbation theory~\cite{Cox1980} and later supplemented by post-Newtonian and relativistic corrections~\cite{Hinderer:2007mb,Damour:2009vw,Binnington:2009bb,Vines:2011ud,Bini:2012gu,Steinhoff:2016rfi,HegadeKR:2025qwj,Pitre:2023xsr,Andersson:2025iyd,Apostolidis:2026qsg,Saketh:2026trm,Apostolidis:2026qsg,Saketh:2024juq}. Detecting such signatures can constrain the neutron-star equation of state and reveal otherwise inaccessible information about the composition, stratification, and collective dynamics of ultradense matter~\cite{Lai:1993di,Flanagan:2007ix,Hinderer:2016eia,Xu:2017hqo,Yu:2017cxe,LIGOScientific:2017vwq,Miao:2023jqe,HegadeKR:2026iou,Counsell:2025hcv,HegadeKR:2024slr,Lau:2020bfq,Constantinou:2021hba}.

In Newtonian stellar perturbation theory, the mode expansion is built on a self-adjoint eigenvalue problem whose eigenfunctions are orthogonal and, under suitable conditions, complete~\cite{Chandrasekhar1964}. This structure allows the tidal force to be projected onto independent oscillator equations for individual modes. No straightforward analogue exists for a radiating relativistic star. The external tide must first be defined in a gauge-consistent manner within the perturbed spacetime; radiative boundary conditions render the mode problem non-Hermitian, so the standard notions of orthogonality and completeness no longer apply; more importantly, the full response contains nonmodal prompt and tail contributions in addition to quasinormal modes. Consequently, although the relativistic perturbation equations can be evolved numerically for a prescribed external source, a causal analytical framework that decomposes the coupled spacetime-fluid response into separately driven modal and nonmodal sectors has remained unavailable.

Several recent theoretical developments now make such a framework possible. First, the retarded Green function for perturbations of a Schwarzschild black hole can be decomposed into a direct contribution, a quasinormal-mode sector, and a late-time tail~\cite{Su:2026fvj}. In the construction used here, the direct response is generated by branch-cut integrals in the complex-frequency plane, rather than by the technically delicate large-arc contribution appearing in Leaver's original contour prescription~\cite{Leaver:1986gd}. Second, complex bilinear forms for radiative perturbations can be used to define quasinormal-mode orthogonality, normalization, and source projection~\cite{Leung:1997was,Zimmerman:2014aha,Mark:2014aja,London:2020uva,Green:2022htq,Ma:2024qcv,Lestingi:2026peq,Minucci:2026dgo}. Third, tidal driving of a relativistic star can be mapped onto an equivalent gravitational-wave scattering problem, allowing the resonant energy deposited as a binary inspiral sweeps through a stellar mode \cite{Gualtieri:2001cm,Pons:2001xs,Berti:2002ry} to be read from the analytic structure of the scattering response~\cite{Miao:2025utd}. Together, these advances provide the essential ingredients for constructing a causal relativistic theory of stellar response under generic time-dependent tides~\cite{Kokkotas:1999bd}.

In this work we construct such a framework for linear, polar perturbations of a nonrotating perfect-fluid star.  We formulate tidal driving as a causal scattering problem (see Fig.~\ref{fig:scatter} for illustration), identify different components of the retarded Green function of the coupled interior--exterior system based on the spectral decomposition method, and deform its inverse-frequency contour into pole and branch-cut sectors. Physically the star response contains four distinct types of signals: the quasinormal mode (QNM), the tail, the surface-reflected wave and the prompt wave. In particular, the prompt wave coincides with that of a Schwarzschild black hole of the same mass, whereas the surface-reflected wave starts when the gravitational wave emitted from the source bounces off the star surface and receives by the observer. In this sense, the surface-reflected wave also encodes information about the star interior. We define the ``direct part'' as a combination of both the prompt wave and the surface-reflected wave, following the terminology in~\cite{Su:2026fvj}.  
In addition, we utilize a complex bilinear form on the gravitational and fluid phase space that supplies the normalization and source projection for the stellar quasinormal modes. The same source also projects onto a continuum of branch-cut states, producing the direct and late-time responses.  This gives first-order evolution equations for the dynamics of a discrete and a continuous-frequency set of oscillators, as a complete representation of the spacetime and fluid perturbations.

The paper is organized as follows.  Section~\ref{sec:theoframe} introduces the tidal scattering problem, the stellar Green function, and the bilinear projection.  Section~\ref{sec:wave-decomposition} derives the waveform decomposition and summarizes its numerical implementation and validation. Section~\ref{sec:conclusion} discusses the physical interpretation, limitations, and applications. We use geometrized units $G=c=1$ and metric signature $(-,+,+,+)$. Our Fourier convention is $e^{-i\omega t}$; time-domain perturbations are not assumed to be monochromatic unless otherwise stated.

\section{Theoretical Framework}\label{sec:theoframe}

In this section, we develop the main theoretical framework for decomposing the metric and fluid perturbations of a star under linear external tidal driving. This includes a discussion about how this tidally driven star can be mapped to a scattering scenario (Sec.~\ref{sec:scattering-formalism}), with ingoing gravitational waves acting as an an effective tidal source for various parts of the star's response. We also explicitly decompose star's Green function using the spectral method (Sec.~\ref{sec:stellar-green-function}), which can be physically identified as the quasinormal mode, the tail, the surface-reflected wave and the prompt wave. In Sec.~\ref{sec:bilinear-form} we describe the bilinear form used for projecting the source into the effective driving force for each piece of signals.

\subsection{Scattering formalism for tidal driving}\label{sec:scattering-formalism}

\begin{figure}[t]
 \centering
 \includegraphics[width=\columnwidth]{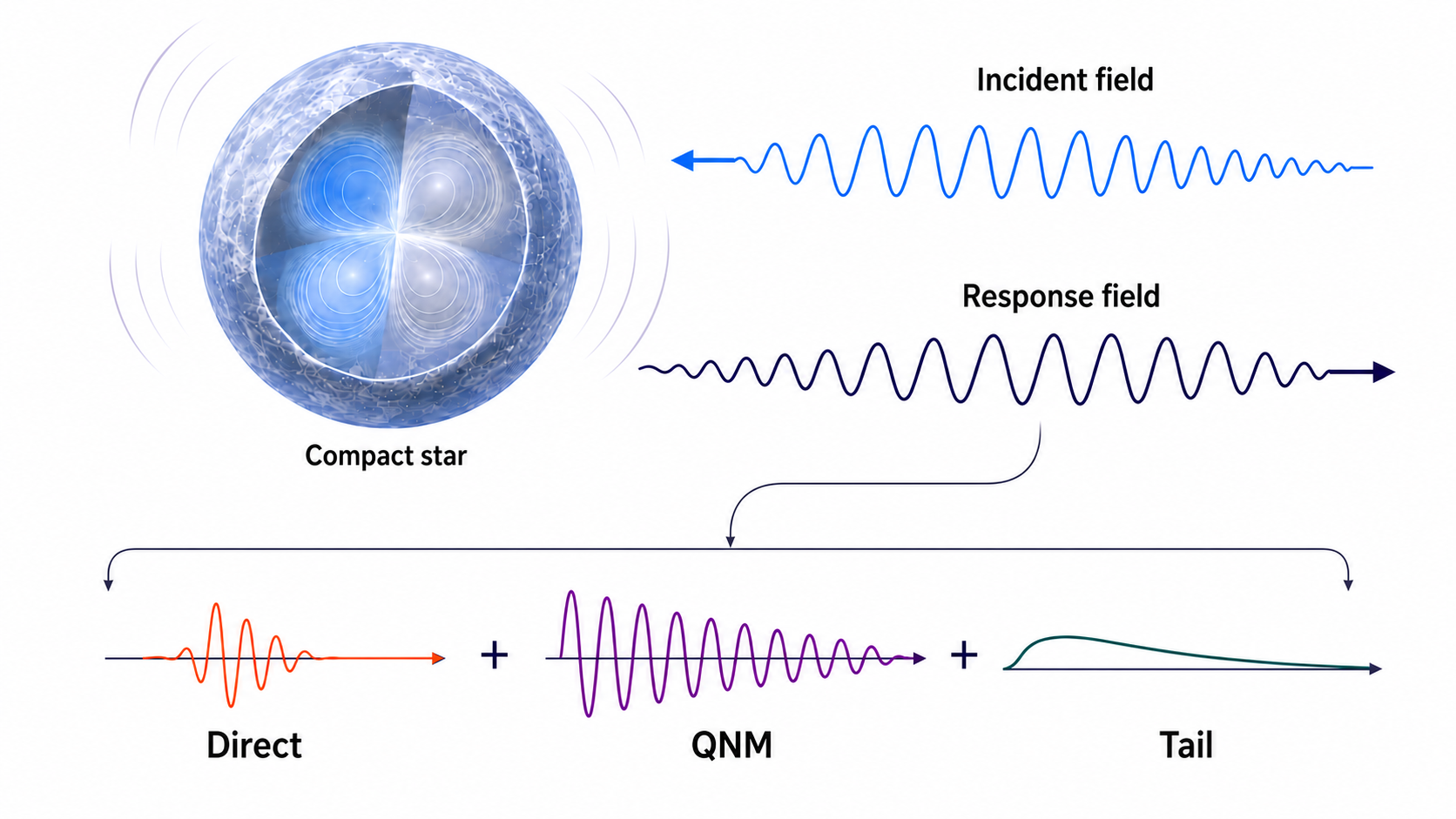}
 \caption{Scattering description of the tidal driving and stellar response.}
 \label{fig:scatter}
\end{figure}

We describe tidal driving as a scattering problem~\cite{Creci:2021rkz,Miao:2025utd}, in which a gravitational wave is injected from past null infinity $\mathscr I^-$ and scattered by the star toward future null infinity $\mathscr I^+$, as illustrated in Fig.~\ref{fig:scatter}. Consider a nonrotating perfect-fluid star of mass $M$ and radius $R$, its unperturbed exterior is described by the Schwarzschild geometry, where an even-parity perturbation with spherical harmonic indices $(\ell,m)$ can be represented by the Zerilli master variable $Z_{\ell m}$~\cite{Zerilli:1970se,Zerilli:1970wzz}.
For a Fourier component of the solution matched to a center-regular stellar interior, the asymptotic form is
\begin{equation}
Z_{\ell m\omega}(r)=
A_{\ell m}^{\rm out}(\omega)e^{i\omega r_*}
+
A_{\ell m}^{\rm in}(\omega)e^{-i\omega r_*}\,,\quad r\rightarrow\infty\,,
\label{eq:scattering-asymptotics}
\end{equation}
where
\begin{equation}
r_*=r+2M\ln\left(\frac{r}{2M}-1\right)
\end{equation}
is the Schwarzschild tortoise coordinate. Restoring the time dependence gives
\begin{equation}
Z_{\ell m\omega}(r)e^{-i\omega t}
=
A_{\ell m}^{\rm out}(\omega)e^{-i\omega u}
+
A_{\ell m}^{\rm in}(\omega)e^{-i\omega v}\,,
\end{equation}
where $u=t-r_*$ and $v=t+r_*$. 

As shown in Ref.~\cite{Miao:2025utd}, this scattering scenario is equivalent, from the viewpoint of the star, to tidal driving by an orbiting point mass. For a point mass on a circular orbit of radius $r_0$, the region between the stellar surface and the orbit, $R<r<r_0$, is a vacuum region. The Zerilli function there can therefore be written as
\begin{equation}
Z_{\ell m\omega}^{\rm pp}(r)
=
A_{\ell m}^{\rm out}Z_{\ell m\omega}^{\rm out}(r)
+
A_{\ell m}^{\rm in}Z_{\ell m\omega}^{\rm in}(r)\,,
\end{equation}
with $\omega=m\Omega$, where $\Omega$ is the orbital frequency and the superscript pp refers to ``point particle''. Here $Z_{\ell m\omega}^{\rm out}$ and $Z_{\ell m\omega}^{\rm in}$ are the homogeneous solutions satisfying $Z_{\ell m\omega}^{\rm out}\sim e^{i\omega r_*}$ and $Z_{\ell m\omega}^{\rm in}\sim e^{-i\omega r_*}$ as $r_*\to \infty$.
The incoming and outgoing amplitudes in the scattering problem can therefore be chosen so that the resulting field agrees with the point-particle solution throughout $R<r<r_0$. Matching this common exterior solution to the stellar interior then produces identical metric and fluid perturbations inside the star. Thus, although the two descriptions obey different boundary conditions outside the orbital radius, they are indistinguishable from the viewpoint of the star.

The ingoing amplitude $A_{\ell m}^{\rm in}$ characterizes the external tidal field, while the stellar response is encoded in the scattering coefficient
\begin{equation}
\mathcal R_\ell(\omega)
=
\frac{A_{\ell m}^{\rm out}(\omega)}
     {A_{\ell m}^{\rm in}(\omega)}\,.
\label{eq:reflection-coefficient}
\end{equation}
This framework provides an operational definition of the external tidal field and allows the energy deposited into a transiently excited mode to be determined from the residue of $\mathcal R_\ell$ at the corresponding mode frequency~\cite{Miao:2025utd}.

The frequency-domain scattering description above applies directly to monochromatic driving. In this work, we generalize it to an arbitrary time-dependent incident waveform and focus on the resulting response in the time domain. To this end, we denote the complete perturbation by a phase-space vector $\boldsymbol{\Psi}_{\ell m}(t,r)$ and decompose it as
\begin{equation}
\boldsymbol{\Psi}_{\ell m}
=
\boldsymbol{\Psi}_{\ell m}^{\rm d}
+
\boldsymbol{\Psi}_{\ell m}^{\rm r}\,,
\label{eq:driver-response-split}
\end{equation}
where $\boldsymbol{\Psi}_{\ell m}^{\rm d}$ is a prescribed incident field carrying the time-dependent driving waveform, while $\boldsymbol{\Psi}_{\ell m}^{\rm r}$ is the response generated by the star. 

The phase-space vector $\boldsymbol{\Psi}_{\ell m}$ is defined on a Cauchy slice of the complete interior--exterior matched spacetime. In the stellar interior, it contains both metric and fluid perturbation variables.
In Appendix~\ref{sec:hamiltonian-appendix}, we construct an explicit representation of this phase space by taking the second variation of the Einstein--Hilbert action coupled to Brown's action~\cite{Brown:1992kc} for a perfect fluid and casting the resulting theory into ADM Hamiltonian form~\cite{Arnowitt:1962hi}.
After imposing the linearized constraints, the phase-space vector takes the canonical form
\begin{equation}
\boldsymbol{\Psi}_{\ell m}
=
\begin{pmatrix}
\boldsymbol Q_{\ell m}\\
\boldsymbol P_{\ell m}
\end{pmatrix}\,,
\label{eq:full-phase-space}
\end{equation}
where, in Regge-Wheeler gauge~\cite{Regge:1957td} adopted in Appendix~\ref{sec:hamiltonian-appendix},
\begin{equation}
\boldsymbol Q_{\ell m}
=
\begin{pmatrix}
H_{2,\ell m}\\
K_{\ell m}\\
W_{\ell m}\\
V_{\ell m}
\end{pmatrix}\,,
\quad
\boldsymbol P_{\ell m}
=
\begin{pmatrix}
P_{H_2,\ell m}\\
P_{K,\ell m}\\
P_{W,\ell m}\\
P_{V,\ell m}
\end{pmatrix}\,.
\label{eq:interior-canonical-variables}
\end{equation}
Here $H_2$ and $K$ are the even-parity metric perturbation
amplitudes, while $W$ and $V$ describe the fluid perturbations and have support only inside the star. In the Schwarzschild exterior, where the fluid perturbations vanish, the remaining gravitational phase space can be reduced to the Zerilli canonical pair,
\begin{equation}
\boldsymbol{\Psi}_{\ell m}
\longrightarrow
\begin{pmatrix}
Z_{\ell m}\\
\Pi_{Z,\ell m}
\end{pmatrix}\,,
\quad
\Pi_{Z,\ell m}
=
C_\ell\dot Z_{\ell m}\,,
\label{eq:exterior-zerilli-canonical-pair}
\end{equation}
with
\begin{equation}
C_\ell
=
\frac{1}{64\pi}
\frac{(\ell+2)!}{(\ell-2)!}\,.
\label{eq:zerilli-canonical-normalization}
\end{equation}

The linear perturbations satisfy the Hamiltonian evolution equation
\begin{equation}
\partial_t\boldsymbol{\Psi}_{\ell m}
=
\mathbb A_\ell\boldsymbol{\Psi}_{\ell m}\,,
\label{eq:hamiltonian-evolution}
\end{equation}
where the explicit construction of $\mathbb A_\ell$ is given in Appendix~\ref{sec:hamiltonian-appendix}. 
Substituting Eq.~\eqref{eq:driver-response-split} into Eq.~\eqref{eq:hamiltonian-evolution}, we find that the response satisfies
\begin{equation}
\left(\partial_t-\mathbb A_\ell\right)
\boldsymbol{\Psi}_{\ell m}^{\rm r}
=
\boldsymbol S_{\ell m}^{\rm d}\,,
\label{eq:driven-response}
\end{equation}
where
\begin{equation}
\boldsymbol S_{\ell m}^{\rm d}
:=
-\left(\partial_t-\mathbb A_\ell\right)
\boldsymbol{\Psi}_{\ell m}^{\rm d}\,.
\label{eq:effective-driving-source}
\end{equation}
Thus, the prescribed incident field $\boldsymbol{\Psi}_{\ell m}^{\rm d}$ determines an effective source for the response field $\boldsymbol{\Psi}_{\ell m}^{\rm r}$.

We may choose $\boldsymbol{\Psi}_{\ell m}^{\rm d}$ to carry the prescribed incoming radiation and satisfy a purely ingoing boundary condition at infinity, while
$\boldsymbol{\Psi}_{\ell m}^{\rm r}$ satisfies a purely outgoing boundary condition there.
This split is not unique in the finite-radius region: different continuations of the same incident data lead to different effective sources and response fields.
Nevertheless, for fixed physical initial and boundary conditions, the total field $\boldsymbol{\Psi}_{\ell m}$ is unique.
Consequently, the response field at infinity is likewise unique and independent of this freedom in the decomposition.

In practice, it is convenient to choose $\boldsymbol{\Psi}_{\ell m}^{\rm d}$ to have support only in the stellar exterior and to vanish at the stellar surface. The stellar interior is then driven entirely through the response field and the matching conditions at $r=R$. In the Schwarzschild exterior, Eq.~\eqref{eq:driven-response} reduces to the sourced Zerilli equation
\begin{equation}
\left(
-\partial_t^2+\partial_{r_*}^2-V_\ell^Z
\right)Z_{\ell m}^{\rm r}
=
S_{\ell m}^{Z,{\rm d}}\,,
\label{eq:sourced-zerilli-response}
\end{equation}
where the effective source is determined by the prescribed exterior driver:
\begin{equation}
S_{\ell m}^{Z,{\rm d}}
=
\left(
\partial_t^2-\partial_{r_*}^2+V_\ell^Z
\right)Z_{\ell m}^{\rm d}\,.
\label{eq:zerilli-effective-source}
\end{equation}
The incident field at past null infinity can be specified as
\begin{equation}
Z_{\ell m}^{\rm d}\big|_{\mathscr I^-}
=
\mathcal A_{\ell m}(v)e^{-i\Phi_{\ell m}(v)}\,,
\quad
\omega_{\ell m}(v)
=
\frac{d\Phi_{\ell m}}{dv}\,.
\label{eq:time-dependent-driver}
\end{equation}
and then extended into the finite-radius exterior through a
chosen continuation.

\subsection{Green's function for stars}\label{sec:stellar-green-function}
In this section, we introduce the Green's function of the Zerilli equation for relativistic stars and generalize the Green-function decomposition and contour-construction strategy developed for black holes in Ref.~\cite{Su:2026fvj} to the stellar case.

\subsubsection{Frequency domain Green's function and its analytical structure}
In the frequency domain, the Green's function {is defined as}
\begin{equation}
\left[
\frac{d^2}{dr_*^2}
+\omega^2
-V_\ell^Z(r)
\right]
\widetilde G_\ell(\omega;r,r')
=
\delta(r_*-r_*')\,,
\label{eq:green-definition}
\end{equation}
where the $r_*$ is the tortoise coordinate of the observer and $r_*'$ is that of the source point. $V_\ell^Z$ is the Zerilli potential. 
It is specified by two physically motivated boundary conditions. At spatial infinity, it must be purely outgoing. At the stellar surface, it must satisfy the junction conditions with a solution of the coupled metric--fluid perturbation equations that is regular at the stellar center. We denote the radial profile of this complete center-regular solution by $\boldsymbol{\psi}_{\ell m\omega}^{\rm reg}(r)$ and its exterior Zerilli component by $Z_{\ell m\omega}^{\rm reg}(r)$. In the Schwarzschild exterior, the latter can be expanded in the incoming and outgoing solutions introduced in Sec.~\ref{sec:scattering-formalism}:
\begin{equation}
Z_{\ell m\omega}^{\rm reg}(r)
=
A_{\ell m}^{\rm in}(\omega)
Z_{\ell m\omega}^{\rm in}(r)
+
A_{\ell m}^{\rm out}(\omega)
Z_{\ell m\omega}^{\rm out}(r)\,.
\label{eq:center-regular-jost}
\end{equation}
The coefficients $A_{\ell m}^{\rm in/out}$ are those used in Eq.~\eqref{eq:reflection-coefficient}.

For $r>r'>R$, the frequency-domain Green's function satisfying these boundary conditions is
\begin{equation}
\widetilde G_\ell(\omega;r,r')
=
\frac{
Z_{\ell m\omega}^{\rm out}(r)
Z_{\ell m\omega}^{\rm reg}(r')
}{
W_\ell(\omega)
}\,,
\label{eq:stellar-green-rprime-greater}
\end{equation}
where the Wronskian is defined by
\begin{align}
W_\ell(\omega)
&:=
Z_{\ell m\omega}^{\rm reg}
\frac{dZ_{\ell m\omega}^{\rm out}}{dr_*}
-
Z_{\ell m\omega}^{\rm out}
\frac{dZ_{\ell m\omega}^{\rm reg}}{dr_*}
\nonumber\\
&=
2i\omega A_{\ell m}^{\rm in}(\omega).
\label{eq:stellar-wronskian}
\end{align}
The Wronskian is independent of the radial coordinate. Its zeros, equivalently the zeros of $A_{\ell m}^{\rm in}(\omega)$, determine the stellar QNM frequencies.

Using Eq.~\eqref{eq:center-regular-jost}, the Green's function can be decomposed into two contributions,
\begin{equation}
\widetilde G_\ell(\omega;r,r')
=
\widetilde G_\ell^-(\omega;r,r')
+
\widetilde G_\ell^+(\omega;r,r')\,,
\label{eq:Gpm-split}
\end{equation}
where
\begin{align}
\widetilde G_\ell^-(\omega;r,r')
&=
\frac{
Z_{\ell m\omega}^{\rm out}(r)
Z_{\ell m\omega}^{\rm in}(r')
}{
2i\omega
}\,,
\label{eq:Gminus}\\
\widetilde G_\ell^+(\omega;r,r')
&=
\mathcal R_\ell(\omega)
\frac{
Z_{\ell m\omega}^{\rm out}(r)
Z_{\ell m\omega}^{\rm out}(r')
}{
2i\omega
}\,.
\label{eq:Gplus}
\end{align}
Here $\mathcal R_\ell=A_{\ell m}^{\rm out}/A_{\ell m}^{\rm in}$ is the reflection coefficient introduced in Sec.~\ref{sec:scattering-formalism}.
Thus, $\widetilde G_\ell^-$ describes propagation in the Schwarzschild exterior and is independent of the stellar boundary condition, whereas all information about the star is encoded in $\widetilde G_\ell^+$. In particular, the stellar QNM frequencies appear as poles of $\widetilde G_\ell^+$, while $\widetilde G_\ell^-$ contains no stellar QNM poles.

Although Eqs.~\eqref{eq:Gpm-split}--\eqref{eq:Gplus} has the same algebraic form as the corresponding black-hole decomposition, the analytic structure of $\mathcal R_\ell(\omega)$ is qualitatively different. This difference arises from the nature of the inner boundary. For a black hole, the horizon acts as an absorbing boundary located at $r_*\rightarrow-\infty$. 
A star, by contrast, behaves as a finite cavity: an incident wave reaching the vicinity of the stellar surface is split into two components. One is promptly reflected at the surface, while the other enters the star, propagates through the coupled metric--fluid interior, and may reemerge into the exterior after one or more internal reflections~\footnote{Surface reflection has also been investigated in scalar-wave scattering by compact stars in
Ref.~\cite{OuldElHadj:2026ewh}, where an explicit separation of direct surface reflection from interior-transmission contributions is provided.}.

Motivated by this physical picture, we decompose the stellar reflection coefficient as
\begin{equation}
\mathcal R_\ell
=
\mathcal R_\ell^{\rm pr}
+
\mathcal R_\ell^{\rm cent}\,.
\label{eq:reflection-prompt-center}
\end{equation}
Here $\mathcal R_\ell^{\rm pr}$ describes the prompt reflection generated locally at the stellar surface, whereas $\mathcal R_\ell^{\rm cent}$ describes waves that enter the star, interact with the metric--fluid interior, and subsequently return to the exterior. Schematically, the two contributions can be written as
\begin{align}
\mathcal R_\ell^{\rm pr}
&=
e^{-2i\omega R_*}\rho_{\rm out}\,,
\label{eq:Rprompt-schematic}
\\
\mathcal R_\ell^{\rm cent}
&=
e^{-2i\omega R_*}
\boldsymbol{\tau}_{\rm out}^{T}
\mathbf C_\ell
\left[
\mathbf I-\boldsymbol{\rho}_{\rm in}
\mathbf C_\ell
\right]^{-1}
\boldsymbol{\tau}_{\rm in}\,,
\label{eq:Rcenter-schematic}
\end{align}
where $R_*=r_*(R)$. The quantity $\rho_{\rm out}$ is the local reflection amplitude for an exterior wave incident on the stellar surface, while $\boldsymbol{\tau}_{\rm in}$ and $\boldsymbol{\tau}_{\rm out}$ describe transmission into and out of the star, respectively. The matrix $\boldsymbol{\rho}_{\rm in}$ describes reflection at the surface from the interior side, while $\mathbf C_\ell$ describes propagation through the stellar interior, including reflection from the regular center. The vector and matrix structure accounts for the coupled metric--fluid propagation channels. Expanding the inverse in Eq.~\eqref{eq:Rcenter-schematic} gives the successive internal round trips of the wave.
In Appendix~\ref{sec:prompt-center-appendix}, we present a detailed analysis of scattering at the stellar surface and provide an explicit procedure for computing the prompt and center-return contributions. Below, we briefly summarize the analytic properties of $\mathcal R_\ell^{\rm pr}$ and $\mathcal R_\ell^{\rm cent}$ relevant to the contour analysis.

Equation~\eqref{eq:Rprompt-schematic} shows that $\mathcal R_\ell^{\rm pr}$ contains only the local surface reflection and is therefore free of global stellar QNM poles. It nevertheless inherits a branch cut along the positive imaginary axis (PIA) through its matching to the exterior Schwarzschild solutions. By contrast, the poles of $\mathcal R_\ell^{\rm cent}$ are determined by
\begin{equation}
\det\left[
\mathbf I-\boldsymbol{\rho}_{\rm in}(\omega)
\mathbf C_\ell(\omega)
\right]
=0\,,
\label{eq:stellar-resonance-condition}
\end{equation}
which yields the complete stellar QNM spectrum, including both fluid and spacetime modes. The center-return contribution $\mathcal R_\ell^{\rm cent}$ has no branch cut along the PIA.

In the large-$|\omega|$ limit, the two contributions behave as
\begin{equation}
\mathcal R_\ell^{\rm pr}
\sim
e^{-2i\omega R_*}\,,
\quad
\mathcal R_\ell^{\rm cent}
\sim
e^{-2i\omega(R_*-x_R)}\,,
\label{eq:Rpromp-Rcenter-asymptotic}
\end{equation}
up to prefactors that vary algebraically with $\omega$, where
\begin{equation}
x_R
=
\int_0^R \frac{A(r)}{N(r)}\,dr
\label{eq:interior-optical-distance}
\end{equation}
is the gravitational optical distance from the stellar center to its surface for the background metric
\begin{equation}
ds^2=-N^2dt^2+A^2dr^2+r^2d\Omega^2\,.
\end{equation}
The additional phase $e^{2i\omega x_R}$ in $\mathcal R_\ell^{\rm cent}$ represents propagation from the surface to the center and back.

\subsubsection{Time domain Green's function and the contour strategy}
\label{sec:time-domain-green-function}
\begin{figure*}[t]
\centering
\includegraphics[width=\textwidth]
{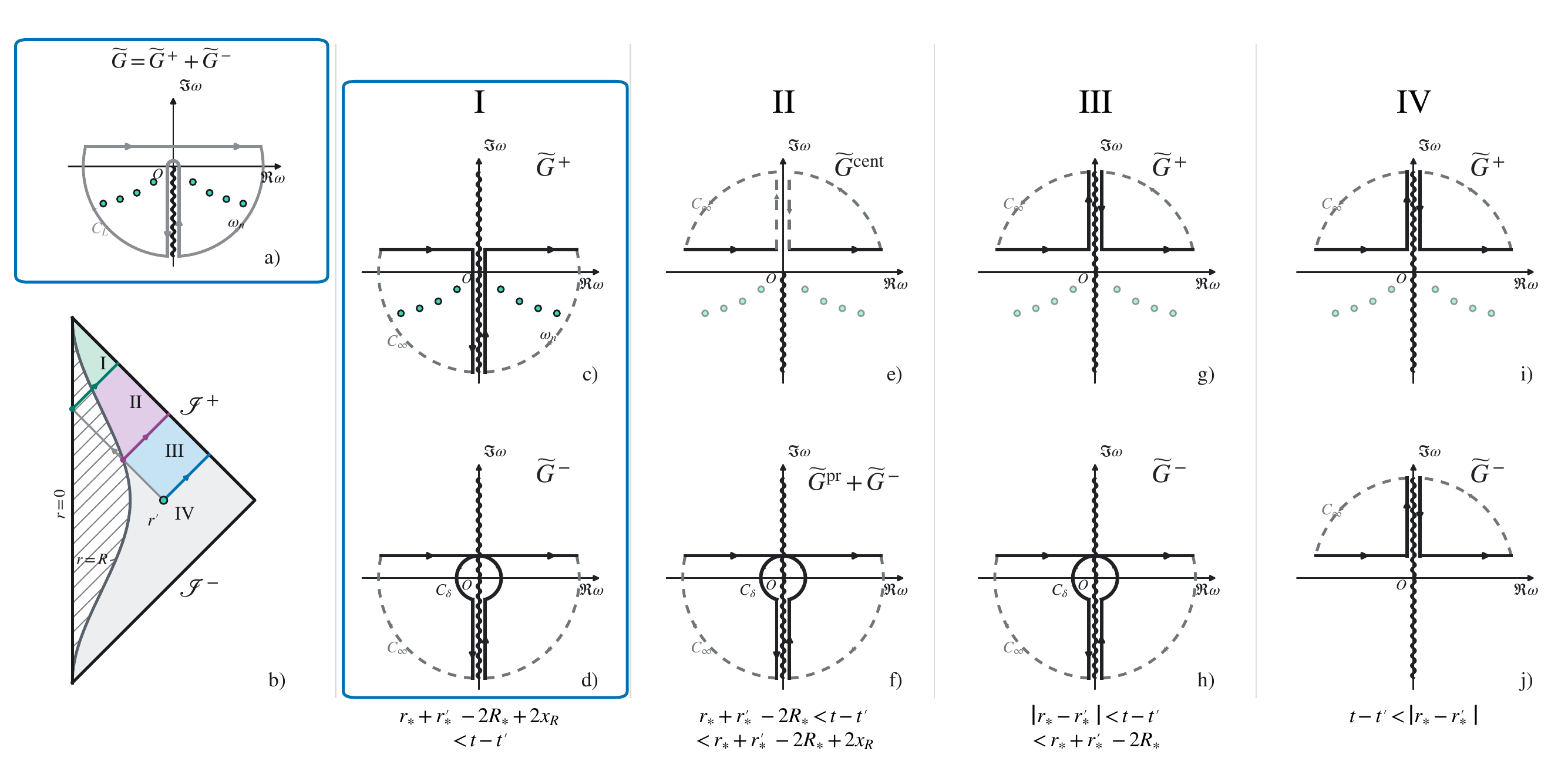}
\caption{Causal regions and frequency contours for the exterior retarded Green's function of a spherical star, with $r_*>r_*'>R_*$. Panel (a) shows the standard Leaver contour, and panel (b) shows the four causal regions in the Penrose diagrams. In Region I [panels (c) and (d)], the contours for both $\widetilde G^+$ and $\widetilde G^-$ close in the lower half-plane, reproducing the Leaver contour. In Region II [panels (e) and (f)], $\widetilde G^{\rm cent}$ closes above and vanishes {since it is analytic there}, whereas $\widetilde G^{\rm pr}+\widetilde G^-$ closes below. In Region III [panels (g) and (h)], $\widetilde G^+$ closes above and $\widetilde G^-$ below. In Region IV [panels (i) and (j)], both contours close above, and their contributions cancel with each other. The small circular detours in Regions II and III avoid the singular point at $\omega=0$; the dashed large arcs have vanishing contributions in their respective regions.}
\label{fig:stellar-green-contours}
\end{figure*}

We now obtain the time-domain Green's function by an inverse Fourier transform:
\begin{equation}
G_\ell(\tau;r,r')=
\frac{1}{2\pi}
\int_{-\infty}^{+\infty}
d\omega\,
\widetilde G_\ell(\omega;r,r')e^{-i\omega\tau},\quad
\tau=t-t'\,.
\label{eq:time-domain-green}
\end{equation}
The integral can be evaluated by deforming the contour in the complex-$\omega$ plane. The appropriate deformation depends on the causal relation between the source and observation points, as illustrated in Fig.~\ref{fig:stellar-green-contours}. The Penrose diagram in lower-left panel shows the corresponding causal structure, in which the exterior spacetime is divided into four regions, labeled I--IV.

{\textbf{Region I}}: $\tau>r_*+r_*'-2R_*+2x_R$. 
In this region, we use the standard Leaver contour~\cite{Leaver:1986gd}, closed in the lower half-plane. The branch cut along the negative imaginary axis (NIA) of the complete Green's function, $\widetilde G_\ell^++\widetilde G_\ell^-$, generates the late-time tail, while the enclosed poles give the QNM contribution. 
Causally, this region allows waves to enter the star through its surface, return from the center (possibly undergo further internal reflections), and emerge through the surface to reach the observer.

{\textbf{Region II}}: $r_*+r_*'-2R_*<\tau<r_*+r_*'-2R_*+2x_R$. 
The \emph{surface-reflected wave} has arrived, whereas the center-returned wave has not. It is therefore useful to further decompose
\begin{equation}
\widetilde G_\ell^+
=
\widetilde G_\ell^{\rm pr}
+
\widetilde G_\ell^{\rm cent}\,,
\end{equation}
according to the decomposition of $\mathcal{R}_\ell$ in Eq.~\eqref{eq:reflection-prompt-center}.
At large $|\omega|$, these two components behave schematically as
\begin{equation}
\widetilde G_\ell^{\rm pr}
\sim
e^{i\omega(r_*+r_*'-2R_*)}\,,\quad
\widetilde G_\ell^{\rm cent}
\sim
e^{i\omega(r_*+r_*'-2R_*+2x_R)}\,.
\end{equation}
We close the contours for $\widetilde G_\ell^-$ and $\widetilde G_\ell^{\rm pr}$ in the lower half-plane, and that for $\widetilde G_\ell^{\rm cent}$ in the upper half-plane, so that the Fourier factor suppresses their large-arc contributions. Since $\widetilde G_\ell^{\rm cent}$ is analytic in the upper half-plane and has no PIA branch cut, its contribution vanishes in this region, consistent with the absence of a center-returned signal. 

{\textbf{Region III}}: $r_*-r_*'<\tau<r_*+r_*'-2R_*$. 
This region contains the \emph{prompt wave} that has not yet interacted with the stellar surface. It is identical to that for a black hole of the same mass, owing to the shared Schwarzschild exterior geometry.
As in the black-hole case~\cite{Su:2026fvj}, we close the contour for $\widetilde G_\ell^+$ in the upper half-plane and that for $\widetilde G_\ell^-$ in the lower half-plane. The resulting Green's function consists of the PIA branch-cut contribution of $\widetilde G_\ell^+$ and the NIA branch-cut contribution of $\widetilde G_\ell^-$. A small circular detour is again required around the singular point at $\omega=0$. 

{\textbf{Region IV}}: $\tau<r_*-r_*'$. 
The observation point lies outside the causal future of the source. The contours for both $\widetilde G_\ell^+$ and $\widetilde G_\ell^-$ can be closed in the upper half-plane. Their PIA contributions cancel in the sum, and the retarded Green's function consequently vanishes.

Following these contour deformations, we decompose the retarded Green's function as
\begin{equation}
G_\ell
=
G_{\ell,\rm dir}
+
G_{\ell,\rm QNM}
+
G_{\ell,\rm tail}\,,
\label{eq:green-time-domain-decomposition}
\end{equation}
where we refer to the sum of the surface-reflected wave in
Region II and the prompt wave in Region III as the
\emph{direct part}:
\begin{equation}
G_{\ell,\rm dir}
=
G_{\ell,\rm II}+G_{\ell,\rm III}\,.
\end{equation}
For convenience, we define the causal windows
\begin{subequations}\label{eq:green-causal-windows}
\begin{align}
\chi_{\rm I}
&=\Theta\!\left(\tau-\tau_{\rm c}\right)\,,\\
\chi_{\rm II}
&=\Theta\!\left(\tau-\tau_{\rm s}\right)
  \Theta\!\left(\tau_{\rm c}-\tau\right)\,,\\
\chi_{\rm III}
&=\Theta\!\left(\tau-\tau_{\rm f}\right)
  \Theta\!\left(\tau_{\rm s}-\tau\right)\,,\\
\chi_{\rm IV}
&=\Theta\!\left(\tau_{\rm f}-\tau\right)\,.
\end{align}
\end{subequations}
where $\Theta$ is the Heaviside function, and
\begin{subequations}\label{eq:green-arrival-times}
\begin{align}
\tau_{\rm f}&=r_*-r_*'\,,\\
\tau_{\rm s}&=r_*+r_*'-2R_*\,,\\
\tau_{\rm c}&=r_*+r_*'-2R_*+2x_R\,.
\end{align}
\end{subequations}
%%%
%%%
With spatial arguments suppressed in the frequency-domain Green's functions, the different contributions are given by
\begin{align}
G_{\ell,\rm QNM}
&=-\chi_{\rm I}\sum_n
\frac{A_{\ell m}^{\rm out}(\omega_n)}
{2\omega_n\,\partial_\omega A_{\ell m}^{\rm in}(\omega_n)}
\nonumber\\
&\hspace{1.1cm}\times
Z_{\ell m\omega_n}^{\rm out}(r)
Z_{\ell m\omega_n}^{\rm out}(r')
e^{-i\omega_n\tau}\,,
\label{eq:green-qnm-residue-sum}
\end{align}

\begin{align}
G_{\ell,\rm tail}
&=\frac{\chi_{\rm I}}{\pi}
\int_0^\infty\mathrm d\sigma\,
\frac{\sigma q_\ell^Z(\sigma)e^{-\sigma\tau}}
{|W_\ell(-i\sigma+0)|^2}
\nonumber\\
&\hspace{1.1cm}\times
Z_{\ell m,-i\sigma}^{\rm reg}(r)
Z_{\ell m,-i\sigma}^{\rm reg}(r')\,,
\label{eq:green-tail-cut-integral}
\end{align}

\begin{align}
G_{\ell,\rm II}
={}&\frac{\chi_{\rm II}}{\pi}\Bigg\{
\int_\delta^\infty\mathrm d\sigma\,e^{-\sigma\tau}
\nonumber\\
&\quad\times\Im\left[
\widetilde G_\ell^-(\epsilon-i\sigma)
+\widetilde G_\ell^{\rm pr}(\epsilon-i\sigma)\right]
\nonumber\\
&+\delta\,\Im
\int_{-\pi/2}^{\pi/2}\mathrm d\phi\,
e^{i\phi-i\delta e^{i\phi}\tau}
\nonumber\\
&\quad\times\left[
\widetilde G_\ell^-(\delta e^{i\phi})
+\widetilde G_\ell^{\rm pr}(\delta e^{i\phi})\right]
\Bigg\}\,,
\label{eq:green-direct-region-II}
\end{align}
and 
\begin{align}
G_{\ell,\rm III}
={}&\frac{\chi_{\rm III}}{\pi}\Bigg\{
-\int_\delta^\infty\mathrm d\sigma\,e^{\sigma\tau}
\Im\left[\widetilde G_\ell^+(\epsilon+i\sigma)\right]
\nonumber\\
&+\int_\delta^\infty\mathrm d\sigma\,e^{-\sigma\tau}
\Im\left[\widetilde G_\ell^-(\epsilon-i\sigma)\right]
\nonumber\\
&+\delta\,\Im
\int_{-\pi/2}^{\pi/2}\mathrm d\phi\,
e^{i\phi-i\delta e^{i\phi}\tau}
\widetilde G_\ell^-(\delta e^{i\phi})
\Bigg\}\,.
\label{eq:green-direct-region-III}
\end{align}

We compute the radial solutions $Z_{\ell m\omega}(r)$ using a high-performance {\sc Julia}-based implementation of the MST formalism~\cite{Mano:1996vt,Mano:1996mf}, as developed in~\cite{Su:2026fvj}.
In numerical evaluations, we set $\epsilon=10^{-80}$. The parameter $\delta$ is the radius of the small circle around the origin, chosen not to enclose any nonzero poles. 
The radius-independent Zerilli branch-cut strength can be evaluated at any exterior radius $r>R$ as~\cite{Casals_2012,Casals_2013}
\begin{equation}
q_\ell^Z(\sigma)
=
\frac{
2\Im\left[Z_{\ell m,-i\sigma+0}^{\rm out}(r)\right]
}{ Z_{\ell m,i\sigma}^{\rm out}(r) },
\quad \sigma>0,
\label{eq:zerilli-branch-cut-strength-evaluation}
\end{equation}
where $-i\sigma+0$ denotes the limit from the right of the NIA.

\subsubsection{Numerical validation}
\label{sec:green-numerical-validation}

To validate the Green-function construction, we compare the contour decomposition with an independent time-domain evolution of the coupled metric--fluid Hamiltonian equations. We cast these equations in the AAKS variables~\cite{Allen:1997xj}, evolving two metric fields together with the fluid enthalpy perturbation. The evolution covers both the stellar interior and Schwarzschild exterior, with regularity imposed at the center and matching enforced at the stellar surface. We reconstruct the exterior Zerilli field from the evolved metric fields. Details evolving equations are given in Appendix~\ref{sec:aaks-appendix}~\&~\ref{sec:time-domain-validation-appendix}. 

As a representative example, we consider the $\ell=2$ perturbation of a star described by the energy-polytrope equation of state
\begin{equation}
p=K\varepsilon^2,
\label{eq:validation-eos}
\end{equation}
where $p$ and $\varepsilon$ denote the pressure and energy density, respectively. 
To facilitate comparison with our previous study, throughout this work, we adopt the stellar model with
$K=86\,M_\odot^2$ and central energy density
$\varepsilon_c=1.36\times10^{-3}\,M_\odot^{-2}$~\cite{Miao:2025utd}.
The corresponding stellar mass and radius are 
$M=1.4\,M_\odot$ and $R=13.84\,{\rm km}$, respectively.

We place the source at $r'=2R$ and extract the response at $r=3R$. Starting from vanishing perturbations, we approximate a unit spacetime impulse by
\begin{align}
\delta(t-t',r_*-r_*')
\approx{}&
\frac{1}{2\pi\sigma_t\sigma_{r_*}}
\exp\left[
-\frac{(t-t')^2}{2\sigma_t^2}
-\frac{(r_*-r_*')^2}{2\sigma_{r_*}^2}
\right]\,.
\label{eq:green-validation-source}
\end{align}
In the simulation, we set ${\sigma_t}={\sigma_{r_*}}= 0.0715\,M$ and $t'=5\sigma_t$.
The computational domain spans $r\in[0,95.6]\,M$ and is discretized uniformly in the optical coordinate $x$, with
${\Delta x}\simeq1.789\times10^{-3}\,M$ and $\Delta t\simeq1.431\times10^{-3}\,M$,
where
\begin{equation}
x(r)=
\begin{cases}
\displaystyle
\int_0^r\frac{A(\bar r)}{N(\bar r)}\,d\bar r\,,
& r\leq R\,,\\
x_R+r_*(r)-R_*\,,
& r\geq R\,.
\end{cases}
\label{eq:numerical-optical-coordinate}
\end{equation}
Thus, the full computational domain contains $58\,665$ grid points, of which $5\,659$ cover the stellar interior.

\begin{figure}[t]
\centering
\includegraphics[width=\columnwidth]{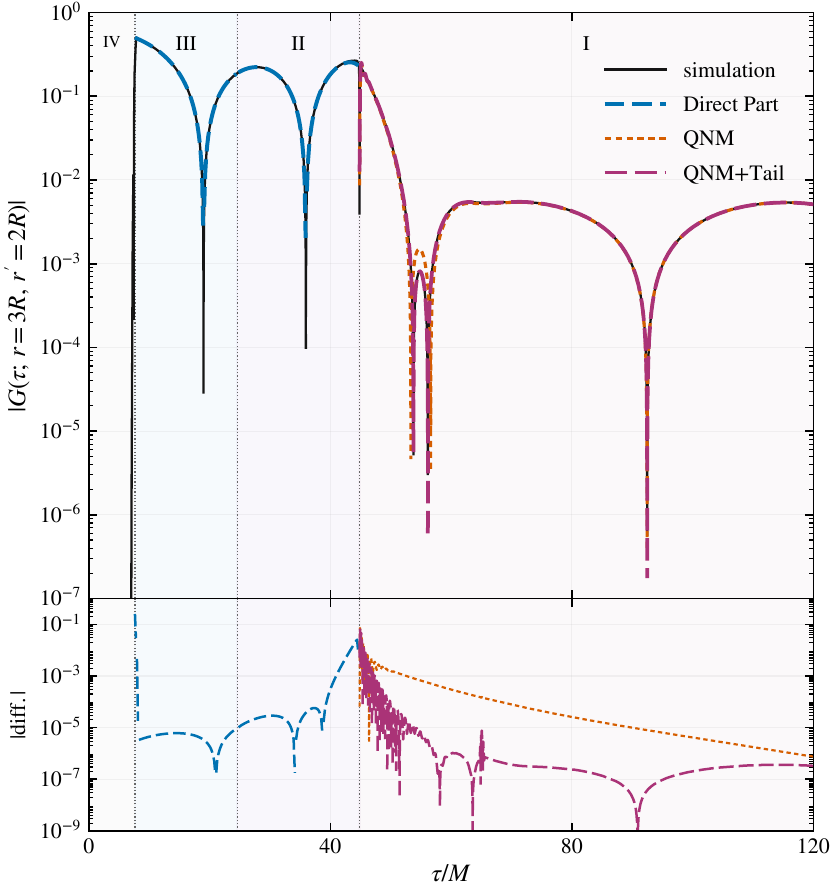}
\caption{Numerical validation of the $\ell =2$ Green function for a source at $r'=2R$ and an observer at $r=3R$. The black curve is obtained from an independent time-domain evolution driven by the narrow Gaussian source in Eq.~\eqref{eq:green-validation-source}. The colored curves show the contour calculation, including the direct contribution in Regions III and II and the QNM and QNM-plus-tail contributions in Region I. The vertical dotted lines mark the times $\tau_{\rm f}=r_*-r_*^\prime$, $\tau_{\rm s}=r_*+r_*^\prime-2R_*$ and $\tau_{\rm c}=r_*+r_*^\prime-2R_*+2x_R$.}
\label{fig:green-numerical-validation}
\end{figure}

Figure~\ref{fig:green-numerical-validation} shows the comparison between the contour construction with the independent time-domain evolution. The QNM contribution includes the $n=0,\ldots,31$ $w$-modes~\footnote{In our labeling, $w_0$ belongs to the $w_{\rm II}$ family (interface modes), whereas $w_1,\ldots,w_{31}$ belong to the $w_{\rm I}$ family (curvature modes), following the terminology of Ref.~\cite{Kokkotas:1999bd}. No additional $w_{\rm II}$ modes were found in our numerical search for the stellar model considered here.}, together with the contributions from the $f$, $p_1$, $p_2$, and $p_3$ modes. As shown in the figure, the Green function obtained from the contour calculation agrees excellently with the numerical simulation in all four causal regions.

\subsection{Bilinear form}
\label{sec:bilinear-form}

For a Newtonian star, the normal-mode operator is self-adjoint with respect to a positive-definite inner product, which provides both mode orthogonality and the projection of an external force onto individual modes~\cite{Chandrasekhar1964}. The corresponding construction is less direct for a relativistic star. Radiative boundary conditions render the QNM problem non-Hermitian, and the QNMs alone do not form a complete basis for the response~\cite{Kokkotas:1999bd,Berti:2025hly}. Nevertheless, black-hole QNMs have been shown to be orthogonal with respect to suitable choices of complex bilinear forms~\cite{Ma:2024qcv,Lestingi:2026peq,Minucci:2026dgo}. Here we extend the construction in~\cite{Minucci:2026dgo} to the complete metric--fluid phase space of a relativistic star. The resulting bilinear form provides the normalization and source projection of individual stellar QNMs. The same structure applies to the continuum states associated with the branch cuts of the Green's function, which generate the direct and late-time-tail responses.

With the harmonic indices suppressed, we write the Hamiltonian
evolution equation as
\begin{equation}
\mathcal O\boldsymbol\Psi=0\,,
\qquad
\mathcal O:=\partial_t-\mathbb A\,.
\label{eq:bilinear-homogeneous-equation}
\end{equation}
The Hamiltonian equations satisfy the local symplectic Green identity
\begin{align}
&
(\mathcal O\boldsymbol\Psi_1)^T
\mathbb J\boldsymbol\Psi_2
+
\boldsymbol\Psi_1^T\mathbb J
(\mathcal O\boldsymbol\Psi_2)
\nonumber\\
&\qquad=
\partial_t\varpi^t
[\boldsymbol\Psi_1,\boldsymbol\Psi_2]
+
\partial_r\varpi^r
[\boldsymbol\Psi_1,\boldsymbol\Psi_2]\,,
\label{eq:local-symplectic-green-identity}
\end{align}
where
\begin{equation}
\mathbb J=
\begin{pmatrix}
0&\mathbf I\\
-\mathbf I&0
\end{pmatrix}
\label{eq:canonical-symplectic-matrix}
\end{equation}
is the canonical symplectic matrix and $\varpi^{t,r}$ are the angularly integrated symplectic-current densities. For $\boldsymbol\Psi=(\boldsymbol Q,\boldsymbol P)^T$, the temporal component is
\begin{equation}
\varpi^t[\boldsymbol\Psi_1,\boldsymbol\Psi_2]
=
\boldsymbol\Psi_1^T\mathbb J\boldsymbol\Psi_2
=
\boldsymbol Q_1^T\boldsymbol P_2
-
\boldsymbol P_1^T\boldsymbol Q_2\,.
\label{eq:symplectic-current-temporal}
\end{equation}
The derivation of Eq.~\eqref{eq:local-symplectic-green-identity} and the explicit radial current are given in Appendix~\ref{sec:symplectic-appendix}.

We first define the symplectic form on a constant-time slice,
\begin{equation}
\Omega
(\boldsymbol\Psi_1,\boldsymbol\Psi_2)
:=
\int_{\mathcal C}\mathrm dr\,
\varpi^t[\boldsymbol\Psi_1,\boldsymbol\Psi_2]\,,
\label{eq:symplectic-form-contour}
\end{equation}
where $\mathcal C$ is a radial integration contour and the angular
dependence has already been integrated out.
For two solutions of Eq.~\eqref{eq:bilinear-homogeneous-equation}, integration of Eq.~\eqref{eq:local-symplectic-green-identity} gives
\begin{equation}
\frac{\mathrm d}{\mathrm dt}
\Omega
(\boldsymbol\Psi_1,\boldsymbol\Psi_2)
= -\left.\varpi^r[\boldsymbol\Psi_1,\boldsymbol\Psi_2]
\right|_{\partial\mathcal C}\,.
\label{eq:symplectic-flux-balance}
\end{equation}
Thus, $\Omega$ is generally not conserved when the
symplectic current has a nonvanishing flux at the boundary $\partial \mathcal C$. To incorporate
this flux, we introduce
\begin{equation}
\mathcal F_{\partial\mathcal C}
(\boldsymbol\Psi_1,\boldsymbol\Psi_2)
:=\int^t\mathrm dt'\,
\left.\varpi^r
[\boldsymbol\Psi_1(t'),\boldsymbol\Psi_2(t')]
\right|_{\partial\mathcal C}\,,
\label{eq:boundary-flux-primitive}
\end{equation}
and define the extended symplectic form
\begin{equation}
\widetilde\Omega
(\boldsymbol\Psi_1,\boldsymbol\Psi_2)
:=\Omega(\boldsymbol\Psi_1,\boldsymbol\Psi_2)
+\mathcal F_{\partial\mathcal C}
(\boldsymbol\Psi_1,\boldsymbol\Psi_2)\,.
\label{eq:extended-symplectic-form}
\end{equation}

Using Eqs.~\eqref{eq:bilinear-homogeneous-equation}
and~\eqref{eq:symplectic-flux-balance}, we obtain
\begin{equation}
    \Omega
    (\mathbb A\boldsymbol\Psi_1,\boldsymbol\Psi_2)
    +\Omega
    (\boldsymbol\Psi_1,\mathbb A\boldsymbol\Psi_2)
    =-\left.
    \varpi^r[\boldsymbol\Psi_1,\boldsymbol\Psi_2]
    \right|_{\partial\mathcal C}\,,
    \label{eq:bulk-generator-identity}
\end{equation}
Similarly, the action of $\mathbb A$ on the boundary term gives
\begin{equation}
    \mathcal F_{\partial\mathcal C}
    (\mathbb A\boldsymbol\Psi_1,\boldsymbol\Psi_2)
    +
    \mathcal F_{\partial\mathcal C}
    (\boldsymbol\Psi_1,\mathbb A\boldsymbol\Psi_2)
    =\left.
    \varpi^r[\boldsymbol\Psi_1,\boldsymbol\Psi_2]
    \right|_{\partial\mathcal C}\,.
    \label{eq:flux-generator-identity}
\end{equation}
Combining Eqs.~\eqref{eq:bulk-generator-identity}
and~\eqref{eq:flux-generator-identity}, we obtain
\begin{equation}
\widetilde\Omega
(\mathbb A\boldsymbol\Psi_1,\boldsymbol\Psi_2)
=
-\widetilde\Omega
(\boldsymbol\Psi_1,\mathbb A\boldsymbol\Psi_2)\,.
\label{eq:extended-symplectic-generator}
\end{equation}

The antisymmetry of the symplectic current makes $\widetilde\Omega$ antisymmetric, i.e., $\widetilde\Omega(\boldsymbol{\Psi}_1,\boldsymbol{\Psi}_2)=-\widetilde\Omega(\boldsymbol{\Psi}_2,\boldsymbol{\Psi}_1)$. To construct a symmetric pairing, we use the time-reflection symmetry of the static stellar background.
Let
\begin{equation}
(\mathcal T\boldsymbol\Psi)(t,r)
:=
\mathbb R\boldsymbol\Psi(-t,r)\,,
\quad
\mathbb R=
\begin{pmatrix}
\mathbf I&0\\
0&-\mathbf I
\end{pmatrix}\,.
\label{eq:time-reflection-map}
\end{equation}
Time-reversal invariance of the canonical Hamiltonian gives
$\mathbb R\mathbb A\mathbb R=-\mathbb A$, and hence
\begin{equation}
\mathcal O\mathcal T
=
-\mathcal T\mathcal O\,,
\quad
\mathcal T\mathbb A
=
-\mathbb A\mathcal T\,,
\label{eq:time-reflection-identities}
\end{equation}
Thus $\mathcal T$ maps $\ker\mathcal O$ into itself.

We then define the bilinear form associated with
$\widetilde\Omega$ by
\begin{equation}
\left\langle
\boldsymbol\Psi_1,\boldsymbol\Psi_2
\right\rangle
:=
\widetilde\Omega
(\mathcal T\boldsymbol\Psi_1,\boldsymbol\Psi_2)\,.
\label{eq:extended-complex-bilinear}
\end{equation}
With the same boundary-flux prescription, this pairing is symmetric
under the interchange of $\boldsymbol{\Psi}_1$, $\boldsymbol{\Psi}_2$.
Equations~\eqref{eq:extended-symplectic-generator}
and~\eqref{eq:time-reflection-identities} give
\begin{equation}
\left\langle
\mathbb A\boldsymbol\Psi_1,\boldsymbol\Psi_2
\right\rangle
=
\left\langle
\boldsymbol\Psi_1,\mathbb A\boldsymbol\Psi_2
\right\rangle\,.
\label{eq:generator-bilinear-symmetry}
\end{equation}

\begin{table*}[t]
\centering
\footnotesize
\setlength{\tabcolsep}{3pt}
\renewcommand{\arraystretch}{1.25}
\caption{Numerical test of QNM orthogonality under the bilinear form
for the $\ell=2$ stellar modes.
Each entry is $\langle\boldsymbol\psi_i,\boldsymbol\psi_j\rangle/
\sqrt{\mathcal N_i\mathcal N_j}$, where $\mathcal N_i=\langle\boldsymbol\psi_i,\boldsymbol\psi_i\rangle$,
with a consistent square-root convention such that the diagonal
entries equal unity.
All off-diagonal entries are consistent with zero within numerical accuracy.}
\label{tab:qnm-bilinear-orthogonality}
\resizebox{\textwidth}{!}{%
\begin{tabular}{c|cccccccc}
\toprule
$i\backslash j$ & $f$ & $p_{1}$ & $p_{2}$ & $p_{3}$ & $w_{0}$ & $w_{1}$ & $w_{2}$ & $w_{3}$ \\
\midrule
$f$ & $1$ & $(-1.9+0.1i)\times10^{-13}$ & $(-3.1+1.4i)\times10^{-17}$ & $(-1.0+1.1i)\times10^{-16}$ & $(0.7-4.9i)\times10^{-14}$ & $(-1.6+1.8i)\times10^{-14}$ & $(-2.1-7.5i)\times10^{-14}$ & $(2.1+3.6i)\times10^{-13}$ \\[5pt]
$p_{1}$ & $(-1.9+0.1i)\times10^{-13}$ & $1$ & $(4.3+3.9i)\times10^{-16}$ & $(-0.6+4.6i)\times10^{-17}$ & $(-2.7-1.6i)\times10^{-12}$ & $(6.0+6.7i)\times10^{-12}$ & $(-1.0-3.3i)\times10^{-11}$ & $(0.8+8.6i)\times10^{-11}$ \\[5pt]
$p_{2}$ & $(-3.1+1.4i)\times10^{-17}$ & $(4.3+3.9i)\times10^{-16}$ & $1$ & $(-4.1+0.4i)\times10^{-17}$ & $(-1.9-0.6i)\times10^{-15}$ & $(4.8+2.8i)\times10^{-15}$ & $(-1.1-1.7i)\times10^{-14}$ & $(1.8+4.6i)\times10^{-14}$ \\[5pt]
$p_{3}$ & $(-1.0+1.1i)\times10^{-16}$ & $(-0.6+4.6i)\times10^{-17}$ & $(-4.1+0.4i)\times10^{-17}$ & $1$ & $(-3.1+1.2i)\times10^{-14}$ & $(8.5-0.3i)\times10^{-14}$ & $(-2.6-1.1i)\times10^{-13}$ & $(5.5+3.9i)\times10^{-13}$ \\[5pt]
$w_{0}$ & $(0.7-4.9i)\times10^{-14}$ & $(-2.7-1.6i)\times10^{-12}$ & $(-1.9-0.6i)\times10^{-15}$ & $(-3.1+1.2i)\times10^{-14}$ & $1$ & $(4.1+1.9i)\times10^{-12}$ & $(-4.7-8.0i)\times10^{-12}$ & $(-0.6+8.8i)\times10^{-12}$ \\[5pt]
$w_{1}$ & $(-1.6+1.8i)\times10^{-14}$ & $(6.0+6.7i)\times10^{-12}$ & $(4.8+2.8i)\times10^{-15}$ & $(8.5-0.3i)\times10^{-14}$ & $(4.1+1.9i)\times10^{-12}$ & $1$ & $(2.4+7.3i)\times10^{-12}$ & $(0.2+1.0i)\times10^{-11}$ \\[5pt]
$w_{2}$ & $(-2.1-7.5i)\times10^{-14}$ & $(-1.0-3.3i)\times10^{-11}$ & $(-1.1-1.7i)\times10^{-14}$ & $(-2.6-1.1i)\times10^{-13}$ & $(-4.7-8.0i)\times10^{-12}$ & $(2.4+7.3i)\times10^{-12}$ & $1$ & $(7.8-6.2i)\times10^{-11}$ \\[5pt]
$w_{3}$ & $(2.1+3.6i)\times10^{-13}$ & $(0.8+8.6i)\times10^{-11}$ & $(1.8+4.6i)\times10^{-14}$ & $(5.5+3.9i)\times10^{-13}$ & $(-0.6+8.8i)\times10^{-12}$ & $(0.2+1.0i)\times10^{-11}$ & $(7.8-6.2i)\times10^{-11}$ & $1$ \\[5pt]
\bottomrule
\end{tabular}%
}
\end{table*}

\subsubsection{Orthogonality}

Consider any two homogeneous perturbations with separated time
dependence,
\begin{equation}
\boldsymbol\Psi_a(t,r)
=e^{-i\omega_a t}\boldsymbol\psi_a(r)\,,
\quad
\mathbb A\boldsymbol\psi_a
=-i\omega_a\boldsymbol\psi_a\,,
\quad a=1,2\,.
\label{eq:separated-phase-space-modes}
\end{equation}
we have
\begin{equation}
\left\langle\boldsymbol\Psi_1,\boldsymbol\Psi_2\right\rangle
=e^{i(\omega_1-\omega_2)t}
\left\langle\boldsymbol\psi_1,\boldsymbol\psi_2
\right\rangle\,,
\label{eq:bilinear-mode-factorization}
\end{equation}
which defines the radial pairing on the chosen contour
$\mathcal C$. 
Equation~\eqref{eq:generator-bilinear-symmetry} then implies
\begin{equation}
(\omega_1-\omega_2)
\left\langle
\boldsymbol\psi_1,\boldsymbol\psi_2
\right\rangle=0\,.
\label{eq:mode-orthogonality}
\end{equation}
Thus, any two solutions satisfying the conditions of Eq.~\eqref{eq:separated-phase-space-modes} with distinct frequencies
are orthogonal under this bilinear form.
In particular, stellar QNMs with distinct frequencies are orthogonal.

\subsubsection{Explicit expression for center-regular solutions}

We now specialize to center-regular solutions that satisfy Eq.~\eqref{eq:separated-phase-space-modes} and are regular at
the stellar center.
In terms of the canonical variables defined in
Eq.~\eqref{eq:interior-canonical-variables}, the bilinear form reads
\begin{align}
\left\langle\boldsymbol\psi_1,\boldsymbol\psi_2
\right\rangle
&={}\int_{\mathcal C}\mathrm dr\,\Big[
H_{2,1}P_{H_2,2}+P_{H_2,1}H_{2,2}
\nonumber\\
&
+K_1P_{K,2}+P_{K,1}K_2+W_1P_{W,2}+P_{W,1}W_2\nonumber\\
&+V_1P_{V,2}+P_{V,1}V_2
\Big]
+\left.
\frac{\varpi^r[\mathbb R\boldsymbol\psi_1,\boldsymbol\psi_2]}
{i(\omega_1-\omega_2)}
\right|_{\partial\mathcal C}\,.
\label{eq:bilinear-form-canonical}
\end{align}
All canonical components $Q_a$ and $P_a$ appearing here depend
only on $r$; the time dependence can be restored through
Eq.~\eqref{eq:bilinear-mode-factorization}.

For a pair of center-regular solutions, we may take the real-axis contour $\mathcal C=[0,\infty)$. Regularity makes the center flux
vanish, $\varpi^r[\mathbb R\boldsymbol\psi_1,\boldsymbol\psi_2]|_{r=0}=0$,
so the flux contribution comes only from spatial infinity.
In the Schwarzschild exterior, we take advantage of the Zerilli canonical pair
$(Z,\Pi_Z)$, which reduce the radial symplectic flux to
\begin{equation}
\varpi^r[\mathbb R\boldsymbol\psi_1,\boldsymbol\psi_2]
=C_\ell\left[Z_2\partial_{r_*}Z_1-Z_1\partial_{r_*}Z_2\right]\,,
\end{equation}
up to a total time derivative whose contribution to the bilinear
form cancels the corresponding radial total derivative in
$\varpi^t$ at infinity, as shown in Appendix~\ref{sec:symplectic-appendix}.
Putting everything together, we obtain
\begin{widetext}
\begin{align}
\left\langle\boldsymbol\psi_1,\boldsymbol\psi_2
\right\rangle
={}&\int_0^R\mathrm dr\,\Big[
H_{2,1}P_{H_2,2}+P_{H_2,1}H_{2,2}
+K_1P_{K,2}+P_{K,1}K_2+W_1P_{W,2}+P_{W,1}W_2
+V_1P_{V,2}+P_{V,1}V_2\Big]
\nonumber\\
&
+\int_{R_*}^{\infty}\mathrm dr_*\,
\left(Z_1\Pi_{Z,2}+\Pi_{Z,1}Z_2\right)
+\mathcal B_R[\boldsymbol\psi_1,\boldsymbol\psi_2]
+\left.\frac{C_\ell}{i(\omega_1-\omega_2)}
\left[Z_2\partial_{r_*}Z_1-Z_1\partial_{r_*}Z_2
\right]\right|_{r_*\to\infty}\,,
\label{eq:bilinear-form-zerilli-real}
\end{align}
\end{widetext}
Here $\mathcal B_R$ is a surface term generated by the canonical transformation between the phase-space variables and the Zerilli pair. Its explicit expression is given in Appendix~\ref{sec:symplectic-appendix}.

\begin{figure}[t]
\centering
\includegraphics[width=\columnwidth]
{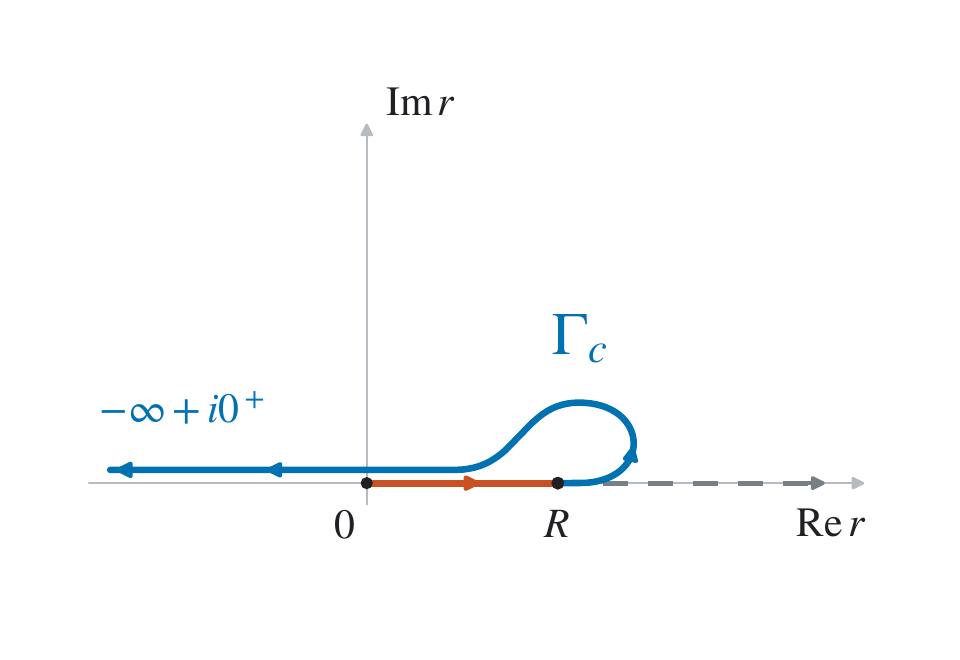}
\caption{Radial integration contour used to evaluate QNM bilinear
products. The stellar interior remains on the real interval
$[0,R)$, while the exterior is continued along $\Gamma_c$ in the
complex-$r$ plane. The asymptotic direction is chosen so that the
outgoing QNM products decay at the endpoint.}
\label{fig:bilinear-contour}
\end{figure}
\subsubsection{Complex contour for QNMs}
The real-axis contour used in Eq.~\eqref{eq:bilinear-form-zerilli-real} is not well suited to QNM calculations: outgoing QNM profiles grow exponentially along the real radial axis, so the exterior integral and its asymptotic flux generally diverge. We therefore keep the stellar interior on the real axis and deform the exterior segment into
the complex-$r$ plane, choosing
\begin{equation}
\mathcal C_{\rm QNM}=[0,R)\cup\Gamma_c,
\label{eq:qnm-radial-contour}
\end{equation}
as illustrated in Fig.~\ref{fig:bilinear-contour}. Here $\Gamma_c$
starts at the stellar surface and enters an asymptotic sector in
which, for the QNM pair under consideration,
\begin{equation}
e^{i(\omega_1+\omega_2)r_*(r)}
\to0
\qquad
\text{at the endpoint of }\Gamma_c\,.
\label{eq:qnm-contour-decay}
\end{equation}
On this contour, the radial bilinear product is given by convergent
integrals and the surface contribution $\mathcal B_R$:
\begin{widetext}
\begin{align}
\left\langle\boldsymbol\psi_1,\boldsymbol\psi_2
\right\rangle
={}&\int_0^R\mathrm dr\,\Big[
H_{2,1}P_{H_2,2}+P_{H_2,1}H_{2,2}
+K_1P_{K,2}+P_{K,1}K_2+W_1P_{W,2}+P_{W,1}W_2
\nonumber\\
&+V_1P_{V,2}+P_{V,1}V_2\Big]
+\int_{\Gamma_c}\mathrm dr_*\,
\left(Z_1\Pi_{Z,2}+\Pi_{Z,1}Z_2\right)
+\mathcal B_R[\boldsymbol\psi_1,\boldsymbol\psi_2]\,.
\label{eq:bilinear-form-zerilli}
\end{align}
\end{widetext}
To validate the orthogonality of stellar QNMs under this bilinear form, we perform numerical tests for the $\ell=2$ modes of the stellar model used in Sec.~\ref{sec:green-numerical-validation}.
The results are presented in Table~\ref{tab:qnm-bilinear-orthogonality}.

\section{Gravitational Wave Decomposition}
\label{sec:wave-decomposition}
Given the formalism developed in Sec.~\ref{sec:theoframe}, we are now ready to solve individual components of the star's response following the time-dependent ingoing gravitational wave at past infinity. We  first present two equivalent representations of the signals using Green's function, which is convoluted with the source, and using the bilinear form, where the dynamics of decomposed waves obey the oscillator-type equation of motion. After that we specify a sample chirping source and explicitly compute the quasinormal modes, the tail and the direct part.  

\subsection{Wave generation}
\label{sec:wave-generation}
In this section, we construct the response field
$\boldsymbol{\Psi}^{\rm r}$, which satisfies the first-order
driven equation~\eqref{eq:driven-response}.
Its retarded solution can be written formally as
\begin{equation}
\boldsymbol{\Psi}^{\rm r}(t)
=
\int_{-\infty}^{t}\mathrm dt'\,
e^{\mathbb A(t-t')}
\boldsymbol S^{\rm d}(t')\,,
\label{eq:retarded-phase-space-response}
\end{equation}
We focus below on the exterior response, particularly the waveform observed at future null infinity. We first use the time-domain Zerilli Green's function of Sec.~\ref{sec:stellar-green-function}, and then express its causal pole and branch-cut contributions as bilinear source projections.

\subsubsection{Green-function representation}

In the Schwarzschild exterior, Eq.~\eqref{eq:driven-response} reduces to the sourced Zerilli equation
\eqref{eq:sourced-zerilli-response}.
Its retarded solution is
\begin{equation}
Z^{\rm r}(t,r)
=
\int_{-\infty}^{t}\mathrm dt'
\int_{R_*}^{\infty}\mathrm dr_*'\,
G(t-t';r,r')S^{Z,{\rm d}}(t',r')\,.
\label{eq:zerilli-response-convolution}
\end{equation}
where the retarded green's function $G(t-t';r,r')$ has already been obtained in Sec.~\ref{sec:time-domain-green-function} and can be divided into direct, QNM and tail components as shown in Eq.~\eqref{eq:green-time-domain-decomposition}.

To get the waveform at future null infinity, we taking $r\to\infty$ at fixed retarded time $u=t-r_*$, and define
\begin{equation}
G^{\mathscr I^+}(\eta;r')
:=
\lim_{r\to\infty}G(\eta+r_*;r,r')\,.
\label{eq:zerilli-green-null-infinity}
\end{equation}
We then obtain
\begin{align}
Z^{\mathscr I^+}(u)
&=
\int_{-\infty}^{\infty}\mathrm dt'
\int_{R_*}^{\infty}\mathrm dr_*'\,
G^{\mathscr I^+}(u-t';r')S^{Z,{\rm d}}(t',r')\,.
\label{eq:waveform-null-infinity}
\end{align}
The causal support of $G^{\mathscr I^+}$ restricts the source events contributing to the waveform. Its decomposition into direct, QNM, and tail contributions induces the corresponding decomposition of $Z^{\mathscr I^+}(u)$.
Explicitly,
\begin{equation}
Z^{\mathscr I^+}(u)
=Z_{\rm dir}^{\mathscr I^+}(u)
+Z_{\rm QNM}^{\mathscr I^+}(u)
+Z_{\rm tail}^{\mathscr I^+}(u)\,,
\label{eq:waveform-component-decomposition}
\end{equation}
where the three components are
\begin{subequations}
\label{eq:waveform-component-convolutions}
\begin{align}
Z_{\rm dir}^{\mathscr I^+}(u)
={}&\int_{-\infty}^{\infty}\mathrm dt'
\int_{R_*}^{\infty}\mathrm dr_*'\,
G_{\rm dir}^{\mathscr I^+}(u-t';r')
S^{Z,{\rm d}}(t',r')\,,
\label{eq:waveform-direct-convolution}\\
Z_{\rm QNM}^{\mathscr I^+}(u)
={}&\int_{-\infty}^{\infty}\mathrm dt'
\int_{R_*}^{\infty}\mathrm dr_*'\,
G_{\rm QNM}^{\mathscr I^+}(u-t';r')
S^{Z,{\rm d}}(t',r')\,,
\label{eq:waveform-qnm-convolution}\\
Z_{\rm tail}^{\mathscr I^+}(u)
={}&\int_{-\infty}^{\infty}\mathrm dt'
\int_{R_*}^{\infty}\mathrm dr_*'\,
G_{\rm tail}^{\mathscr I^+}(u-t';r')
S^{Z,{\rm d}}(t',r')\,.
\label{eq:waveform-tail-convolution}
\end{align}
\end{subequations}
Here each null-infinity kernel is obtained from the corresponding component of the Green function in Eq.~\eqref{eq:green-time-domain-decomposition} by the limit in Eq.~\eqref{eq:zerilli-green-null-infinity}.

\begin{figure}[t]
\centering
\includegraphics[width=0.37\columnwidth]
{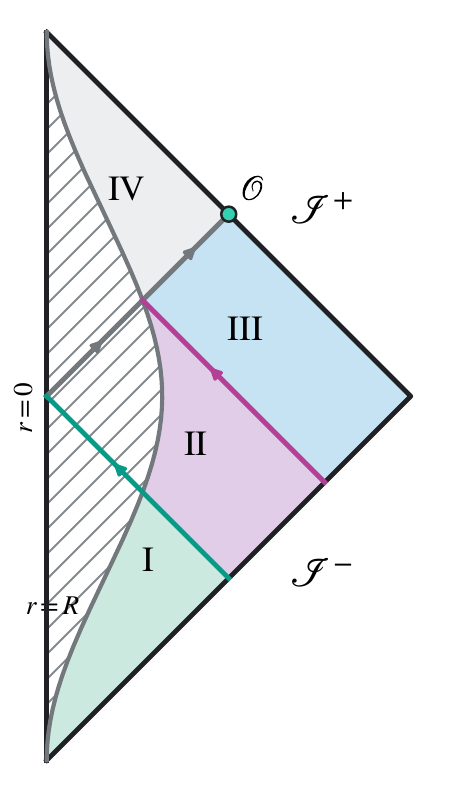}
\caption{Penrose diagram showing the causal partition of exterior source events relative to a fixed observer $\mathscr O$ at $\mathscr I^+$.
The incident waveform is prescribed on $\mathscr I^-$ and extended into the finite-radius exterior (colored regions).
}
\label{fig:observer-causal-penrose}
\end{figure}

It is worth noting that, although the Green-function decomposition itself is unique, the resulting decomposition of the observed waveform is source dependent.
This is illustrated in Fig.~\ref{fig:observer-causal-penrose}: while the incident data at $\mathscr I^-$ are fixed, the different waveform components observed at $\mathscr O$ receive contributions from effective sources in the corresponding causal regions, and these sources depend on how the incident field is extended from $\mathscr {I}^-$ into the finite-radius exterior.

\subsubsection{Bilinear form representation}

Alternatively, we reorganize the Green-function representation in terms of source projections defined by the bilinear form. We use these projections to describe the discrete and continuum contributions to the waveform and their driven evolution.

For the prescribed incident field, supported in the stellar exterior and vanishing at the surface, the effective source has only a Zerilli momentum component:
\begin{equation}
\left.\boldsymbol S^{\rm d}(t',r')\right|_{r'>R}
=
\begin{pmatrix}
0\\
-C_\ell S^{Z,\rm d}(t',r')
\end{pmatrix}\,.
\end{equation}
Its bilinear projection onto a field $\boldsymbol X$ is therefore
\begin{equation}
\left\langle\boldsymbol X,\boldsymbol S^{\rm d}(t')\right\rangle
=
-C_\ell\int_{R_*}^{\infty}\mathrm dr_*'\,
Z_X(r')S^{Z,\rm d}(t',r')\,,
\label{eq:exterior-bilinear-source-projection}
\end{equation}
where $Z_X$ is the exterior Zerilli component of $\boldsymbol X$. To apply this projection to the causal regions identified in Sec.~\ref{sec:stellar-green-function}, we partition the source using the windows in Eq.~\eqref{eq:green-causal-windows}:
\begin{equation}
\boldsymbol S_A^{\rm d}(t',r';t,r)
=
\chi_A(t-t';r,r')\boldsymbol S^{\rm d}(t',r')\,,
\quad A={\rm I},\ldots,{\rm IV}\,.
\label{eq:causal-source-sector-definition}
\end{equation}

\textbf{Region I.}
In this region, the bilinear source projections organize the response as
\begin{equation}
Z_{\rm I}^{\rm r}(t,r)
=
\sum_n a_{n,{\rm I}}(t,r)Z_n(r)
+
\int_0^{-i\infty}\mathrm d\sigma\,
a_{\sigma,{\rm I}}^{\rm tail}(t,r)
Z_\sigma^{\rm reg}(r)\,,
\label{eq:region-I-spectral-decomposition}
\end{equation}
where
\begin{align}
a_{n,{\rm I}}(t,r)
&=
\frac{1}{\mathcal N_n}
\int_{-\infty}^{t}\mathrm dt'\,
e^{-i\omega_n(t-t')}
\left\langle
\boldsymbol\psi_n,
\boldsymbol S_{\rm I}^{\rm d}(t';t,r)
\right\rangle\,,
\label{eq:qnm-driven-amplitude}\\
a_{\sigma,{\rm I}}^{\rm tail}(t,r)
&=
\frac{\sigma\,q_\ell^Z(i\sigma)}
{\pi C_\ell|W_\ell(\sigma+0)|^2}\nonumber\\
&\times
\int_{-\infty}^{t}\mathrm dt'\,
e^{-i\sigma(t-t')}
\left\langle
\boldsymbol\psi_\sigma^{\rm reg},
\boldsymbol S_{\rm I}^{\rm d}(t';t,r)
\right\rangle\,.
\label{eq:tail-driven-amplitude}
\end{align}
Equation~\eqref{eq:region-I-spectral-decomposition} implies that the Region-I response can be interpreted as a discrete QNM sum and a continuum integral along the NIA.
Note that Eq.~\eqref{eq:qnm-driven-amplitude} expresses the QNM contribution using the bilinear norm $\mathcal N_n=\langle\boldsymbol\psi_n,\boldsymbol\psi_n\rangle$ rather than the Green-function residue. Equating the two expressions gives
\begin{equation}
\frac{C_\ell}{\mathcal N_n}
=
\frac{A_{\ell m}^{\rm out}(\omega_n)}
{\left.2\omega_n\,\partial_\omega A_{\ell m}^{\rm in}(\omega)\right|_{\omega_n}}.
\label{eq:qnm-bilinear-residue-identity}
\end{equation}
We verify this relation numerically and present representative results in Table~\ref{tab:qnm-bilinear-residue-check}.
It is also notable that the right-hand side of Eq.~\eqref{eq:qnm-bilinear-residue-identity} is the stellar excitation factor introduced in Ref.~\cite{Miao:2025utd}, where its magnitude was shown to be approximately the inverse quality factor $Q_n=|\Re\omega_n|/|\Im\omega_n|$ for weakly damped fluid modes.
Consequently, $|C_\ell/\mathcal N_n|\approx Q_n^{-1}$ for such fluid modes ($f$ and $p$-modes), whereas our numerical results for the highly damped $w$-modes give $|C_\ell/\mathcal N_n|\sim\mathcal O(1)\times Q_n^{-1}$.

\begin{table*}[t]
\centering
\footnotesize
\setlength{\tabcolsep}{3pt}
\renewcommand{\arraystretch}{1.18}
\caption{Independent numerical check of Eq.~\eqref{eq:qnm-bilinear-residue-identity} for the $\ell=2$ stellar modes.
Both $\Re\omega_n$ and $\Im\omega_n$ are given in units of
$M_\odot^{-1}$.
The bilinear norm $\mathcal N_n$ is evaluated along the QNM contour
described in Sec.~\ref{sec:bilinear-form}.
RHS denotes the right-hand side of Eq.~\eqref{eq:qnm-bilinear-residue-identity}, evaluated independently
from the frequency derivative of the scattering coefficients.
For all modes listed, $C_\ell/\mathcal N_n$ agrees with the RHS
to all six significant figures displayed.
We also report their absolute difference, which is consistent
with the estimated numerical errors.
The last column gives $|C_\ell/\mathcal N_n|Q_n$, which is expected
to be close to unity for the weakly damped $f$- and $p$-modes.}
\label{tab:qnm-bilinear-residue-check}
\begin{tabular}{ccccccc}
\toprule
Mode & $\Re\omega_n$ & $\Im\omega_n$ & $C_\ell/\mathcal N_n$ & RHS & $\left|C_\ell/\mathcal N_n-\mathrm{RHS}\right|$ & $|C_\ell/\mathcal N_n|Q_n$ \\[5pt]
\midrule
$f$ & $0.0493960$ & $-1.72676\times10^{-5}$ & $(1.84919+2.96624i)\times10^{-4}$ & $(1.84919+2.96624i)\times10^{-4}$ & $1.9\times10^{-16}$ & $0.999913$ \\[5pt]
$p_{1}$ & $0.120526$ & $-2.81587\times10^{-6}$ & $(1.42031+1.85502i)\times10^{-5}$ & $(1.42031+1.85502i)\times10^{-5}$ & $2.4\times10^{-15}$ & $1.00000$ \\[5pt]
$p_{2}$ & $0.185771$ & $-6.82288\times10^{-8}$ & $(1.54807+3.33053i)\times10^{-7}$ & $(1.54807+3.33053i)\times10^{-7}$ & $8.5\times10^{-20}$ & $1.00000$ \\[5pt]
$p_{3}$ & $0.247825$ & $-3.61884\times10^{-10}$ & $(0.109896+1.45610i)\times10^{-9}$ & $(0.109896+1.45610i)\times10^{-9}$ & $1.0\times10^{-19}$ & $1.00000$ \\[5pt]
$w_{0}$ & $0.140543$ & $-2.32067\times10^{-1}$ & $(-1.39546-1.54193i)\times10^{-1}$ & $(-1.39546-1.54193i)\times10^{-1}$ & $8.8\times10^{-13}$ & $0.125945$ \\[5pt]
$w_{1}$ & $0.316480$ & $-2.18624\times10^{-1}$ & $(2.38439-1.42131i)\times10^{-1}$ & $(2.38439-1.42131i)\times10^{-1}$ & $1.7\times10^{-12}$ & $0.401834$ \\[5pt]
$w_{2}$ & $0.562700$ & $-2.72139\times10^{-1}$ & $(0.935637+1.07601i)\times10^{-1}$ & $(0.935637+1.07601i)\times10^{-1}$ & $3.0\times10^{-12}$ & $0.294834$ \\[5pt]
$w_{3}$ & $0.794516$ & $-3.06060\times10^{-1}$ & $(-4.98394+9.60088i)\times10^{-2}$ & $(-4.98394+9.60088i)\times10^{-2}$ & $4.0\times10^{-11}$ & $0.280815$ \\[5pt]
\bottomrule
\end{tabular}
\end{table*}

It is also instructive to examine the evolution of the mode amplitudes. Differentiating Eqs.~\eqref{eq:qnm-driven-amplitude} and \eqref{eq:tail-driven-amplitude} gives
\begin{align}
&(\partial_t+i\omega_n) a_{n,{\rm I}}(t,r)
=
\frac{1}{\mathcal N_n}
\int_{-\infty}^{t}\mathrm dt'\,
e^{-i\omega_n(t-t')}\nonumber\\
&\times
\left\langle
\boldsymbol\psi_n,
\delta\!\left(t-t'-\tau_{\rm c}\right)
\boldsymbol S^{\rm d}(t',r')
\right\rangle
\nonumber\\
&=
\frac{1}{\mathcal N_n}
\left\langle
\boldsymbol\psi_n,
e^{-i\omega_n\tau_{\rm c}}
\boldsymbol S^{\rm d}\!\left(t-\tau_{\rm c},r'\right)
\right\rangle,
\label{eq:sectoral-amplitude-evolution}\\
&(\partial_t+i\sigma) a_{\sigma,{\rm I}}^{\rm tail}(t,r)
=
\frac{\sigma\,q_\ell^Z(i\sigma)}
{\pi C_\ell|W_\ell(\sigma+0)|^2}
\int_{-\infty}^{t}\mathrm dt'\,
e^{-i\sigma(t-t')}\nonumber\\
&\times
\left\langle
\boldsymbol\psi_\sigma^{\rm reg},
\delta\!\left(t-t'-\tau_{\rm c}\right)
\boldsymbol S^{\rm d}(t',r')
\right\rangle
\nonumber\\
&=
\frac{\sigma\,q_\ell^Z(i\sigma)}
{\pi C_\ell|W_\ell(\sigma+0)|^2}
\left\langle
\boldsymbol\psi_\sigma^{\rm reg},
e^{-i\sigma\tau_{\rm c}}
\boldsymbol S^{\rm d}\!\left(t-\tau_{\rm c},r'\right)
\right\rangle\,.
\label{eq:sectoral-tail-amplitude-evolution}
\end{align} 
These equations show that the discrete- and continuum-mode amplitudes observed at radius $r$ are driven by bilinear projections of source events as they enter Region I after the corresponding propagation delay.

\begin{figure}[t]
\centering
\includegraphics[width=0.7\columnwidth]
{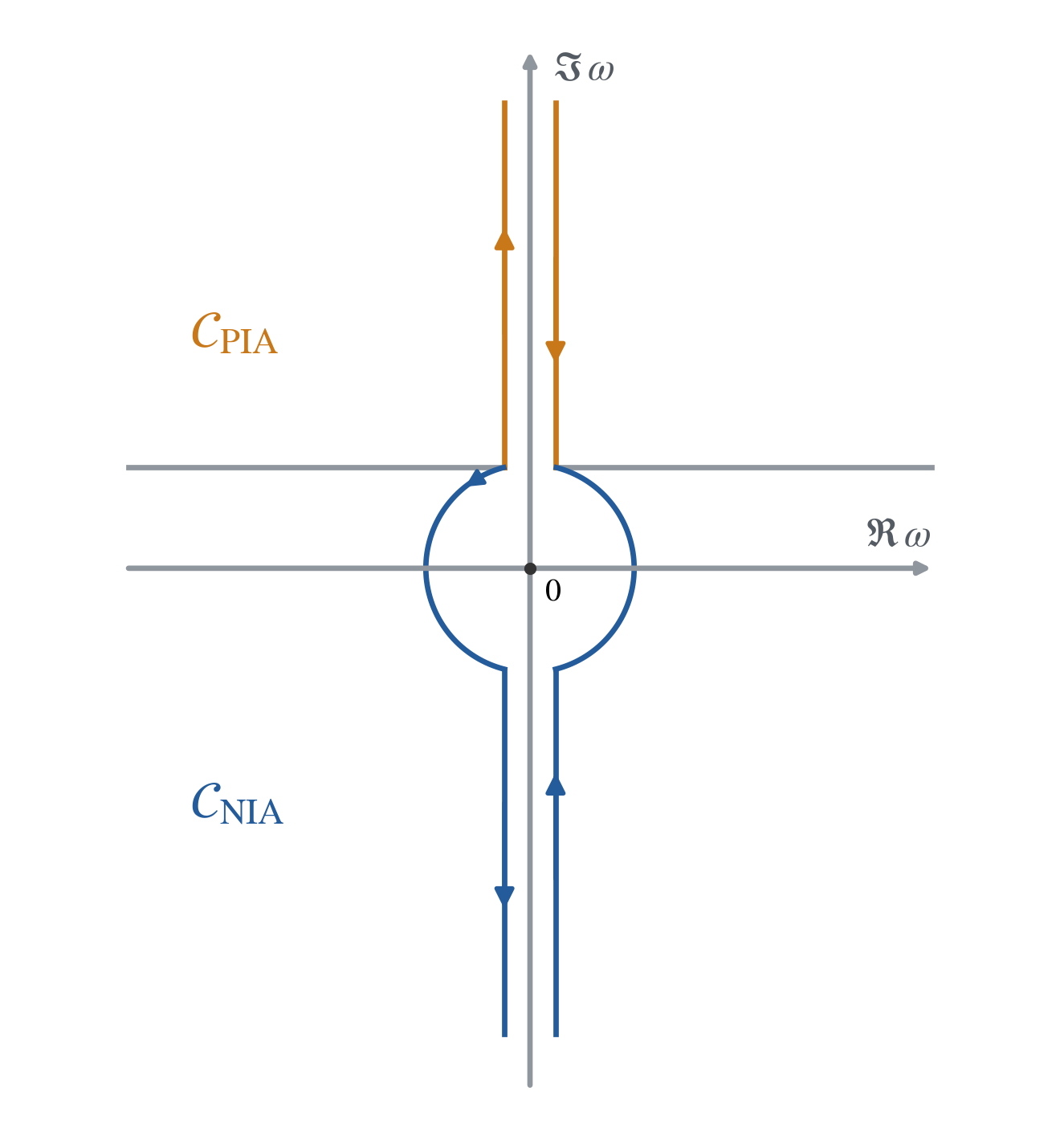}
\caption{
Schematic complex-frequency contours for the spectral
decomposition used in Eqs.~\eqref{eq:region-II-spectral-decomposition} and~\eqref{eq:region-III-spectral-decomposition}.
}
\label{fig:nia-pia-contours}
\end{figure}

\textbf{Region II.}
Let $\boldsymbol\psi_\sigma^{-}$ and $\boldsymbol\psi_\sigma^{+}$ be phase-space vectors with Zerilli components
\begin{equation}
Z_\sigma^{-}=Z_\sigma^{\rm in},
\qquad
Z_\sigma^{+}=Z_\sigma^{\rm out},
\qquad r>R.
\end{equation}
The Region-II response can then be written as
\begin{align}
Z_{\rm II}^{\rm r}(t,r)
={}&
\int_{\mathcal{C}_{\rm NIA}}\mathrm d\sigma\,
a_{\sigma,{\rm II}}^{\rm dir,-}(t,r)Z_\sigma^{\rm out}(r)
\nonumber\\
&+
\int_{\mathcal{C}_{\rm NIA}}\mathrm d\sigma\,
a_{\sigma,{\rm II}}^{{\rm dir,pr}}(t,r)Z_\sigma^{\rm out}(r),
\label{eq:region-II-spectral-decomposition}
\end{align}
where
\begin{align}
a_{\sigma,{\rm II}}^{\rm dir,-}(t,r)
&={}
\frac{1}{4\pi i C_\ell\sigma}\nonumber\\
&\times\int_{-\infty}^{t}\mathrm dt'\,
e^{-i\sigma(t-t')}
\left\langle
\boldsymbol\psi_\sigma^{-},
\boldsymbol S_{\rm II}^{\rm d}(t';t,r)
\right\rangle\,,\\
a_{\sigma,{\rm II}}^{\rm dir,pr}(t,r)
&={}
\frac{\mathcal{R}_\ell^{\rm pr}(\sigma)}{4\pi i C_\ell\sigma}\nonumber\\
&\times\int_{-\infty}^{t}\mathrm dt'\,
e^{-i\sigma(t-t')}
\left\langle
\boldsymbol\psi_\sigma^{+},
\boldsymbol S_{\rm II}^{\rm d}(t';t,r)
\right\rangle\,.
\label{eq:region-II-direct-amplitude}
\end{align}
In Eq.~\eqref{eq:region-II-spectral-decomposition}, $\mathcal C_{\rm NIA}$ follows both banks of the NIA branch cut in opposite directions and includes the small detour around the origin.
These amplitudes satisfy
\begin{align}
&(\partial_t+i\sigma)a_{\sigma,{\rm II}}^{\rm dir,-}(t,r)
={}\frac{1}{4\pi i C_\ell\sigma}
\nonumber\\
&\hspace{1.2cm}\times\Bigg[
\left\langle
\boldsymbol\psi_\sigma^{-},
e^{-i\sigma\tau_{\rm s}}
\boldsymbol S^{\rm d}(t-\tau_{\rm s},r')
\right\rangle\nonumber\\
&\hspace{1.2cm}-\left\langle
\boldsymbol\psi_\sigma^{-},
e^{-i\sigma\tau_{\rm c}}
\boldsymbol S^{\rm d}(t-\tau_{\rm c},r')
\right\rangle
\Bigg].\\
&(\partial_t+i\sigma)a_{\sigma,{\rm II}}^{\rm dir,pr}(t,r)
={}\frac{\mathcal{R}_\ell^{\rm pr}(\sigma)}{4\pi i C_\ell\sigma}
\nonumber\\
&\hspace{1.2cm}\times\Bigg[
\left\langle
\boldsymbol\psi_\sigma^{+},
e^{-i\sigma\tau_{\rm s}}
\boldsymbol S^{\rm d}(t-\tau_{\rm s},r')
\right\rangle\nonumber\\
&\hspace{1.2cm}-\left\langle
\boldsymbol\psi_\sigma^{+},
e^{-i\sigma\tau_{\rm c}}
\boldsymbol S^{\rm d}(t-\tau_{\rm c},r')
\right\rangle
\Bigg].
\label{eq:region-II-direct-amplitude-evolution}
\end{align}
Thus, the continuum-mode amplitudes are driven by source events as they enter and leave the observer’s Region-II causal window.

\textbf{Region III.} 
Similar to region II, the response of region III can be written as
\begin{align}
Z_{\rm III}^{\rm r}(t,r)
={}&
\int_{\mathcal C_{\rm NIA}}\mathrm d\sigma\,
a_{\sigma,{\rm III}}^{{\rm dir},-}(t,r)Z_\sigma^{\rm out}(r)
\nonumber\\
&+
\int_{\mathcal C_{\rm PIA}}\mathrm d\sigma\,
a_{\sigma,{\rm III}}^{{\rm dir},+}(t,r)Z_\sigma^{\rm out}(r),
\label{eq:region-III-spectral-decomposition}
\end{align}
where 
\begin{align}
&a_{\sigma,{\rm III}}^{{\rm dir},-}(t,r)
={}
\frac{1}{4\pi iC_\ell\sigma}\nonumber\\
&\times
\int_{-\infty}^{t}\mathrm dt'\,
e^{-i\sigma(t-t')}
\left\langle
\boldsymbol\psi_\sigma^{-},
\boldsymbol S_{\rm III}^{\rm d}(t';t,r)
\right\rangle,
\label{eq:region-III-minus-amplitude}\\
&a_{\sigma,{\rm III}}^{{\rm dir},+}(t,r)
=
\frac{\mathcal R_\ell(\sigma)}{4\pi iC_\ell\sigma}\nonumber\\
&\times\int_{-\infty}^{t}\mathrm dt'\,
e^{-i\sigma(t-t')}
\left\langle
\boldsymbol\psi_\sigma^{+},
\boldsymbol S_{\rm III}^{\rm d}(t';t,r)
\right\rangle.
\label{eq:region-III-plus-amplitude}
\end{align}
Again, the corresponding amplitudes satisfy
\begin{align}
&(\partial_t+i\sigma)a_{\sigma,{\rm III}}^{{\rm dir},-}(t,r)
={}\frac{1}{4\pi iC_\ell\sigma}\nonumber\\
&\hspace{1.2cm}\times\Bigg[
\left\langle
\boldsymbol\psi_\sigma^{-},
e^{-i\sigma\tau_{\rm f}}
\boldsymbol S^{\rm d}(t-\tau_{\rm f},r')
\right\rangle\nonumber\\
&\hspace{1.2cm}-
\left\langle
\boldsymbol\psi_\sigma^{-},
e^{-i\sigma\tau_{\rm s}}
\boldsymbol S^{\rm d}(t-\tau_{\rm s},r')
\right\rangle
\Bigg],
\label{eq:region-III-minus-evolution}
\end{align}
\begin{align}
    &(\partial_t+i\sigma)a_{\sigma,{\rm III}}^{{\rm dir},+}(t,r)
={}\frac{\mathcal R_\ell(\sigma)}{4\pi iC_\ell\sigma}\nonumber\\
&\hspace{1.2cm}\times\Bigg[
\left\langle
\boldsymbol\psi_\sigma^{+},
e^{-i\sigma\tau_{\rm f}}
\boldsymbol S^{\rm d}(t-\tau_{\rm f},r')
\right\rangle\nonumber\\
&\hspace{1.2cm}-
\left\langle
\boldsymbol\psi_\sigma^{+},
e^{-i\sigma\tau_{\rm s}}
\boldsymbol S^{\rm d}(t-\tau_{\rm s},r')
\right\rangle
\Bigg]\,,
\label{eq:region-III-plus-evolution}
\end{align}
which shows that they are driven by source events as they enter and leave the observer’s Region-III causal window.

In summary, we use the bilinear form to provide a source-projection framework for a formal spectral representation of the full retarded response. After accounting for causality, we identify the discrete pole contributions with stellar QNMs and the remaining non-pole continuum integrals with the direct and late-time-tail responses. 
Both the discrete and continuum contributions can be described by
a set of driven oscillator-type evolution equations.
\emph{Therefore a tidally driven star spacetime may be described by the dynamics of a set of discrete and continuous modes.}

It is also straightforward to get the waveform at future null infinity by taking the $r\to\infty$ limit in Eq.~\eqref{eq:region-I-spectral-decomposition},~\eqref{eq:region-II-spectral-decomposition} and ~\eqref{eq:region-III-spectral-decomposition}. The different contributions read
\begin{align}
&Z_{\rm I}^{\mathscr I^+}(u)
=
\sum_n\frac{1}{\mathcal N_n}
\int_{-\infty}^{\infty}\mathrm dt'\,
e^{-i\omega_n(u-t')}
\left\langle
\boldsymbol\psi_n,\boldsymbol S_{\rm I}^{\rm d}(t')
\right\rangle\nonumber\\
&\hspace{1.2cm}+\int_{0}^{-i\infty}\mathrm d\sigma\,
\frac{\sigma q_\ell^Z(i\sigma)
A_{\ell m}^{\rm out}(\sigma)}
{\pi C_\ell|W_\ell(\sigma+0)|^2}\nonumber\\
&\hspace{1.2cm}\times
\int_{-\infty}^{\infty}\mathrm dt'\,
e^{-i\sigma(u-t')}
\left\langle
\boldsymbol\psi_\sigma^{\rm reg},\boldsymbol S_{\rm I}^{\rm d}(t')
\right\rangle\,.
\label{eq:bilinear-tail-null-infinity}
\end{align}
\begin{align}
    &Z_{\rm II}^{\mathscr I^+}(u)
=\int_{\mathcal{C}_{\rm NIA}}\frac{\mathrm d\sigma}
{4\pi i C_\ell\sigma}
\int_{-\infty}^{\infty}\mathrm dt'\,
e^{-i\sigma(u-t')}\nonumber\\
&\hspace{1.2cm}\times\Big[
\left\langle
\boldsymbol\psi_\sigma^-,
\boldsymbol S_{\rm II}^{\rm d}(t')
\right\rangle
+\mathcal R_\ell^{\rm pr}(\sigma)
\left\langle
\boldsymbol\psi_\sigma^+,
\boldsymbol S_{\rm II}^{\rm d}(t')
\right\rangle
\Big]\,,
\label{eq:bilinear-region-II-null-infinity}
\end{align}
\begin{align}
    &Z_{\rm III}^{\mathscr I^+}(u)
=\int_{\mathcal{C}_{\rm NIA}}\frac{\mathrm d\sigma}
{4\pi i C_\ell\sigma}
\int_{-\infty}^{\infty}\mathrm dt'\,
e^{-i\sigma(u-t')}
\left\langle
\boldsymbol\psi_\sigma^-,
\boldsymbol S_{\rm III}^{\rm d}(t')
\right\rangle
\nonumber\\
&\hspace{0.6cm}+\int_{\mathcal{C}_{\rm PIA}}\frac{\mathrm d\sigma}
{4\pi i C_\ell\sigma}\,
\mathcal R_\ell(\sigma)
\int_{-\infty}^{\infty}\mathrm dt'\,
e^{-i\sigma(u-t')}
\left\langle
\boldsymbol\psi_\sigma^+,
\boldsymbol S_{\rm III}^{\rm d}(t')
\right\rangle\,.
\label{eq:bilinear-region-III-null-infinity}
\end{align}

\subsubsection{Monochromatic driving and the decomposition of the
scattering coefficient}
\label{sec:monochromatic-scattering-decomposition}

The preceding two subsections describe how to calculate the
response waveform and its direct, QNM, and tail components at
an observation point $(t,r)$ for a given driving source, using
either the Green-function or the bilinear form representation.
We now specialize these expressions to a unit-amplitude monochromatic wave incident from $\mathscr I^-$, setting the instantaneous driving frequency in Eq.~\eqref{eq:time-dependent-driver} to a constant $\omega$, and extract the corresponding decomposition of the scattering coefficient $\mathcal R(\omega)$ from
the outgoing waveform at $\mathscr I^+$.

As discussed above, the waveform decomposition generally depends on how the prescribed incident field is extended into the finite-radius exterior.
To fix this freedom, we adopt an auxiliary injection-surface construction: the effective source is supported on a finite-radius injection surface, which is then taken to $\mathscr I^-$ at fixed incoming time $v$.
With this construction, the waveform observed at $\mathscr I^+$
is related to the incident waveform on $\mathscr I^-$ through the convolution
\begin{equation}
Z^{\mathscr I^+}(u)
=
\int\mathrm dv\,
\mathcal{K}(u-v)Z^{\rm d}(v)\,.
\end{equation}
In this sense, the observed waveform and its components are determined solely by the incident data on $\mathscr I^-$ and do not depend on an arbitrary continuation into the finite-radius exterior.

We first introduce an auxiliary injection surface at
$y=y_s>R_*$, where we denote $y=r_*(r)$ for convenience, and choose the auxiliary incident field
\begin{equation}
Z_s^{\rm d}(t,y)
=\Theta(y-y_s)I_\omega(t,y)\,.
\label{eq:monochromatic-boundary-driver}
\end{equation}
with
\begin{equation}
I_\omega(t,y)
=e^{-i\omega t}Z_\omega^{\rm in}(y),
\label{eq:monochromatic-exact-incoming-field}
\end{equation}
The effective source then reads
\begin{align}
S_s^{Z,\rm d}(t,y)
={}&-I_\omega(t,y_s)\delta'(y-y_s)
\nonumber\\
&-\partial_y I_\omega(t,y_s)\delta(y-y_s)\,,
\label{eq:monochromatic-boundary-source}
\end{align}
where $\delta'$ denotes differentiation with respect to $y$.

Substituting this source into the Green-function representation
and performing the spatial integral gives
\begin{align}
Z_{X,s}^{\mathscr I^+}(u)
={}&\int_{-\infty}^{\infty}\mathrm dt'\,
\Big[
I_\omega(t',y)
\partial_y G_X^{\mathscr I^+}(u-t';y)
\nonumber\\
&\qquad
-\partial_y I_\omega(t',y)
G_X^{\mathscr I^+}(u-t';y)
\Big]_{y=y_s}\,.
\label{eq:monochromatic-boundary-green-response}
\end{align}
where $X\in\{{\rm dir},{\rm QNM},{\rm tail}\}$.

We then take $y_s\to\infty$ at fixed incoming time $v=t'+y_s$,
thereby recovering the unit-amplitude monochromatic incident wave at $\mathscr I^-$, i.e., $I_\omega(t',y_s)\to e^{-i\omega v}$.
The outgoing components may take the form
\begin{equation}
Z_X^{\mathscr I^+}(u)
:=\lim_{y_s\to\infty}Z_{X,s}^{\mathscr I^+}(u)
=\mathcal R_X(\omega)e^{-i\omega u}\,.
\label{eq:monochromatic-component-response}
\end{equation}
Consequently, we obtain the decomposition of the scattering coefficient, 
\begin{equation}
\mathcal R(\omega)
=
\mathcal R_{\rm dir}(\omega)
+\mathcal R_{\rm QNM}(\omega)
+\mathcal R_{\rm tail}(\omega)\,.
\label{eq:monochromatic-reflection-decomposition}
\end{equation}
Hereafter we evaluate these contributions separately.
For convenience, we shall use $d_{\rm s}:=-2R_*$ and $d_{\rm c}:=2x_R-2R_*$. 

\paragraph{QNM contribution.}
Applying Eq.~\eqref{eq:monochromatic-boundary-green-response}
to the pole contribution with Eq.\eqref{eq:green-qnm-residue-sum} gives
\begin{align}
Z_{\rm QNM}^{\mathscr I^+}(u)
={}&\sum_n\Bigg[-\frac{2i\omega_nC_\ell}{\mathcal N_n}
\int_{-\infty}^{u-d_{\rm c}}\mathrm dv\,
e^{-i\omega_n(u-v)}e^{-i \omega v}
\nonumber\\
&+\frac{2C_\ell}{\mathcal N_n}
e^{-i\omega_nd_{\rm c}}e^{-i \omega (u-d_{\rm c})}\Bigg]\\
={}&e^{-i \omega u} \sum_n \frac{2\omega C_\ell}{\mathcal N_n}\frac{e^{i(\omega-\omega_n)d_{\rm c}}}{\omega-\omega_n}
\label{eq:monochromatic-boundary-qnm-response}
\end{align}
where we have used the identity Eq.\eqref{eq:qnm-bilinear-residue-identity}.
It follows that
\begin{equation}
\mathcal R_{\rm QNM}(\omega)
=
\sum_n
\frac{2\omega C_\ell}{\mathcal N_n}
\frac{e^{i(\omega-\omega_n)d_{\rm c}}}{\omega-\omega_n}\,.
\label{eq:monochromatic-qnm-reflection}
\end{equation}

\paragraph{Tail contribution.}
Similar to the QNM part, substituting Eq.~\eqref{eq:green-tail-cut-integral} into Eq.~\eqref{eq:monochromatic-boundary-green-response} and taking the $y_s\to \infty$ limit gives 
\begin{align}
&Z_{\rm tail}^{\mathscr I^+}(u)
={}
\frac{2}{\pi}\int_0^\infty \mathrm d\sigma\,
\frac{
\sigma q_\ell^Z(\sigma)
\bigl[A_{\ell m}^{\rm out}(-i\sigma)\bigr]^2
}{
|W_\ell(-i\sigma+0)|^2
}
\nonumber\\
&\hspace{.2cm}\times
\left[
\sigma\int_{-\infty}^{u-d_{\rm c}}\mathrm dv\,
e^{-\sigma(u-v)}e^{-i \omega v}
-e^{-\sigma d_{\rm c}}e^{-i \omega (u-d_{\rm c})}
\right]\nonumber\\
&\hspace{.2cm}={}\frac{2i\omega e^{-i\omega u}}{\pi}
\int_0^\infty\mathrm d\sigma\,
\frac{
\sigma q_\ell^Z(\sigma)
\left[A_{\ell m}^{\rm out}(-i\sigma)\right]^2
}{
|W_\ell(-i\sigma+0)|^2(\sigma-i\omega)
}
e^{(i\omega-\sigma)d_{\rm c}}
\label{eq:monochromatic-boundary-qnm-response}
\end{align}
and hence 
\begin{equation}
\mathcal R_{\rm tail}(\omega)
=
\frac{2i\omega}{\pi}
\int_0^\infty\mathrm d\sigma\,
\frac{
\sigma q_\ell^Z(\sigma)
\left[A_{\ell m}^{\rm out}(-i\sigma)\right]^2
}{
|W_\ell(-i\sigma+0)|^2(\sigma-i\omega)
}
e^{(i\omega-\sigma)d_{\rm c}}.
\label{eq:monochromatic-tail-reflection}
\end{equation}

\paragraph{Direct contribution.}
At $\mathscr I^+$, $\widetilde G^-$ has no branch cut along
the NIA, so its contour integrals in Regions II and III reduce
to contributions from a small arc around $\omega=0$~\cite{Su:2026fvj}.
By contrast, $\widetilde G^{\rm pr}$ retains a branch-cut
contribution along the NIA, in addition to its small-arc
contribution. We denote the resulting time-domain contributions
from these small-arc integrals by $G^{-}_0$ and
$G_{0}^{\rm pr}$, respectively.
Then the direct part of the waveform reads
\begin{align}
&Z_{\rm dir}^{\mathscr I^+}(u)
={}
\frac1\pi\int_0^\infty\mathrm d\sigma\,
\frac{\Im\mathcal R_\ell^{\rm pr}(-i\sigma+0)}{\sigma}
\nonumber\\
&\hspace{0.3cm}\times\Bigg[
\sigma\int_{u-d_{\rm c}}^{u-d_{\rm s}}\mathrm dv\,
e^{-\sigma(u-v)}e^{-i\omega v}
\nonumber\\
&\hspace{0.3cm}
-e^{-\sigma d_{\rm s}}e^{-i\omega(u-d_{\rm s})}
+e^{-\sigma d_{\rm c}}e^{-i\omega(u-d_{\rm c})}
\Bigg]
\nonumber\\
&\hspace{0.3cm}+\frac1\pi\int_0^\infty\mathrm d\sigma\,
\frac{\Im\mathcal R_\ell(i\sigma+0)}{\sigma}
\nonumber\\
&\hspace{0.3cm}\times\Bigg[
-\sigma\int_{u-d_{\rm s}}^\infty\mathrm dv\,
e^{\sigma(u-v)}e^{-i\omega v}
+e^{\sigma d_{\rm s}}e^{-i\omega(u-d_{\rm s})}
\Bigg]
\nonumber\\
&\hspace{0.3cm}+2G^-_0e^{-i\omega(u-d_{\rm c})}
+2G_0^{\rm pr}\left[
e^{-i\omega(u-d_{\rm c})}
-e^{-i\omega(u-d_{\rm s})}
\right]
\nonumber\\
&\hspace{0.3cm}={}
e^{-i\omega u}\Bigg\{
2G^-_0e^{i\omega d_{\rm c}}
+2G_0^{\rm pr}\left[
e^{i\omega d_{\rm c}}-e^{i\omega d_{\rm s}}
\right]
\nonumber\\
&\hspace{0.3cm}+\frac{i\omega}{\pi}
\int_0^\infty\mathrm d\sigma\,
\frac{\Im\mathcal R_\ell^{\rm pr}(-i\sigma+0)}{\sigma}
\frac{
e^{(i\omega-\sigma)d_{\rm c}}
-e^{(i\omega-\sigma)d_{\rm s}}
}{i\omega-\sigma}
\nonumber\\
&\hspace{0.3cm}+\frac{i\omega}{\pi}
\int_0^\infty\mathrm d\sigma\,
\frac{\Im\mathcal R_\ell(i\sigma+0)}{\sigma}
\frac{e^{(i\omega+\sigma)d_{\rm s}}}{i\omega+\sigma}
\Bigg\}\,,
\label{eq:monochromatic-direct-waveform}
\end{align}
Therefore, 
\begin{align}
&\mathcal{R}_{\rm dir}(\omega) ={}
2G^-_0e^{i\omega d_{\rm c}}
+2G_0^{\rm pr}\left[
e^{i\omega d_{\rm c}}-e^{i\omega d_{\rm s}}
\right]
\nonumber\\
&\hspace{0.3cm}+\frac{i\omega}{\pi}
\int_0^\infty\mathrm d\sigma\,
\frac{\Im\mathcal R_\ell^{\rm pr}(-i\sigma+0)}{\sigma}
\frac{
e^{(i\omega-\sigma)d_{\rm c}}
-e^{(i\omega-\sigma)d_{\rm s}}
}{i\omega-\sigma}
\nonumber\\
&\hspace{0.3cm}+\frac{i\omega}{\pi}
\int_0^\infty\mathrm d\sigma\,
\frac{\Im\mathcal R_\ell(i\sigma+0)}{\sigma}
\frac{e^{(i\omega+\sigma)d_{\rm s}}}{i\omega+\sigma}\,.
\end{align}

In summary, the auxiliary injection-surface construction allows us to obtain the response at $\mathscr I^+$ to a monochromatic wave incident from $\mathscr I^-$ and to decompose the reflection coefficient into direct, QNM, and tail contributions.
The scattering coefficient can be further related to the stellar dynamical tidal Love number through the tidal phase shift~\cite{Saketh:2026trm}. For $\ell=2$, the relation reads
\begin{equation}
k_2(\omega)
=
\frac{15\,\tan\delta_2^{\rm tidal}(\omega)}
{4(R\omega)^5\left(1+2\pi M|\omega|\right)}\,,
\label{eq:tidal-response-scattering-phase}
\end{equation}
where the scattering phase is defined by $\mathcal R_2(\omega)=-e^{2i\delta_2(\omega)}$.
The tidal phase shift is obtained by removing the ``free-propagation'' phase that unrelated to the tidal response~\cite{Miao:2025utd}.
At leading tidal order, this subtraction can be written as
$\delta_2^{\rm tidal}(\omega)=\delta_2^\star(\omega)-\delta_2^{\rm BH}(\omega)$, where the black hole has the same mass as the star and its higher-order absorptive contribution is neglected.
Our decomposition of $\mathcal R(\omega)$ may therefore provide a framework for investigating how the direct, QNM, and tail contributions combine to determine the dynamical tidal Love number and its static limit.

\subsection{Numerical implementation}
\label{sec:waveform-numerical-implementation}
In this section, we apply the framework developed above to a representative example involving a continuously chirping source. We focus in particular on the excitation of the stellar $f$-mode as the driving frequency sweeps through resonance. In a realistic binary merger, the relatively high $f$-mode frequency generally prevents the mode from undergoing a complete resonance crossing. The chirping source considered here therefore serves as an illustrative example, while the framework itself applies to arbitrary time-dependent source.

\subsubsection{A chirping source}
\label{sec:chirping-source}

As discussed in Sec.~\ref{sec:scattering-formalism}, it is convenient to introduce a prescribed incident field of the form given in Eq.~\eqref{eq:time-dependent-driver}. We extend this field into the stellar exterior as
\begin{equation}
Z^{\rm d}(t,r)
=
\chi(r)F(v)\,,
\label{eq:chirp-exterior-driver}
\end{equation}
where $\chi(r)$ smoothly interpolates from zero at the stellar surface to unity at spatial infinity. Substitution into Eq.~\eqref{eq:zerilli-effective-source} gives
\begin{equation}
S^{Z,{\rm d}}(t,r)
=
\left[
V^Z\chi-\partial_{r_*}^2\chi
\right]F(v)
-
2\left(\partial_{r_*}\chi\right)F'(v)\,.
\label{eq:chirp-effective-source}
\end{equation}

To focus on the resonant excitation of the stellar $f$-mode, we consider an incident wave whose frequency chirps through the $f$-mode frequency:
\begin{equation}
F(v) = w(v)e^{-i\Phi(v)}\,, \quad 
\Phi(v) = \omega_{fR}v+\frac{1}{2}\alpha v^2\,.
\label{eq:chirp-incident-waveform}
\end{equation}
Here $\omega_{fR}:=\Re\omega_f$ is the real part of the $f$-mode frequency, and $\alpha$ is the frequency-sweep rate. The instantaneous driving frequency is therefore
\begin{equation}
\omega(v)
= \frac{\mathrm d\Phi}{\mathrm dv}
= \omega_{fR}+\alpha v\,,
\label{eq:chirp-instantaneous-frequency}
\end{equation}
so that the driving frequency crosses the $f$-mode resonance at $v=0$.

The function $w(v)$ is a smooth compact window that controls the switch-on and switch-off of the incident wave. 
After the driving is switched off, the excited modes continue
to oscillate freely with their intrinsic damping.
Introducing the resonance timescale $T_{\rm res}:=\alpha^{-1/2}$, we choose
\begin{equation}
w(v)
=
s\!\left(\frac{v-v_{\rm on}}{2T_{\rm res}}\right)
s\!\left(\frac{v_{\rm off}-v}{2T_{\rm res}}\right)\,,
\label{eq:chirp-window}
\end{equation}
with the smooth step
\begin{equation}
s(y)
=
\begin{cases}
0\,, & y\leq 0\,,\\
\displaystyle
\frac{e^{-1/y}}
{e^{-1/y}+e^{-1/(1-y)}}\,,
& 0<y<1\,,\\
1, & y\geq 1\,.
\end{cases}
\label{eq:smooth-step-function}
\end{equation}
In practice, we take $v_{\rm on}=-12\,T_{\rm res}$ and
$v_{\rm off}=7\,T_{\rm res}$.

For the numerical calculation, we use the stellar model introduced in Sec.~\ref{sec:green-numerical-validation} and focus on $\ell=2$. Its $f$-mode frequency is $\omega_f = (0.049396 - 1.72676\times10^{-5}i)\,M_\odot^{-1}$.
We also take the sweep rate of $\alpha=10^{-5}\,M_\odot^{-2}$,
for which ${\alpha}/{\omega_{fR}^2}\simeq 4.1\times10^{-3}\ll1$.
The frequency sweep is therefore sufficiently slow near resonance for the stationary-phase approximation (SPA) to apply. The corresponding resonance timescale is $T_{\rm res}\approx 316.2\,M_\odot \approx226.3\,M$.

Finally, we choose the radial extension function in Eq.~\eqref{eq:chirp-exterior-driver} analytically as
\begin{equation}
\chi_p(r)
= \left(1-\frac{R}{r}\right)^p\,,
\qquad p\geq3\,.
\label{eq:radial-extension-function}
\end{equation}
The condition $p\geq3$ ensures that $\chi_p$, $\partial_{r_*}\chi_p$, and $\partial_{r_*}^2\chi_p$ vanish at the stellar surface, so that the effective source in Eq.~\eqref{eq:chirp-effective-source} is smooth there. In the calculation presented below, we take $p=5$. 
We caution that different values of $p$ correspond to different extensions of the same incident waveform from $\mathscr I^-$ into the finite-radius exterior.
The resulting effective sources, and hence the decomposition of the observed waveform into direct, QNM, and tail contributions, depend on $p$.
Nevertheless, the total outgoing waveform at $\mathscr I^+$ is independent of this choice, as we have also verified numerically.

\begin{figure*}[t]
\centering
\includegraphics[width=0.98\textwidth]{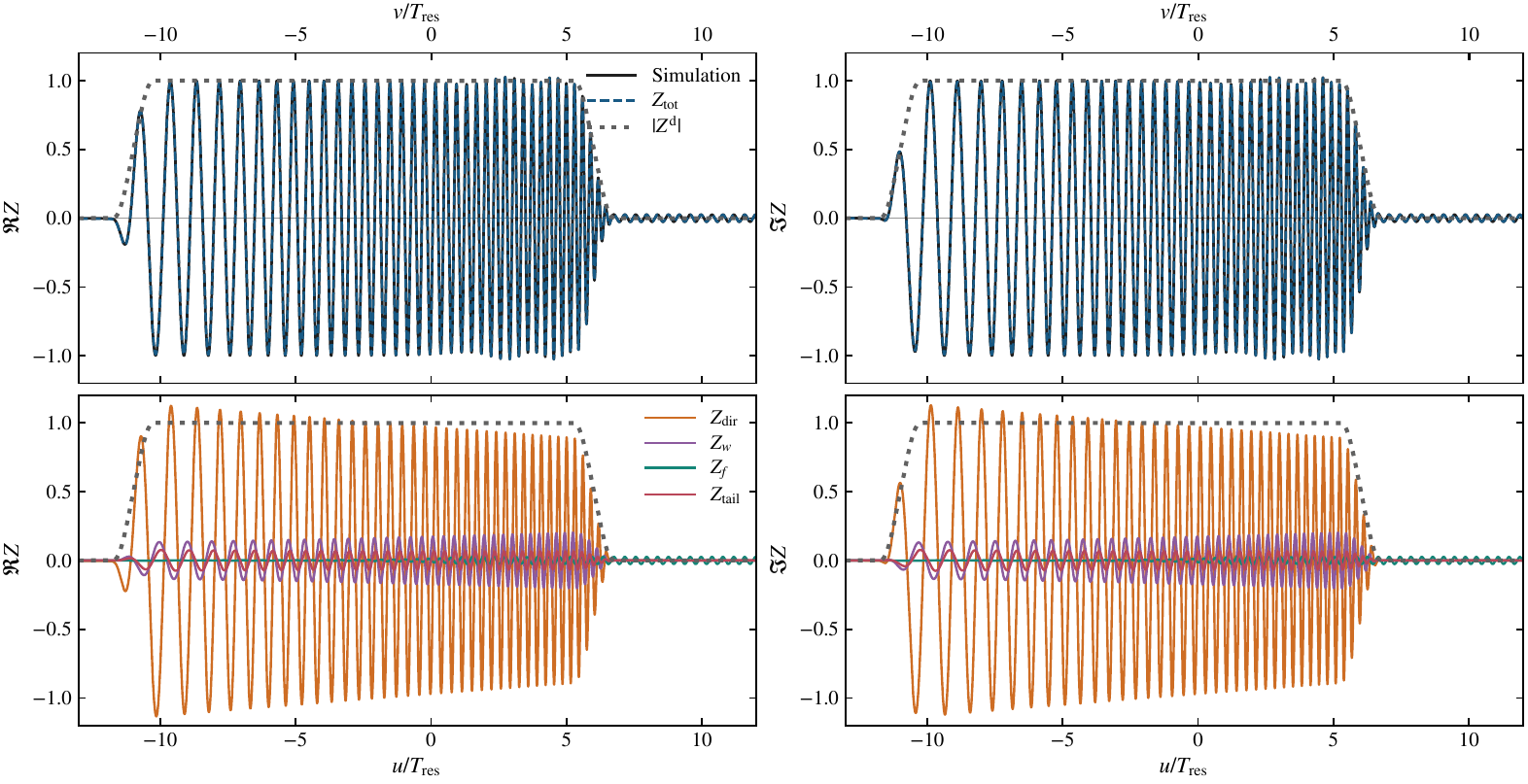}
\caption{Response to the incident chirp given in Sec.~\ref{sec:chirping-source}. The left and right columns show the real and imaginary parts, respectively. The upper panels compare the independent time-domain simulation with the reconstructed total response $Z_{\rm tot}=Z_{\rm dir}+Z_w+Z_f+Z_{\rm tail}$, while the lower panels show the four contributions separately. 
The simulation curve is mapped from $r_{\rm obs}$ to $\mathscr I^+$ using Eq.~\eqref{eq:chirp-finite-radius-map}. 
The gray dotted curves show the incident-wave amplitude envelope
evaluated at $v=u-d_{\rm c}$ (upper axis), aligned with the earliest arrival of the corresponding QNM response.}
\label{fig:chirp-waveform-decomposition}
\end{figure*}

\begin{figure}[t]
\centering
\includegraphics[width=\columnwidth]{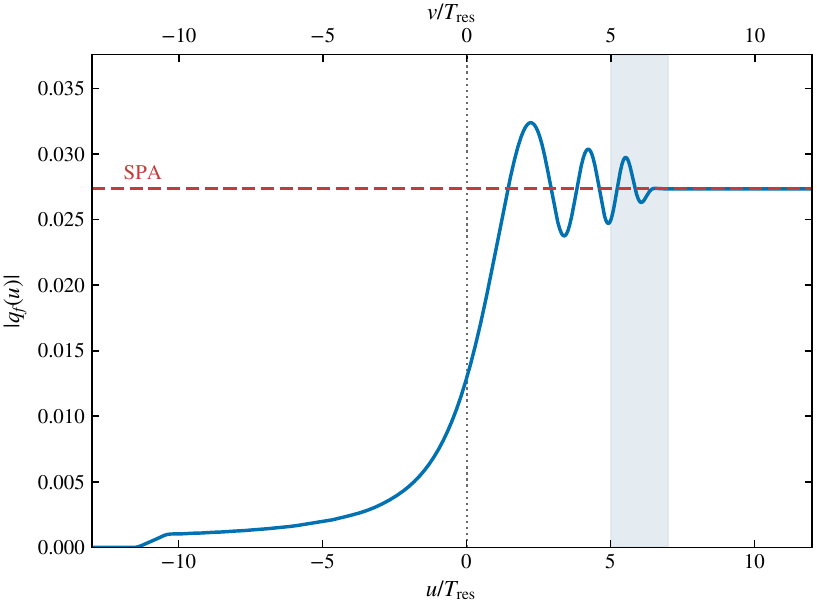}
\caption{Evolution of the damping-removed $f$-mode amplitude $q_f(u)$. 
The lower and upper axes are the same as in Fig.~\ref{fig:chirp-waveform-decomposition}.
The dotted vertical line marks the resonance crossing at $v=0$, while the shaded band marks the smooth switch-off of the incident wave during $v_{\rm off}-2T_{\rm res}\leq v\leq v_{\rm off}$.
The red dashed line shows the post-resonance amplitude predicted by the SPA calculation using Eq.~\eqref{eq:fmode-spa-amplitude}.
}
\label{fig:fmode-amplitude-spa}
\end{figure}

\subsubsection{Results}
\label{sec:chirp-results}

We calculate the response to the chirping source introduced above by convolving it with the direct, QNM, and tail components of the retarded Green function, as in Eqs.~\eqref{eq:waveform-direct-convolution}--\eqref{eq:waveform-tail-convolution}. To examine the $f$-mode excitation separately, we split the QNM waveform into
\begin{equation}
Z_{\rm QNM}^{\mathscr I^+}=Z_w^{\mathscr I^+}+Z_f^{\mathscr I^+}\,.
\label{eq:chirp-qnm-split}
\end{equation}
For $Z_w^{\mathscr I^+}$, we sum the $w$-modes with $n=0,\ldots,31$. We neglect the $p$-modes, whose contribution is much smaller for this source. 
We also compute $Z_f^{\mathscr I^+}$ independently using the mode-amplitude evolution equation, Eq.~\eqref{eq:sectoral-amplitude-evolution}, and find agreement with the Green-function convolution.

As an independent validation, we evolve the same incident chirp using the coupled metric--fluid equations in the AAKS variables, similar to that conducted in Sec.~\ref{sec:green-numerical-validation}. The initial data are generated by a characteristic evolution from $\mathscr I^-$. 
We extract the resulting waveform at $r_{\rm obs}=7497.9\,M_\odot\simeq800\,R$. 
The chirp-specific evolution setup is summarized in Appendix~\ref{sec:time-domain-validation-appendix}, while the AAKS variables are reviewed in Appendix~\ref{sec:aaks-appendix}.

To compare directly with the waveform calculated at future null
infinity, we map the time-domain simulation from the finite
extraction radius $r_{\rm obs}$ to $\mathscr I^+$ using~\cite{Nakano:2015rda}
\begin{align}
&Z^{\mathscr I^+}(u)
=
Z(u+r_{*\rm obs},r_{\rm obs})\nonumber\\
&\hspace{0.5cm}
-\frac{3}{r_{\rm obs}}
\int_{-\infty}^{u}\mathrm du'\,
Z(u'+r_{*\rm obs},r_{\rm obs})
+\mathcal O(r_{\rm obs}^{-2})\,,
\label{eq:chirp-finite-radius-map}
\end{align}
where $r_{*\rm obs}=r_*(r_{\rm obs})$.
Figure~\ref{fig:chirp-waveform-decomposition} shows the comparison with the independent time-domain evolution. The direct term dominates while the incident wave is present and closely follows its chirping phase. The $w$-mode sector produces a smaller response during the non-resonant driving phase and decays rapidly during the subsequent free-oscillation phase. By contrast, the $f$-mode is resonantly excited as the driving frequency passes through $\omega_{fR}$ and remains as a long-lived oscillation after the source has been switched off.  
For the source considered here, the combined $w$-mode contribution can exceed the $f$-mode contribution during driving despite the absence of a $w$-mode resonance, which is consistent with the relatively large excitation factors of the $w$-modes.
Overall, the reconstructed waveform agrees closely with the simulation in both its real and imaginary parts.

To examine the resonant excitation of the $f$-mode more directly, let $a_f(u)$ denote its amplitude and define the damping-removed coefficient
\begin{equation}
q_f(u)
:=
e^{i\omega_f u}a_f(u)\,,
\quad
|q_f(u)|
=
|a_f(u)|e^{\gamma_f u}\,,
\label{eq:fmode-damping-removed-amplitude}
\end{equation}
where $\gamma_f=-\Im\omega_f$.
For an adiabatic linear chirp, the post-resonance amplitude can be evaluated using SPA as 
\begin{equation}
q_f^{\rm SPA}
=
-\frac{2i\omega_f C_\ell}{\mathcal N_f}
\sqrt{\frac{2\pi}{\alpha}}
\exp\left[
\frac{i(\omega_f-\omega_{fR})^2}{2\alpha}
-\frac{i\pi}{4}
\right]\,.
\label{eq:fmode-spa-amplitude}
\end{equation}

Figure~\ref{fig:fmode-amplitude-spa} shows the evolution of $|q_f(u)|$. The amplitude remains small while the driving frequency is well below resonance. The chirp crosses resonance at $v=0$, and the Region-I contribution sourced at this crossing reaches the observer at $u=v+d_{\rm c}\approx 3.41\,M\approx0.015\,T_{\rm res}$. Around this time, the $f$-mode is rapidly excited and exhibits a finite-time oscillatory transient. 
After the incident wave has switched off and its final contribution has reached the observer at $u=v_{\rm off}+d_{\rm c}\approx7.015\,T_{\rm res}$, the $f$-mode oscillates freely, with constant $|q_f|$. 
The resulting late-time plateau agrees closely with the SPA
prediction and is also consistent with the mode-excitation amplitude
obtained in our previous work [cf.~Eq.~(36) of Ref.~\cite{Miao:2025utd}].

\section{Conclusion}
\label{sec:conclusion}

In this work, we have shown that the spacetime and matter response of a relativistic star, to an external tidal driving force, can be described by the dynamical evolution of a set of discrete modes and a continuous mode spectrum. The formalism is rather general, which is still applicable if the effective tidal-driving source in Eq.~\eqref{eq:driven-response} is replaced by a generic matter source. In addition, although the spectral decomposition of black hole perturbations has received significant progress in recent developments~\cite{DeAmicis:2025xuh,Arnaudo:2025kit,Arnaudo:2025uos,Su:2026fvj,Kuntz:2025gdq,Su:2026gmp,Zhang:2026xgv,Ma:2026qbq}, the effective oscillator representation of the metric perturbations has long been absent. In Appendix~\ref{sec:black-hole-bilinear-appendix}, assuming a Schwarzschild background, we adopt the same approach discussed in the main text to explicitly write down the complete decomposition into effective oscillators and their corresponding equations of motion.

There are at least several possible avenues for future developments. First, the neutron star oscillation during a typical binary inspiral stage can now be decomposed into a mode and a nonmodal (or a continuous spectrum of modes) basis, where the energy transfer and gravitational wave radiation can be consistently modelled in the relativistic setting. This is useful for modelling transient resonances for quasi-circular binaries as the orbital frequency sweeps through certain mode frequencies, but probably also instructive for the understanding of dynamic tides, which should include direct part, off-resonance $w$, $f$, and other sub-dominant modes according to the discussion in Sec.~\ref{sec:chirp-results}. 
It is also possible to produce eccentric binaries containing neutron stars through either multi-body interactions in dense stellar clusters~\cite{Samsing:2013kua,Clausen:2012zu,Lee:2009ca,Kocsis:2011jy,Petrovich:2017otm} or the Kozai--Lidov process in hierarchical triple systems~\cite{Wen:2002km,Liu:2018nrf,Stegmann:2025clo}.
In such systems, the mode excitation of stars becomes dynamically important during pericenter passages. The formalism developed in this work may help consistently evolve relevant star modes in this systems. 

Second, the formalism developed in this work provides a way to analytically decompose and understand generic star perturbations. Similar to the black hole case, it may be interesting to further consider stars under nonlinear tidal driving, where the coupling between different parts of the linear signal can also generate secondary signals, e.g., quadratic $f$ modes, $w$ modes, $f$-$w$ modes, etc. Since star-scattering calculations can also be performed in the Post-Minkowskian regime with diagram techniques~\cite{Ivanov:2024sds,Ivanov:2026icp}, it might be instructive to compare their results within regimes of overlap.

Third, although the primary target of this study is compact (fluid) stars, a similar construction may also apply for more general exotic compact objects, e.g., boson stars~\cite{Kaup:1968zz,Ruffini:1969qy,Liebling:2012fv}, gravastars~\cite{Mazur:2001fv,Mazur:2004fk}, AdS bubbles~\cite{Danielsson:2017riq,Giri:2024cks}, etc. The matter content and structures of various exotic compact objects generally differ from each other, but their dynamical response may still share the same direct part-quasinormal modes-tail structure. Understanding the physical content of each piece of the signal should help model these objects in dynamical environments and develop strategies for detecting them.

\begin{acknowledgments}
We acknowledge GPT-5.6 Sol and GPT-6 Astra for assistance with the numerical analysis. All results were verified by the authors.
ZM is supported by the Postdoctoral Innovation Talent Support Program of CPSF (No. BX20240223) and CPSF funded project (No. 2024M761948).
HY is supported by the Natural Science Foundation of China (Grant 12573048). NK is supported by the Shuimu fellowship of Tsinghua University. 

\end{acknowledgments}  

\bibliography{References}

\appendix
\section{Hamiltonian formulation of general polar perturbations}
\label{sec:hamiltonian-appendix}

In this appendix, we couple Brown's pullback action for a perfect fluid to the Einstein--Hilbert action and cast the combined system in ADM form. Its second variation about a static stellar background yields the Hamiltonian governing linear perturbations. We focus on even-parity perturbations and derive the associated constraints and Hamilton equations. Finally, we specialize the resulting system to Regge--Wheeler gauge. We take $\kappa =16\pi$ and $L=\ell(\ell+1)$.

\subsection{ADM--Brown action}

We first recall Brown's pullback action for a perfect fluid~\cite{Brown:1992kc}. Let $X^A$ ($A=1,2,3$) be comoving coordinates on matter space. A fixed three-form $N_{ABC}(X)$ defines the identically conserved particle current
\begin{equation}
n^a=\frac{1}{3!}\epsilon^{abcd}N_{ABC}(X)
\nabla_bX^A\nabla_cX^B\nabla_dX^C\,.
\label{eq:app-pullback-current}
\end{equation}
The proper particle density and fluid four-velocity are
\begin{equation}
n=\sqrt{-n^an_a}\,,\qquad u^a=\frac{n^a}{n}.
\end{equation}
For a zero-temperature fluid with frozen composition variables $x^I=x^I(X)$, the energy density is $\varepsilon=\varepsilon(n,x^I)$,
with
\begin{equation}
\mu=\varepsilon_{,n}\,,\quad
p=n\mu-\varepsilon\,,\quad
w=\varepsilon+p=n\mu\,.
\end{equation}
The pullback form of the fluid action is
\begin{equation}
S_{\rm f}[g,X]
=-\int\!\sqrt{-g}\,\varepsilon(n,x^I)\,\mathrm d^4x.
\label{eq:app-covariant-action}
\end{equation}

We add it to the Einstein--Hilbert action $S_{\rm EH}=\frac{1}{\kappa}\int\sqrt{-g}\,R\,\mathrm d^4x$ and use the ADM decomposition~\cite{Arnowitt:1962hi}
\begin{equation}
\mathrm ds^2=-N^2\mathrm dt^2
+h_{ij}(\mathrm dx^i+N^i\mathrm dt)
(\mathrm dx^j+N^j\mathrm dt)\,.
\end{equation}
Up to boundary terms, the full action takes the canonical form
\begin{equation}
S=\int\mathrm dt\,\mathrm d^3x\,
\bigl[\pi^{ij}\dot h_{ij}+P_A\dot X^A
-N\mathcal H-N^i\mathcal H_i\bigr]\,.
\label{eq:app-adm-brown-action}
\end{equation}
Here $\pi^{ij}=\sqrt h(K^{ij}-h^{ij}K)/\kappa$ are the momenta
conjugate to $h_{ij}$, with $K_{ij}=(\dot h_{ij}-D_iN_j-D_jN_i)/(2N)$, and $P_A$ are the momenta conjugate to $X^A$. Variation with respect to the lapse and shift gives the constraint equations
\begin{equation}
\mathcal H=0\,,\qquad \mathcal H_i=0\,.
\label{eq:app-total-constraints}
\end{equation}
These constraints split into gravitational and fluid parts,
\begin{equation}
\mathcal H=\mathcal H_g+\mathcal H_f,
\qquad
\mathcal H_i=\mathcal H_i^g+\mathcal H_i^f.
\end{equation}
The gravitational contributions, with $\pi=h_{ij}\pi^{ij}$, are
\begin{align}
\mathcal H_g&=\frac{\kappa}{\sqrt h}
\left(\pi^{ij}\pi_{ij}-\frac12\pi^2\right)
-\frac{\sqrt h}{\kappa}\,{}^{(3)}R\,,
\nonumber\\
\mathcal H_i^g&=-2D_j\pi^j{}_i\,.
\label{eq:app-gravity-constraints}
\end{align}
Here $h=\det(h_{ij})$, and ${}^{(3)}R$ is the scalar curvature
of $h_{ij}$. Let $U_i$ be the spatial projection of the fluid
four-velocity, and define $\Gamma=\sqrt{1+U_iU^i}$ and
$\Pi=\sqrt h\,n\Gamma$. Writing
$X_A{}^i\partial_iX^B=\delta_A{}^B$, Brown's fluid momentum and
constraints are
\begin{align}
P_A&=-\mu\Pi U_iX_A{}^i\,,
\nonumber\\
\mathcal H_f&=\sqrt h\bigl(\varepsilon+wU_iU^i\bigr)
=\sqrt h\bigl(w\Gamma^2-p\bigr)\,,
\nonumber\\
\mathcal H_i^f&=P_A\partial_iX^A=-\mu\Pi U_i\,.
\label{eq:app-fluid-constraints}
\end{align}

\subsection{First variation and static background}

We assume the unperturbed system to be static and spherically symmetric, and write
\begin{equation}
\mathrm ds^2=-N(r)^2\mathrm dt^2+A(r)^2\mathrm dr^2
+r^2\mathrm d\Omega^2\,.
\end{equation}
The background shift vanishes, and the canonical momenta satisfy
$\pi^{ij}=P_A=0$. For this background, the independent equations
obtained from the first variation read
\begin{align}
\frac{1-A^{-2}}{r^2}+\frac{2A'}{rA^3}
&=8\pi\varepsilon\,,
\nonumber\\
\frac{A^{-2}-1}{r^2}+\frac{2N'}{rNA^2}
&=8\pi p\,,
\nonumber\\
-(\varepsilon+p)\frac{N'}{N}&=p'\,.
\label{eq:app-background-equations}
\end{align}
Here a prime denotes $\mathrm d/\mathrm dr$. The first two equations follow from the variations of $N(r)$ and $A(r)$; the last follows from the fluid variation and is equivalently implied by the Einstein
equations through the Bianchi identity.

Introducing the mass function
\begin{equation}
m(r):=\frac r2\bigl(1-A^{-2}\bigr)\,,
\end{equation}
we obtain the familiar form,
\begin{align}
m'&=4\pi r^2\varepsilon\,,
\nonumber\\
\frac{N'}{N}
&=\frac{m+4\pi r^3p}
{r(r-2m)}\,,
\nonumber\\
p'&=-\frac{(\varepsilon+p)(m+4\pi r^3p)}
{r(r-2m)}\,.
\label{eq:app-tov-equations}
\end{align}
where the third equation is the Tolman--Oppenheimer--Volkoff equation~\cite{Tolman1939,Oppenheimer1939}.

\subsection{Second variation}

We now take the second variation of the ADM--Brown action about the
static solution obtained above. Consider a one-parameter family of
functions
\begin{align}
h_{ij}(e)&=\bar h_{ij}+e q_{ij}\,,
&
\pi^{ij}(e)&=e p^{ij}\,,
\nonumber\\
X^A(e)&=\bar X^A+e Q^A\,,
&
P_A(e)&=e\mathsf P_A\,,
\nonumber\\
N(e)&=\bar N+e n_1\,,
&
N^i(e)&=e n_1^i\,.
\end{align}
Here an overbar denotes the unperturbed solution. The second
variation of $S$ leads to
\begin{align}
J
&:=\left.\frac12\frac{\mathrm d^2S}{\mathrm de^2}
\right|_{e=0}
\nonumber\\
&=\int\mathrm dt\,\mathrm d^3x\,
\Bigl[
p^{ij}\dot q_{ij}+\mathsf P_A\dot Q^A
-\frac{\bar N}{2}\delta^2\mathcal H
-n_1\delta\mathcal H
-n_1^i\delta\mathcal H_i
\Bigr]\,.
\label{eq:app-exact-second-variation}
\end{align}
Boundary integrals that do not contribute to the variation under
the chosen boundary conditions have been discarded, since they do
not alter the Hamilton equations obtained from $J$.

Here and below, a vertical bar denotes the covariant derivative
associated with the background spatial metric $\bar h_{ij}$.
The linearized Hamiltonian constraint is
\begin{align}
\delta\mathcal H
={}&-\frac{\sqrt{\bar h}}{\kappa}
\left(
\delta\,{}^{(3)}R
+\frac12 q\,{}^{(3)}\bar R
\right)
\nonumber\\
&+\sqrt{\bar h}
\left(
\frac12\bar\varepsilon q
+\bar\mu\,\delta n
+\bar\varepsilon_{,I}\,\delta x^I
\right)\,,
\label{eq:app-linear-hamiltonian-constraint}
\end{align}
where $q=\bar h^{ij}q_{ij}$,
$\delta x^I=x^I{}_{,A}Q^A$, and
\begin{equation}
\delta\,{}^{(3)}R
=q^{ij}{}_{|ij}-q^{|i}{}_{|i}
-{}^{(3)}\bar R^{ij}q_{ij}\,.
\end{equation}
The quadratic Hamiltonian constraint has the general structure
\begin{equation}
\delta^2\mathcal H
=
\frac{2\kappa}{\sqrt{\bar h}}
\left(p^{ij}p_{ij}-\frac12p^2\right)
-\frac{1}{\kappa}\delta^2
\!\left(\sqrt h\,{}^{(3)}R\right)
+\delta^2\mathcal H_{\rm f}\,,
\label{eq:app-quadratic-hamiltonian-constraint}
\end{equation}
where $p=\bar h_{ij}p^{ij}$ and
$\delta^2\mathcal H_{\rm f}$ contains the second-order fluid-coordinate
and fluid-momentum contributions. The linearized momentum constraint is
\begin{equation}
\delta\mathcal H_i
=-2p^j{}_{i|j}
+\mathsf P_A\partial_i\bar X^A\,,
\label{eq:app-linear-momentum-constraint}
\end{equation}
where $p^j{}_i=\bar h_{ik}p^{jk}$. Indices are raised and lowered
with $\bar h_{ij}$.

Writing $J$ in canonical form defines the quadratic Hamiltonian
\begin{equation}
\mathscr H
=\int\mathrm d^3x\,
\left[
\frac{\bar N}{2}\delta^2\mathcal H
+n_1\delta\mathcal H
+n_1^i\delta\mathcal H_i
\right]\,.
\label{eq:app-quadratic-hamiltonian}
\end{equation}
Variation with respect to $n_1$ and $n_1^i$ imposes the linearized
constraints $\delta\mathcal H=0$ and $\delta\mathcal H_i=0$,
respectively.

\subsection{Even-parity perturbations}

For the even-parity perturbation of order $(\ell,m)$, we may write
the metric perturbation in Regge--Wheeler harmonic form~\cite{Regge:1957td},
\begin{align}
&q_{ij}
=h_1(\hat f_1)_{ij}
+A^2H_2(\hat f_2)_{ij}
+r^2K(\hat f_3)_{ij}
+r^2G(\hat f_4)_{ij}\,,
\\
&n_1
=-\frac{N}{2}H_0Y_{\ell m}\,,
\\
&\bigl[(n_1)_i\bigr]
=
\begin{pmatrix}
H_1Y_{\ell m}&
h_0\partial_\theta Y_{\ell m}&
h_0\partial_\phi Y_{\ell m}
\end{pmatrix}\,.
\label{eq:app-general-polar-metric}
\end{align}
where
\begin{align}
(\hat f_1)_{ij}
&=
\begin{pmatrix}
0&\partial_\theta&\partial_\phi \\
\rm sym &0&0\\
\rm sym &0&0
\end{pmatrix}Y_{\ell m}\,,
\\
(\hat f_2)_{ij}
&=
\begin{pmatrix}
1&0&0\\
0&0&0\\
0&0&0
\end{pmatrix}Y_{\ell m}\,,
\nonumber\\
(\hat f_3)_{ij}
&=
\begin{pmatrix}
0&0&0\\
0&1&0\\
0&0&\sin^2\theta
\end{pmatrix}Y_{\ell m}\,,
\\
(\hat f_4)_{ij}
&=
\begin{pmatrix}
0&0&0\\
0&\partial_\theta^2
&\partial_\theta\partial_\phi 
 -\cot\theta\,\partial_\phi \\
0&\rm sym
&\partial_\phi^2
 +\sin\theta\cos\theta\,\partial_\theta
\end{pmatrix}Y_{\ell m}\,.
\label{eq:app-even-tensor-harmonics}
\end{align}

The fluid displacement can be written in the coordinate order
$(r,\theta,\phi)$ as
\begin{equation}
\bigl[\xi^i\bigr]
=
\begin{pmatrix}
\dfrac{W}{Ar^2}Y_{\ell m}\\
-\dfrac{V}{r^2}\partial_\theta Y_{\ell m}\\
-\dfrac{V}{r^2\sin^2\theta}\partial_\phi Y_{\ell m}
\end{pmatrix}\,,
\label{eq:app-general-fluid-displacement}
\end{equation}
For frozen composition, $\Delta x^I=0$, and the Eulerian perturbation is
\begin{equation}
\delta x^I
=-\xi^r x^{I\prime}
=-\frac{W}{Ar^2}x^{I\prime}Y_{\ell m}\,.
\end{equation}
Thus, before constraint reduction or gauge fixing, the polar sector
contains six configuration amplitudes
$(H_2,K,h_1,G,W,V)$ and three lapse--shift amplitudes
$(H_0,H_1,h_0)$.

In the harmonic basis introduced above, we obtain
\begin{align}
p_{ij}
=\frac{\sqrt{\bar h}}{2\kappa N}\Bigl[
&\mathsf B(\hat f_1)_{ij}
+A^2(-2\mathsf Z+L\mathsf T)(\hat f_2)_{ij}
\nonumber\\
&+r^2(-\mathsf X-\mathsf Z+L\mathsf T)(\hat f_3)_{ij}
+r^2\mathsf T(\hat f_4)_{ij}
\Bigr]\,,
\label{eq:app-polar-adm-momentum}
\end{align}
where
\begin{subequations}\label{eq:app-velocity-combinations}  
\begin{align}
\mathsf X
&=\dot H_2-\frac{2}{A^2}
\left(H_1'-\frac{A'}{A}H_1\right)\,,\\
\mathsf Z
&=\dot K-\frac{2H_1}{A^2r}\,,\\
\mathsf B
&=\dot h_1-h_0'+\frac{2h_0}{r}-H_1\,,\\
\mathsf T
&=\dot G-\frac{2h_0}{r^2}\,.
\end{align}
\end{subequations}

For the fluid, the Brown momentum relation in
Eq.~\eqref{eq:app-fluid-constraints}, together with
$Q^A=-\xi^i\partial_i\bar X^A$, gives
\begin{subequations}\label{eq:app-linear-fluid-momentum}
\begin{align}
\mathsf P_A
&=-\sqrt{\bar h}\,wU_i\bar X_A{}^i\,,\\
\mathsf P_A\dot Q^A
&=\sqrt{\bar h}\,wU_i\dot\xi^i\,,\\
U^i&=\frac{\dot\xi^i+n_1^i}{N}\,.
\end{align}
\end{subequations}

Using these expansions and integrating over the angles, we obtain
\begin{align}
\Theta
&:=\int\mathrm d^3x\,
\left(p^{ij}\dot q_{ij}+\mathsf P_A\dot Q^A\right)
\nonumber\\
&=\int\mathrm dr\,\bigl(
P_{H_2}\dot H_2+P_K\dot K
+P_{h_1}\dot h_1+P_G\dot G
\nonumber\\
&\hspace{4.7em}
+P_W\dot W+P_V\dot V
\bigr)\,,
\label{eq:app-radial-canonical-term}
\end{align}
where the momenta conjugate to $(H_2,K,h_1,G,W,V)$ are given by
\begin{subequations}\label{eq:app-general-momenta}
\begin{align}
P_{H_2}
&=\frac{Ar^2}{\kappa N}
\left(-\mathsf Z+\frac L2\mathsf T\right)\,,\\
P_K
&=\frac{Ar^2}{\kappa N}
\left(-\mathsf X-\mathsf Z+\frac L2\mathsf T\right)\,,\\
P_{h_1}
&=\frac{L}{\kappa NA}\mathsf B\,,\\
P_G
&=\frac{Ar^2L}{2\kappa N}
\left(\mathsf X+\mathsf Z-\mathsf T\right)\,,\\
P_W&=\frac wN
\left(\frac A{r^2}\dot W+H_1\right)\,,\\
P_V&=\frac{AwL}{N}(\dot V-h_0).
\end{align}
\end{subequations}

\subsection{Constraints and Hamiltonian equations}

Using the canonical coordinates and momenta, we can get
\begin{align}
J=\int\mathrm dt\,\mathrm dr\,
\Bigl[
&\sum_{Q\in\{H_2,K,h_1,G,W,V\}}P_Q\dot Q
\nonumber\\
&-\mathfrak h_{\rm mom}-\mathfrak h_{\rm shift}
-\mathscr V^{(2)}
+\frac N2H_0\mathcal C_0
\Bigr]\,,
\label{eq:app-complete-radial-action}
\end{align}
where $\mathscr V^{(2)}$ is the potential obtained from the second
variation in Eq.~\eqref{eq:app-exact-second-variation}. 
The momentum-dependent part is
\begin{align}
\mathfrak h_{\rm mom}
={}&\frac{\kappa N}{2Ar^2}
\left[
P_{H_2}^2-2P_{H_2}P_K
+\frac{L}{L-2}
\left(P_K+\frac{2P_G}{L}\right)^2
\right]
\nonumber\\
&+\frac{\kappa NA}{2L}P_{h_1}^2
+\frac{Nr^2}{2Aw}P_W^2
+\frac{N}{2AwL}P_V^2\,.
\label{eq:app-hamiltonian-momentum-part}
\end{align}
After radial integration by parts, the shift-dependent part becomes
\begin{equation}
\mathfrak h_{\rm shift}
=H_1\mathcal C_r+h_0\mathcal C_\Omega\,,
\label{eq:app-hamiltonian-shift-part}
\end{equation}
up to a boundary term.

Variation with respect to $H_0$, $H_1$, and $h_0$ imposes the three
linearized constraints
\begin{align}
\mathcal C_0
&=Ar^2\left(
\delta\varepsilon_{\ell m}
-\frac{\delta R_{\ell m}}{\kappa}
\right)=0\,,
\label{eq:app-general-hamiltonian-constraint}\\
\mathcal C_r
&=-\left(\frac{2P_{H_2}}{A^2}\right)'
-\frac{2A'}{A^3}P_{H_2}
+\frac{2P_K}{A^2r}
+P_{h_1}
-\frac{r^2}{A}P_W=0\,,
\label{eq:app-general-radial-constraint}\\
\mathcal C_\Omega
&=-P_{h_1}'-\frac2rP_{h_1}
+\frac{2P_G}{r^2}+P_V=0\,.
\label{eq:app-general-angular-constraint}
\end{align}
Here $\delta\varepsilon_{\ell m}$ and $\delta R_{\ell m}$ are the
$(\ell,m)$ coefficients of the fluid energy-density and spatial
scalar-curvature perturbations, respectively. Their explicit expressions are 
\begin{equation}
\delta\varepsilon_{\ell m}
=w\left[-\frac{W'}{Ar^2}-\frac{L}{r^2}V
-\frac12H_2-K+\frac L2G\right]
-\frac{\varepsilon'}{Ar^2}W\,.
\label{eq:app-general-density-perturbation}
\end{equation}
and
\begin{align}
\delta R_{\ell m}={}&\frac{1}{r^2A^3}
\Bigl[LA^3(H_2+K)-2r^2AK''\nonumber\\
&+2r^2A'K'
+2rAH_2'-6rAK'
\nonumber\\
&-4rA'H_2-2A^3K+2AH_2\nonumber\\
&+Lr^2AG''-Lr^2A'G'+3LrAG'\Bigr]
\nonumber\\
&-\frac{2L}{A^2r^2}
\left[h_1'+\left(\frac1r-\frac{A'}A\right)h_1\right]\,.
\label{eq:app-general-curvature-perturbation}
\end{align}

The Hamiltonian corresponding to Eq.~\eqref{eq:app-complete-radial-action}
is
\begin{equation}
\mathscr H
=\int\mathrm dr\,
\left[
\mathfrak h_{\rm mom}
+\mathscr V^{(2)}
+H_1\mathcal C_r+h_0\mathcal C_\Omega
-\frac N2H_0\mathcal C_0
\right]\,.
\label{eq:app-radial-hamiltonian}
\end{equation}
Variation with respect to $P_Q$ and $Q$ gives the Hamiltonian equations
\begin{align}
    \dot Q&=\frac{\delta\mathscr H}{\delta P_Q}\,,\\
    \dot P_Q &=-\frac{\delta\mathscr H}{\delta Q}\,,
    \label{eq:app-general-hamilton-equations}
\end{align}

\subsection{Explicit expressions in Regge--Wheeler gauge}

We now impose Regge--Wheeler gauge by setting
\begin{equation}
h_0=h_1=G=0\,.
\label{eq:app-rw-gauge}
\end{equation}
The coordinate Hamilton equations then simplify the
momentum--velocity relations to
\begin{align}
P_{H_2}
&=-\frac{Ar^2}{\kappa N}
\left(\dot K-\frac{2H_1}{A^2r}\right)\,,
\nonumber\\
P_K
&=-\frac{Ar^2}{\kappa N}
\left[
\dot H_2+\dot K
-\frac2{A^2}
\left(H_1'-\frac{A'}A H_1\right)
-\frac{2H_1}{A^2r}
\right]\,,
\nonumber\\
P_{h_1}
&=-\frac{L}{\kappa NA}H_1\,,\nonumber\\
P_G&=-\frac L2P_K\,,
\nonumber\\
P_W
&=\frac wN\left(\frac A{r^2}\dot W+H_1\right)\,,\nonumber\\
P_V&=\frac{AwL}{N}\dot V\,.
\label{eq:app-rw-physical-momenta}
\end{align}

For the fluid terms, by defining
\begin{equation}
\Gamma_1
=\frac np\left(\frac{\partial p}{\partial n}\right)_{x^I}\,.
\label{eq:app-rw-compression}
\end{equation}
we can obtain the energy-density and pressure perturbations with
\begin{align}
\delta\varepsilon
&=-w\left(\frac{W'}{Ar^2}+\frac{L}{r^2}V
+\frac12H_2+K\right)-\frac{\varepsilon'}{Ar^2}W\,,\label{eq:app-rw-thermodynamic-perturbations-e}\\
\delta p
&=-\Gamma_1p\left(\frac{W'}{Ar^2}+\frac{L}{r^2}V
+\frac12H_2+K\right)-\frac{p'}{Ar^2}W\,.
\label{eq:app-rw-thermodynamic-perturbations-p}
\end{align}

Evaluating the conjugate Hamilton equations and eliminating the
canonical momenta yields
\begin{align}
&\dot H_1
=N^2\left(H_2'-K'+2\frac{N'}N H_2\right)\,,
\label{eq:app-rw-H1-evolution}\\
&H_0=H_2\,,
\label{eq:app-rw-H0-H2}\\
&\frac{\ddot K}{N^2}
={}fK''
+\left[\frac2r-\frac{2m}{r^2}
+4\pi r(p-\varepsilon)\right]K'\nonumber\\
&\hspace{0.4cm}+\left[\frac{2-L}{r^2}-8\pi w\right]K
+\left[-\frac{2f}{r^2}+4\pi w\right]H_2\nonumber\\
&\hspace{0.4cm}-8\pi\delta p
-\frac{8\pi}{Ar^2}(wW'+\varepsilon'W)
-\frac{8\pi L}{r^2}wV\,,
\label{eq:app-rw-K-evolution}\\
&\frac{\ddot H_2+\ddot K}{N^2}
={}f(H_2''-K'')
+\left[\frac{6m}{r^2}
+4\pi r(5p-\varepsilon)\right]H_2'\nonumber\\
&\hspace{0.4cm}+\left[-\frac{4m}{r^2}
+4\pi r(\varepsilon-3p)\right]K'
\nonumber\\
&\hspace{0.4cm}+\left[
16\pi w-\frac{4m}{r^3}
+\frac{4(m+4\pi r^3p)^2}{r^4f}
\right]H_2
-16\pi\delta p\,,
\label{eq:app-rw-HK-evolution}\\
&\frac A{r^2}\ddot W+\dot H_1
={}N^2\left[
\frac12H_2'
-\frac{\delta p'}w
-\frac{N'}{Nw}(\delta\varepsilon+\delta p)
\right]\,,
\label{eq:app-rw-W-evolution}\\
&\ddot V
={}N^2\left(\frac{\delta p}{w}-\frac12H_2\right)\,.
\label{eq:app-rw-V-evolution}
\end{align}
where $f=A^{-2}=1-2m/r$.

The origin of these equations is worth making explicit. The $h_1$ momentum equation gives the evolution of $H_1$. The $G$ and $K$ momentum equations obey
\begin{equation}
\dot P_G+\frac L2\dot P_K
=\frac{L(L-2)AN}{4\kappa}(H_2-H_0)\,.
\label{eq:app-angular-tracefree-relation}
\end{equation}
The gauge-preserving coordinate equation gives
$P_G=-LP_K/2$, so the left-hand side vanishes and $H_0=H_2$
for $\ell\ge2$. Once this relation is imposed, the remaining
$G$ equation is equivalent to the $K$ equation. The $H_2$ and $K$
momentum equations give
Eqs.~\eqref{eq:app-rw-K-evolution} and
\eqref{eq:app-rw-HK-evolution}, respectively, after using the
momentum--velocity relations and constraints. The $W$ and $V$
momentum equations give the radial and angular fluid equations
\eqref{eq:app-rw-W-evolution} and
\eqref{eq:app-rw-V-evolution}. 

In Regge-Wheeler gauge, and after using $H_0=H_2$, the three linearized
constraints in
Eqs.~\eqref{eq:app-general-hamiltonian-constraint}--
\eqref{eq:app-general-angular-constraint} reduce to
\begin{align}
&H_2'
+\frac{A^2}{r}
\left[\frac L2+1+4\pi r^2(p-\varepsilon)\right]H_2
\nonumber\\
&=rK''
+A^2\left(3-\frac{5m}{r}-4\pi r^2\varepsilon\right)K'\nonumber\\
&-\frac{A^2}{r}
\left(\frac L2-1+8\pi r^2w\right)K
\nonumber\\
&-\frac{8\pi A}{r}(wW'+\varepsilon'W)
-\frac{8\pi LA^2}{r}wV\,,
\label{eq:app-rw-hamiltonian-constraint}\\
&LH_1
=2r^2\dot K'
+2r\left(1-r\frac{N'}{N}\right)\dot K
-2r\dot H_2-\kappa Aw\dot W\,,
\label{eq:app-rw-radial-constraint}\\
&H_1'
=A^2(\dot H_2+\dot K)
+\left(\frac{A'}A-\frac{N'}N\right)H_1
+\kappa A^2w\dot V\,.
\label{eq:app-rw-angular-constraint}
\end{align}

We caution that these equations are not all independent. Let $\mathscr C_{\rm H}$, $\mathscr C_r$, and $\mathscr C_\Omega$ denote the left-hand sides minus the right-hand sides of Eqs.~\eqref{eq:app-rw-hamiltonian-constraint}--\eqref{eq:app-rw-angular-constraint}, respectively. The background equations imply
\begin{equation}
2r\,\partial_t\mathscr C_{\rm H}
=
\partial_r\mathscr C_r
-\left(\frac{A'}A-\frac{N'}N\right)\mathscr C_r
-L\mathscr C_\Omega\,.
\label{eq:app-constraint-dependence}
\end{equation}
Thus, when the Hamiltonian and radial momentum constraints hold at all times, the angular momentum constraint follows for $\ell\ge2$.
Taking the time derivative of the angular momentum constraint and using Eqs.~\eqref{eq:app-rw-H1-evolution} and
\eqref{eq:app-rw-V-evolution} then yields
Eq.~\eqref{eq:app-rw-HK-evolution}. Consequently, after imposing $H_0=H_2$, one may retain the evolution equations for
$H_1$, $K$, $W$, and $V$, together with the Hamiltonian and radial momentum constraints, as a sufficient set of six equations.
It is straightforward to verify the remaining six equations are equivalent to the perturbation equations obtained by Thorne and Campolattaro from the linearized Einstein equations~\cite{Thorne1967}.

Finally, we can collect the RW-gauge canonical variables into the phase-space vector of order $(\ell,m)$,
\begin{equation}
\boldsymbol\Psi
=
\begin{pmatrix}
\boldsymbol Q\\
\boldsymbol P
\end{pmatrix}\,,
\label{eq:app-phase-space-vector}
\end{equation}
with 
\begin{equation}
\boldsymbol Q= \begin{pmatrix}
H_{2}\\ K\\ W\\ V
\end{pmatrix}\,,
\qquad
\boldsymbol P
=
\begin{pmatrix}
P_{H_2}\\ P_{K}\\ P_{W}\\ P_{V}
\end{pmatrix}\,.
\end{equation}
These variables are understood to satisfy the linearized constraints
and the stellar boundary and matching conditions. The quadratic
Hamiltonian then defines a linear evolution operator through
\begin{equation}
\partial_t\boldsymbol\Psi
=
\mathbb J
\frac{\delta \mathscr H}
     {\delta\boldsymbol\Psi}
\equiv
\mathbb A\boldsymbol\Psi\,,
\label{eq:app-hamiltonian-generator}
\end{equation}
where $\mathbb J$ is the canonical symplectic matrix. Since the background is static, $\mathbb A$ is time independent. This is the phase-space evolution equation used in Eq.~\eqref{eq:hamiltonian-evolution} of the main text. 
In the vacuum exterior, the fluid variables are absent. Set
\begin{equation}
\Lambda=\frac{L-2}{2}r+3M\,,
\end{equation}
We introduce the exterior variable
\begin{equation}
X:=\frac{2}{L}
\left[r^2K'+r\left(1-r\frac{N'}{N}\right)K-rH_2\right]\,,
\label{eq:app-exterior-X}
\end{equation}
which obeys $H_1 = \dot X$ [cf. Eq.~\eqref{eq:app-rw-radial-constraint}].
The Zerilli canonical pair is then given by
\begin{align}\label{eq:app-exterior-zerilli-pair}
Z&=\frac{r}{\Lambda}(rK-fX)\,,\\
\Pi_Z&=C_\ell\dot Z=
C_\ell\frac{r}{\Lambda}(r\dot K-fH_1)\,,
\end{align}
where 
\begin{equation}
    C_\ell = \frac{L(L-2)}{4\kappa} = \frac{1}{64\pi}\frac{(\ell+2)!}{(\ell-2)!}\,.
\end{equation}

\section{Local Green identity and the symplectic current}
\label{sec:symplectic-appendix}

The Hamiltonian formulation derived in the preceding section gives, for a fixed $(\ell,m)$ sector with harmonic indices suppressed,
\begin{equation}
J=\int\mathrm dt\,\mathrm dr\,\mathcal L\,,
\qquad
\mathcal L=\boldsymbol P^T\dot{\boldsymbol Q}-\mathfrak h\,,
\label{eq:app-local-canonical-lagrangian}
\end{equation}
where $\mathfrak h$ is the radial integrand of the Hamiltonian in Eq.~\eqref{eq:app-radial-hamiltonian}. 
We may perform a variation on the Lagrangian $\mathcal L$ and integrate by parts to get the equations of motion ($\dot{\boldsymbol Q}=\boldsymbol E_P$ and $\dot{\boldsymbol P}=-\boldsymbol E_Q$), i.e.,
\begin{align}
\delta\mathcal L
={}&
(\dot{\boldsymbol Q}-\boldsymbol E_P)^T\delta\boldsymbol P
+(-\dot{\boldsymbol P}-\boldsymbol E_Q)^T\delta\boldsymbol Q
\nonumber\\
&+\partial_t(\boldsymbol P^T\delta\boldsymbol Q)
-\partial_r\beta[\boldsymbol\Psi;\delta\boldsymbol\Psi]\,.
\label{eq:app-local-lagrangian-variation}
\end{align}
Retaining radial derivatives through second order, we have
\begin{align}
\boldsymbol E_Q
&=
\frac{\partial\mathfrak h}{\partial\boldsymbol Q}
-\partial_r\frac{\partial\mathfrak h}{\partial\boldsymbol Q'}
+\partial_r^2\frac{\partial\mathfrak h}{\partial\boldsymbol Q''}\,,
\nonumber\\
\boldsymbol E_P
&=
\frac{\partial\mathfrak h}{\partial\boldsymbol P}
-\partial_r\frac{\partial\mathfrak h}{\partial\boldsymbol P'}
+\partial_r^2\frac{\partial\mathfrak h}{\partial\boldsymbol P''}\,,
\label{eq:app-local-euler-derivatives}
\end{align}
and
\begin{align}
\beta[\boldsymbol\Psi;\delta\boldsymbol\Psi]
={}&
\left(
\frac{\partial\mathfrak h}{\partial\boldsymbol Q'}
-\partial_r\frac{\partial\mathfrak h}{\partial\boldsymbol Q''}
\right)^T\delta\boldsymbol Q
+
\left(
\frac{\partial\mathfrak h}{\partial\boldsymbol Q''}
\right)^T\delta\boldsymbol Q'
\nonumber\\
&+
\left(
\frac{\partial\mathfrak h}{\partial\boldsymbol P'}
-\partial_r\frac{\partial\mathfrak h}{\partial\boldsymbol P''}
\right)^T\delta\boldsymbol P
+
\left(
\frac{\partial\mathfrak h}{\partial\boldsymbol P''}
\right)^T\delta\boldsymbol P'\,.
\label{eq:app-local-boundary-one-form}
\end{align}
Terms involving derivatives absent from $\mathfrak h$ vanish identically.

Defining the symplectic potential $\Theta^a$ by the total-derivative term in Eq.~\eqref{eq:app-local-lagrangian-variation}, we obtain
\begin{equation}
\Theta^t[\boldsymbol\Psi;\delta\boldsymbol\Psi]
=\boldsymbol P^T\delta\boldsymbol Q\,,
\qquad
\Theta^r[\boldsymbol\Psi;\delta\boldsymbol\Psi]
=-\beta[\boldsymbol\Psi;\delta\boldsymbol\Psi]\,.
\label{eq:app-symplectic-potential}
\end{equation}
With the convention
\begin{equation}
\varpi^a[1,2]
:=
\delta_2\Theta^a[\delta_1]
-\delta_1\Theta^a[\delta_2]\,,
\end{equation}
the symplectic current then reads
\begin{align}
\varpi^t[1,2]
&=
\boldsymbol Q_1^T\boldsymbol P_2
-\boldsymbol P_1^T\boldsymbol Q_2
=
\boldsymbol\Psi_1^T\mathbb J\boldsymbol\Psi_2\,,
\nonumber\\
\varpi^r[1,2]
&=
\beta[\boldsymbol\Psi_1;\boldsymbol\Psi_2]
-\beta[\boldsymbol\Psi_2;\boldsymbol\Psi_1]\,.
\label{eq:app-local-symplectic-current}
\end{align}

Because the second variations of $\mathfrak h$ commute, applying
Eq.~\eqref{eq:app-local-lagrangian-variation} to
$\boldsymbol\Psi_1$ and $\boldsymbol\Psi_2$ gives
\begin{align}
\partial_r\varpi^r[1,2]
={}&
\boldsymbol Q_1^T\boldsymbol E_Q[\boldsymbol\Psi_2]
+\boldsymbol P_1^T\boldsymbol E_P[\boldsymbol\Psi_2]
\nonumber\\
&-
\boldsymbol Q_2^T\boldsymbol E_Q[\boldsymbol\Psi_1]
-\boldsymbol P_2^T\boldsymbol E_P[\boldsymbol\Psi_1]\,.
\label{eq:app-radial-green-identity}
\end{align}
The linearized Hamilton equations identify
\begin{equation}
\mathbb A\boldsymbol\Psi
=
\begin{pmatrix}
\boldsymbol E_P[\boldsymbol\Psi]\\
-\boldsymbol E_Q[\boldsymbol\Psi]
\end{pmatrix}\,,
\end{equation}
and thus
\begin{equation}
\mathcal O\boldsymbol\Psi
=
\begin{pmatrix}
\boldsymbol{\dot Q}-\boldsymbol E_P[\boldsymbol\Psi]\\
\boldsymbol{\dot P}+\boldsymbol E_Q[\boldsymbol\Psi]
\end{pmatrix}\,,
\end{equation}
where $\mathcal O:=\partial_t-\mathbb A$. Combining Eq.~\eqref{eq:app-radial-green-identity} with $\varpi^t[1,2]=\boldsymbol\Psi_1^T\mathbb J\boldsymbol\Psi_2$
yields
\begin{equation}
(\mathcal O\boldsymbol\Psi_1)^T\mathbb J\boldsymbol\Psi_2
+\boldsymbol\Psi_1^T\mathbb J
(\mathcal O\boldsymbol\Psi_2)
=
\partial_t\varpi^t[1,2]+\partial_r\varpi^r[1,2]\,,
\label{eq:app-local-green-identity}
\end{equation}
which is the local Green identity, Eq.\eqref{eq:local-symplectic-green-identity}, used in the main text.

We now give the explicit expressions in the vacuum exterior. In Regge-Wheeler gauge, Eq.~\eqref{eq:app-symplectic-potential} can be written as
\begin{align}
\Theta^{t}
={}&P_{H_2}\delta H_2+P_K\delta K\,,
\nonumber\\
\Theta^{r}
={}&-\frac1\kappa\Bigl[
\bigl(-r^2fK'+4rfH_2-2rfK\bigr)\delta H_2
\nonumber\\
&+\bigl(r^2f(H_2'-K')+(r^2f'-2rf)H_2\bigr)\delta K\nonumber\\
&
-2r^2fH_2\delta K'\Bigr]
+2fH_1\delta P_{H_2}\,.
\label{eq:app-exterior-adm-symplectic-potential}
\end{align}
Using the Zerilli transformation in Eqs.~\eqref{eq:app-exterior-X} and~\eqref{eq:app-exterior-zerilli-pair}, together with the vacuum constraints, we obtain
\begin{align}
\Theta^t
&=\frac{1}{f}\Pi_Z\delta Z+\partial_r\beta_{\rm can}\,,
\nonumber\\
\Theta^r
&=-C_\ell(\partial_{r_*}Z)\delta Z
-\partial_t\beta_{\rm can}
+\mathscr D[Z;\delta Z]\,,
\label{eq:app-exterior-zerilli-potential-reduction}
\end{align}
where
\begin{align}
&\beta_{\rm can}[Z;\delta Z]
=\frac{1}{\kappa C_\ell}\Bigg\{
\Big[
\frac{L\Lambda}{2f}\Pi_Z\nonumber\\
&\hspace{1.2cm}+\left(a+\frac{\Lambda}{rf}\right)
\left((rfa-\Lambda)\Pi_Z
+r^2\partial_{r_*}\Pi_Z\right)
\big]\delta Z
\nonumber\\
&\hspace{1.2cm}+\frac{r}{f}
\left((rfa-\Lambda)\Pi_Z
+r^2\partial_{r_*}\Pi_Z\right)
\delta(\partial_{r_*}Z)
\Bigg\}\,,
\label{eq:canonical-surface-one-form}
\end{align}
The coefficient $a$ is defined as 
\begin{equation}
a(r)=\frac{(L-2) r^2-3(L-2) Mr-6M^2}
        {2\Lambda(r-2M)}.
\end{equation}
The remainder $\mathscr D$ has a lengthy explicit form, but it is symmetric in two vacuum solutions:
\begin{equation}
\mathscr D[Z_1;Z_2]=\mathscr D[Z_2;Z_1]\,.
\end{equation}
Defining
\begin{equation}
k[1,2]
:=\beta_{\rm can}[Z_2;Z_1]
-\beta_{\rm can}[Z_1;Z_2]\,,
\end{equation}
and substituting Eq.~\eqref{eq:app-exterior-zerilli-potential-reduction} into the definition~\eqref{eq:app-local-symplectic-current}, we find
\begin{align}
\varpi^t[1,2]
&=\frac{1}{f}
\bigl(Z_1\Pi_{Z,2}-\Pi_{Z,1}Z_2\bigr)
+\partial_r k[1,2]\,,
\nonumber\\
\varpi^r[1,2]
&=C_\ell\bigl(
Z_2\partial_{r_*}Z_1
-Z_1\partial_{r_*}Z_2
\bigr)
-\partial_t k[1,2]\,.
\label{eq:app-exterior-zerilli-symplectic-current}
\end{align}

Substituting Eq.~\eqref{eq:app-exterior-zerilli-symplectic-current} into the bilinear form defined in Eq.~\eqref{eq:bilinear-form-canonical}, the two $k$  ontributions in $\varpi^r$ and $\varpi^t$ at infinity cancel, leading to
\begin{align}
&\int_R^\infty\mathrm dr\,
\varpi^t[\mathbb R\boldsymbol\psi_1,\boldsymbol\psi_2]
+
\left.
\frac{\varpi^r[\mathbb R\boldsymbol\psi_1,
\boldsymbol\psi_2]}{i\Delta\omega}
\right|_{\infty}
\nonumber\\
&\quad=
\int_{R_*}^{\infty}\mathrm dr_*\,
\left(Z_1\Pi_{Z,2}+\Pi_{Z,1}Z_2\right)
+\mathcal B_R[\boldsymbol\psi_1,\boldsymbol\psi_2]
\nonumber\\
&\qquad\quad+
\left.
\frac{C_\ell}{i\Delta\omega}
\left(
Z_2\partial_{r_*}Z_1-Z_1\partial_{r_*}Z_2
\right)
\right|_{r_*\to\infty}\,,
\label{eq:app-exterior-bilinear-reduction}
\end{align}
where the surface term is
\begin{align}
\mathcal B_R[\boldsymbol\psi_1,\boldsymbol\psi_2]
&:=
-\left.k[\mathbb R\boldsymbol\psi_1,\boldsymbol\psi_2]
\right|_{r=R}
\nonumber\\
&=
-\left[
\beta_{\rm can}[\boldsymbol\psi_1;\boldsymbol\psi_2]
+\beta_{\rm can}[\boldsymbol\psi_2;\boldsymbol\psi_1]
\right]_{r=R}\,.
\label{eq:app-exterior-canonical-surface-term}
\end{align}
This is the explicit form given in Eq.\eqref{eq:bilinear-form-zerilli-real} in the main text.

\section{AAKS reduction of the Hamiltonian system}
\label{sec:aaks-appendix}

In this section, we use the AAKS variables introduced in
Ref.~\cite{Allen:1997xj} to reorganize the first-order Hamiltonian evolution system derived in Appendix~\ref{sec:hamiltonian-appendix} into a constrained pair of second-order wave equations for the gravitational and fluid sectors. We extend the original barotropic
formulation to a fluid with frozen-composition stratification by retaining the radial displacement as an auxiliary dynamical variable.
The resulting system includes buoyancy and can therefore describe $g$-modes. Its barotropic limit is used in the simulations of Secs.~\ref{sec:green-numerical-validation} and
\ref{sec:waveform-numerical-implementation}; the same variables also provide a convenient basis for the surface-reflection problem discussed in Sec.~\ref{sec:stellar-green-function}.

We define
\begin{align}
S&=\frac{N^2}{r}(H_0-K)\,,
&F&=rK\,,
\nonumber\\
H&=\frac{\delta p}{\varepsilon+p}\,,
&\frac{\mathrm dx}{\mathrm dr}&=\frac{A}{N}\,,
\label{eq:appendix-aaks-definitions}
\end{align}
where $x$ is the tortoise coordinate. For a stratified fluid with frozen composition, we further define the adiabatic sound speed, radial displacement amplitude, and Schwarzschild discriminant by
\begin{equation}
c_s^2:=\frac{\Gamma_1p}{w}\,,
\qquad
d:=\frac{W}{Ar^2}\,,
\qquad
\mathcal A_s:=\frac{\varepsilon'}{w}
-\frac{p'}{\Gamma_1p}\,,
\label{eq:appendix-aaks-stratification-definitions}
\end{equation}
The thermodynamic perturbations in
Eqs.~\eqref{eq:app-rw-thermodynamic-perturbations-e} and~\eqref{eq:app-rw-thermodynamic-perturbations-p} then imply
\begin{equation}
\delta\varepsilon
=\frac{w}{c_s^2}H-w\mathcal A_s d.
\label{eq:appendix-aaks-stratified-density}
\end{equation}
In general, $c_s^2$ is the frozen-composition sound speed and need not equal $\mathrm dp/\mathrm d\varepsilon$ along the equilibrium background.

Rewriting the Hamiltonian equations in terms of the AAKS variables, we obtain the following wave equations for $S$ and $H$:
\begin{widetext}
\begin{align}
\partial_t^2 S
={}&\partial_x^2S
-\frac{2N^2}{r^3}
\left[
\frac{Lr}{2}-2\pi r^3(\varepsilon+3p)-m
\right]S+\frac{4N^4}{r^5}
\left[
\frac{(m+4\pi pr^3)^2}{r-2m}
+4\pi\varepsilon r^3-3m
\right]F\,,
\nonumber\\
\partial_t^2 H
={}&c_s^2\Bigg\{
\partial_x^2H
+\frac{AN}{r^2}
\left[
(m+4\pi pr^3)(1-c_s^{-2})+2(r-2m)
\right]\partial_xH
\nonumber\\
&\quad
-\frac{2N^2}{r^2}
\left[
\frac L2-2\pi r^2w(3+c_s^{-2})
\right]H
-\frac{AN(m+4\pi pr^3)(1-c_s^{-2})}{2r^3}
\,\partial_xF
\nonumber\\
&\quad
-\frac{N^2}{r^2}
\left[
\frac{2(m+4\pi pr^3)^2}{r^2(r-2m)c_s^2}
-\frac{(m+4\pi pr^3)(1-c_s^{-2})}{2r^2}
-4\pi r(3p+\varepsilon)
\right]F
\nonumber\\
&\quad
+\frac{A(m+4\pi pr^3)(1-c_s^{-2})}{2Nr}
\,\partial_xS
\nonumber\\
&\quad
-\left[
\frac{(m+4\pi pr^3)^2(1+c_s^{-2})}{r^2(r-2m)}
-\frac{(m+4\pi pr^3)(1-c_s^{-2})}{2r^2}
-4\pi r(3p+\varepsilon)
\right]S
\Bigg\}
+\mathcal Q_{\rm strat}\,,
\label{eq:appendix-aaks-time-domain-equations}
\end{align}
\end{widetext}
where the additional stratification term is
\begin{align}
\mathcal Q_{\rm strat}
={}&-c_s^2\partial_r
\left[
\frac{N^2}{A^2}\mathcal A_s
\left(\frac{N'}N d-H\right)
\right]
\nonumber\\
&-\left[
c_s^2\left(\frac{A'}A+\frac2r\right)
-\frac{N'}N
\right]
\frac{N^2}{A^2}\mathcal A_s
\left(\frac{N'}N d-H\right)
\nonumber\\
&-4\pi c_s^2N^2w\mathcal A_s d\,.
\label{eq:appendix-aaks-buoyancy-correction}
\end{align}
The metric variable $F$ also satisfies an evolution equation,
\begin{widetext}
\begin{align}
\partial_t^2F
={}&\partial_x^2F
-\frac{2N^2}{r^3}
\left[
\frac{Lr}{2}-2\pi r^3(3\varepsilon+p)-m
\right]F
+2\left[
4\pi r^2w-1+\frac{2m}{r}
\right]S
\nonumber\\
&-8\pi rN^2w(1-c_s^{-2})H
-8\pi rN^2w\mathcal A_s d\,.
\label{eq:appendix-aaks-F-evolution}
\end{align}
\end{widetext}
However, $F$ does not represent an additional independent
wave degree of freedom, because it is constrained by the
Hamiltonian constraint. In terms of the AAKS variables,
this constraint reads
\begin{widetext}
\begin{align}
0={}&\partial_x^2F
-\frac{AN(m+4\pi pr^3)}{r^2}\partial_xF
+\frac{N^2}{r^3}
\left[12\pi r^3\varepsilon-m-Lr\right]F
-\frac{r}{AN}\partial_xS
\nonumber\\
&+\left(8\pi r^2w-\frac L2-2+\frac{4m}{r}\right)S
+\frac{8\pi rN^2w}{c_s^2}H
-8\pi rN^2w\mathcal A_s d\,.
\label{eq:appendix-aaks-time-domain-constraint}
\end{align}
\end{widetext}

The radial displacement obeys the auxiliary evolution equation
\begin{equation}
\ddot d
=\frac{N^2}{A^2}
\left[
K'-\frac12H_2'-H'
-2\frac{N'}N H_2
+\mathcal A_s\left(\frac{N'}N d-H\right)
\right]\,,
\label{eq:appendix-aaks-displacement-evolution}
\end{equation}
Notably, this equation contains no radial derivatives of $d$: the displacement evolves in time but does not constitute an independent radial propagation channel.
For a Fourier component proportional
to $e^{-i\omega t}$, it becomes an algebraic relation 
\begin{equation}
d=
-\frac{(N^2/A^2)
\left[
K'-H_2'/2-H'-2(N'/N)H_2-\mathcal A_s H
\right]}
{\omega^2+(N^2/A^2)(N'/N)\mathcal A_s}\,.
\label{eq:appendix-aaks-modal-displacement-elimination}
\end{equation}

The angular displacement can be reconstructed from
\begin{equation}
V=\frac{r^2}{L}
\left[
-\frac{H-(N'/N)d}{c_s^2}
-d'
-\left(\frac{A'}A+\frac2r\right)d
-\frac{H_2}{2}-K
\right]\,.
\label{eq:appendix-aaks-angular-reconstruction}
\end{equation}

Before closing this section, we outline how the reduced equations follow from the RW-gauge Hamiltonian equations of Appendix~\ref{sec:hamiltonian-appendix}. The thermodynamic relation in Eq.~\eqref{eq:app-rw-thermodynamic-perturbations-e} gives the exact source identity
\begin{equation}
wW'+\varepsilon'W+LAwV
=-Ar^2\left[
\delta\varepsilon+w\left(\frac{H_2}{2}+K\right)
\right]\,.
\label{eq:app-aaks-fluid-source-identity}
\end{equation}
Since $H_0=H_2$, the definition of $S$ implies
\begin{equation}
\ddot S
=\frac{N^2}{r}
\left[(\ddot H_2+\ddot K)-2\ddot K\right]\,.
\label{eq:app-aaks-S-kinematic}
\end{equation}
We therefore subtract twice Eq.~\eqref{eq:app-rw-K-evolution} from Eq.~\eqref{eq:app-rw-HK-evolution}. The explicit $\delta p$ terms cancel, while Eq.~\eqref{eq:app-aaks-fluid-source-identity} collects the displacement terms into a single $\delta\varepsilon$ contribution. Eliminating that contribution with the Hamiltonian constraint \eqref{eq:app-rw-hamiltonian-constraint}, and then substituting $K=F/r$ and $H_2=F/r+rS/N^2$, gives the $S$ equation in Eq.~\eqref{eq:appendix-aaks-time-domain-equations}. This cancellation does not require $\mathcal A_s=0$. The same substitutions, together with Eq.~\eqref{eq:appendix-aaks-stratified-density}, turn the RW Hamiltonian constraint into Eq.~\eqref{eq:appendix-aaks-time-domain-constraint}.

For the fluid sector, the pressure perturbation in
Eq.~\eqref{eq:app-rw-thermodynamic-perturbations} can be written as
the compression identity
\begin{equation}
H=-c_s^2\left[
d'+\left(\frac{A'}A+\frac2r\right)d
+\frac{L}{r^2}V+\frac{H_2}{2}+K
\right]
+\frac{N'}N d\,.
\label{eq:app-aaks-compression-identity}
\end{equation}
The background equations imply
\begin{equation}
\frac{w'}w+\frac{N'}N(1+c_s^{-2})
=\mathcal A_s\,,
\end{equation}
so the pressure and density terms in the radial fluid equation
combine into
\begin{equation}
\frac{\delta p'}w
+\frac{N'}{Nw}(\delta\varepsilon+\delta p)
=H'+\mathcal A_s
\left(H-\frac{N'}N d\right)\,.
\label{eq:app-aaks-radial-euler-source}
\end{equation}
Using this identity in Eq.~\eqref{eq:app-rw-W-evolution} and
combining it with Eq.~\eqref{eq:app-rw-H1-evolution} yields the
displacement equation~\eqref{eq:appendix-aaks-displacement-evolution}. The angular fluid equation gives $\ddot V=N^2(H-H_2/2)$.

Finally, differentiating Eq.~\eqref{eq:app-aaks-compression-identity} twice with respect to time yields
\begin{align}
\ddot H={}&-c_s^2\left[
(\ddot d)'
+\left(\frac{A'}A+\frac2r\right)\ddot d
+\frac{L}{r^2}\ddot V
+\frac{\ddot H_2}{2}+\ddot K
\right]
+\frac{N'}N\ddot d\,.
\label{eq:app-aaks-H-intermediate}
\end{align}
Substituting Eqs.~\eqref{eq:appendix-aaks-displacement-evolution}, \eqref{eq:app-rw-V-evolution}, \eqref{eq:app-rw-K-evolution}, and \eqref{eq:app-rw-HK-evolution} into
Eq.~\eqref{eq:app-aaks-H-intermediate}, and using the Hamiltonian constraint~\eqref{eq:appendix-aaks-time-domain-constraint}, yields the $H$ equation in Eq.~\eqref{eq:appendix-aaks-time-domain-equations}.

\section{Surface reflection and center-return decomposition}
\label{sec:prompt-center-appendix}

\begin{figure}[t]
\centering
\includegraphics[width=\columnwidth]
{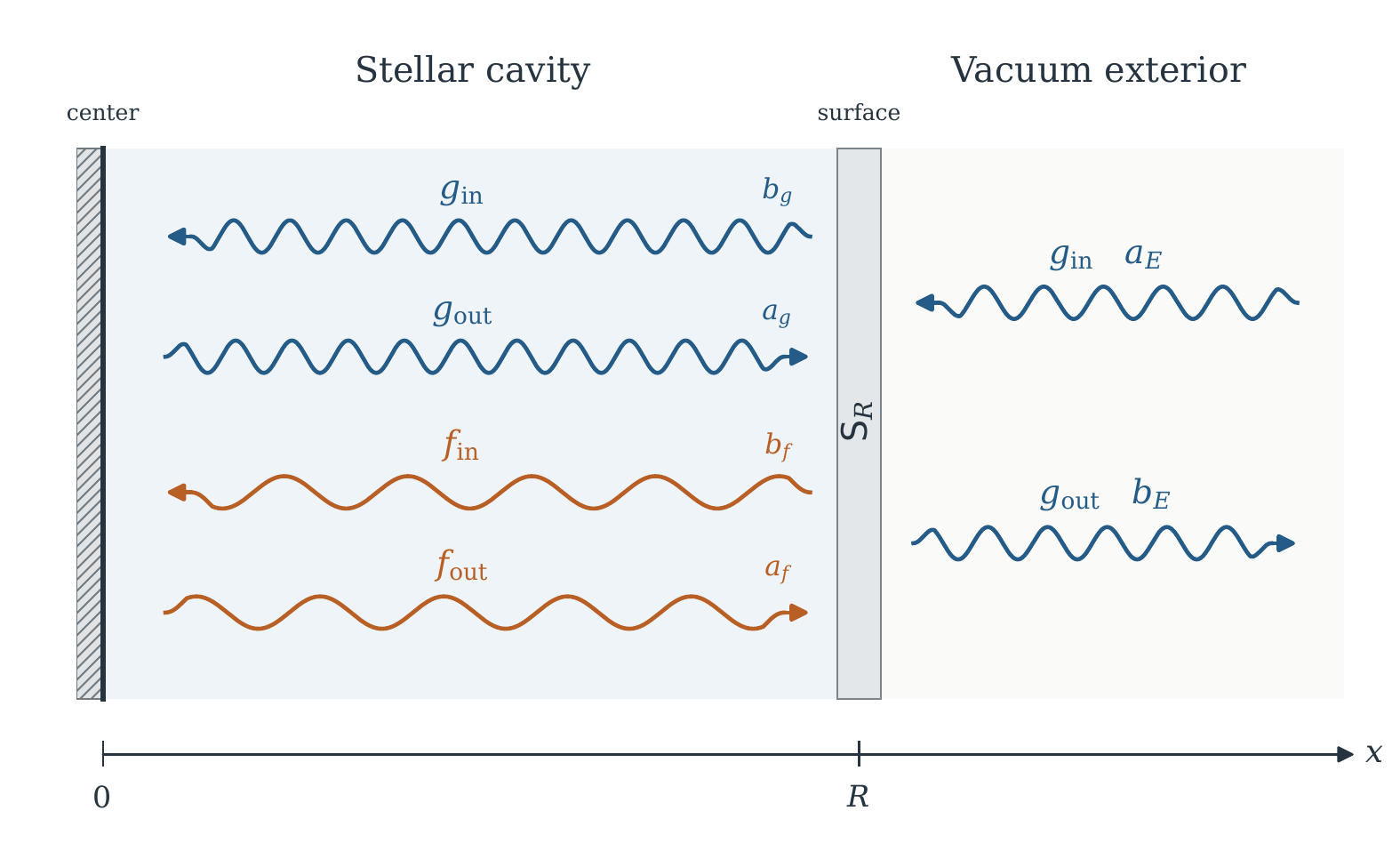}
\caption{Schematic of the cavity-reflection problem.
The stellar interior supports inward and outward gravitational ($g$) and acoustic ($f$) channels, while the vacuum exterior supports only the gravitational channel. The channels are coupled by the surface junction conditions, and regularity at the center determines the center-return map.}
\label{fig:general-cavity-reflection}
\end{figure}

In this section, we present a detailed formulation of the general prompt--center decomposition discussed in Sec.~\ref{sec:stellar-green-function} of the main text.
We then develop a practical approximation for its numerical evaluation using the AAKS formulation derived in the Appendix~\ref{sec:aaks-appendix}.

\subsection{General cavity reflection}

We formulate the problem as a general cavity-reflection problem, as illustrated in Fig.~\ref{fig:general-cavity-reflection}.
In the vacuum exterior, the Zerilli component of the regular solution can be written as
\begin{equation}
Z^{\rm reg}
=
A^{\rm in}Z^{\rm in}
+
A^{\rm out}Z^{\rm out}\,,
\end{equation}
where $Z^{\rm in/out}\sim e^{\mp i\omega r_*}$ at infinity.
The stellar reflection coefficient is therefore
$\mathcal R_\ell=A^{\rm out}/A^{\rm in}$.

In the interior, we identify four solutions describing inward and outward propagation in the gravitational and acoustic channels.
Writing the AAKS fields as
\begin{equation}
\boldsymbol U^{\rm reg}
=
\begin{pmatrix}
S^{\rm reg}\\
H^{\rm reg}
\end{pmatrix}\,,
\end{equation}
we expand the solution as
\begin{equation}
\boldsymbol U^{\rm reg}
=
b_g\boldsymbol g_{\rm in}
+b_f\boldsymbol f_{\rm in}
+a_g\boldsymbol g_{\rm out}
+a_f\boldsymbol f_{\rm out}\,.
\label{eq:appendix-interior-channel-expansion}
\end{equation}
Each channel is a two-component solution of the coupled AAKS equations. The labels $g$ and $f$ distinguish the gravitational and acoustic channels; they do not imply that either component vanishes. Define the interior amplitude vectors by
\begin{equation}
\boldsymbol b_I=
\begin{pmatrix}b_g\\b_f\end{pmatrix}\,,
\qquad
\boldsymbol a_O=
\begin{pmatrix}a_g\\a_f\end{pmatrix}\,.
\end{equation}
For the exterior amplitudes, we remove the surface propagation phases by setting
\begin{equation}
a_E=A^{\rm in}e^{-i\omega R_*}\,,
\qquad
b_E=A^{\rm out}e^{+i\omega R_*}\,,
\end{equation}

The surface junction conditions define the block scattering relation
\begin{equation}
\begin{pmatrix}
b_E\\
\boldsymbol b_I
\end{pmatrix}
=
\begin{pmatrix}
\rho_{\rm out}&\boldsymbol\tau_{\rm out}^{T}\\
\boldsymbol\tau_{\rm in}&\boldsymbol\rho_{\rm in}
\end{pmatrix}
\begin{pmatrix}
a_E\\
\boldsymbol a_O
\end{pmatrix}\,.
\label{eq:appendix-surface-scattering-matrix}
\end{equation}
Here $\rho_{\rm out}$ describes reflection from the exterior side, $\boldsymbol\tau_{\rm in}$ and $\boldsymbol\tau_{\rm out}$ describe transmission into and out of the star, and $\boldsymbol\rho_{\rm in}$ describes reflection from the interior side, including conversion between the two channels.

Regularity at the center relates the returning amplitudes to the inward amplitudes through
\begin{equation}
\boldsymbol a_O
=
\mathbf C_\ell(\omega)\boldsymbol b_I\,.
\label{eq:appendix-center-return-map}
\end{equation}
The center-return matrix $\mathbf C_\ell$ includes propagation
through the stellar interior and reflection at the regular center.

Eliminating $\boldsymbol a_O$ and $\boldsymbol b_I$ from
Eqs.~\eqref{eq:appendix-surface-scattering-matrix} and
\eqref{eq:appendix-center-return-map} gives
\begin{align}
\mathcal R_\ell
&=\mathcal R_\ell^{\rm pr}+\mathcal R_\ell^{\rm cent}\,,
\nonumber\\
\mathcal R_\ell^{\rm pr}
&=e^{-2i\omega R_*}\rho_{\rm out}\,,
\nonumber\\
\mathcal R_\ell^{\rm cent}
&=e^{-2i\omega R_*}
\boldsymbol\tau_{\rm out}^{T}\mathbf C_\ell
\left(\mathbf I-\boldsymbol\rho_{\rm in}\mathbf C_\ell\right)^{-1}
\boldsymbol\tau_{\rm in}\,.
\label{eq:appendix-prompt-center-result}
\end{align}
The prompt term describes reflection without a center return,
while the inverse in the center-return term resums successive
internal round trips.

In numerical practice, we evaluate the complete and prompt reflection coefficients through their surface logarithmic derivatives:
\begin{align}
\mathcal R_\ell(\omega)
&=-\left.
\frac{\partial_{r_*}Z^{\rm in}-m_{\rm reg}Z^{\rm in}}
{\partial_{r_*}Z^{\rm out}-m_{\rm reg}Z^{\rm out}}
\right|_{r=R}\,,
\nonumber\\
\mathcal R_\ell^{\rm pr}(\omega)
&=-\left.
\frac{\partial_{r_*}Z^{\rm in}-m_I Z^{\rm in}}
{\partial_{r_*}Z^{\rm out}-m_I Z^{\rm out}}
\right|_{r=R}\,,
\label{eq:appendix-reflection-impedance}
\end{align}
where
\begin{equation}
m_{\rm reg}
=\left.\frac{\partial_{r_*}Z^{\rm reg}}{Z^{\rm reg}}\right|_{r=R}\,,
\qquad
m_I
=\left.\frac{\partial_{r_*}Z_I}{Z_I}\right|_{r=R}\,.
\end{equation}
Here $Z_I$ is the exterior Zerilli field obtained by matching the inward interior solution
\begin{equation}
\boldsymbol U_I
=b_g\boldsymbol g_{\rm in}+b_f\boldsymbol f_{\rm in}\,,
\end{equation}
with the relative amplitudes fixed by the free-surface condition $\Delta p(R)=0$.
The center-return contribution is then obtained as
$\mathcal R_\ell^{\rm cent}=\mathcal R_\ell-\mathcal R_\ell^{\rm pr}$.

\subsection{AAKS surface channels and numerical prompt impedance}

With Eq.~\eqref{eq:appendix-reflection-impedance}, the remaining task is to identify the inward and outward solutions in the gravitational and acoustic channels. Their exact construction is complicated by the metric--fluid coupling. We therefore adopt an approximation motivated by their leading-order decoupling in the high-frequency limit.

In the frequency domain, with time dependence $e^{-i\omega t}$
and $\mathcal A_s=0$,
Eq.~\eqref{eq:appendix-aaks-time-domain-equations} becomes
\begin{widetext}
\begin{align}
&\partial_x^2S
+\left\{
\omega^2-\frac{2N^2}{r^3}
\left[\frac L2r-2\pi r^3(\varepsilon+3p)-m\right]
\right\}S
={}
-\frac{4N^4}{r^5}
\left[
\frac{(m+4\pi r^3p)^2}{r-2m}
+4\pi\varepsilon r^3-3m
\right]F\,,
\label{eq:appendix-full-S-frequency}\\
&\partial_x^2H
+\frac{AN}{r^2}
\left[
(m+4\pi r^3p)(1-c_s^{-2})+2(r-2m)
\right]\partial_xH
+
\left\{
\frac{\omega^2}{c_s^2}
-\frac{2N^2}{r^2}
\left[\frac L2-2\pi r^2w(3+c_s^{-2})\right]
\right\}H
\nonumber\\
&\hspace{1.2cm}=
\frac{AN(m+4\pi r^3p)(1-c_s^{-2})}{2r^3}\partial_xF+
\frac{N^2}{r^2}
\left[
\frac{2(m+4\pi r^3p)^2}{r^2(r-2m)c_s^2}
-\frac{(m+4\pi r^3p)(1-c_s^{-2})}{2r^2}
-4\pi r(3p+\varepsilon)
\right]F
\nonumber\\
&\hspace{1.2cm}-
\frac{A(m+4\pi r^3p)(1-c_s^{-2})}{2Nr}\partial_xS+
\left[
\frac{(m+4\pi r^3p)^2(1+c_s^{-2})}{r^2(r-2m)}
-\frac{(m+4\pi r^3p)(1-c_s^{-2})}{2r^2}
-4\pi r(3p+\varepsilon)
\right]S\,.
\label{eq:appendix-full-H-frequency}
\end{align}
\end{widetext}

At high frequency, the gravitational and acoustic channels decouple at leading order, while the cross-channel driving is subleading. Motivated by this behavior, we use the constraints to express $F$ and $\partial_xF$ in terms of $S$, $\partial_xS$, $H$, and $\partial_xH$, where $\partial_x=(N/A)\partial_r$. We then retain the terms involving $S$ and $\partial_xS$ in the $S$ equation, and those involving $H$ and $\partial_xH$ in the $H$ equation. For $\mathcal A_s=0$, the resulting decoupled equations are
\begin{equation}
\partial_x^2S+U_g\partial_xS
+\bigl(\omega^2-V_g\bigr)S=0\,,
\label{eq:appendix-decoupled-S-x}
\end{equation}
and
\begin{widetext}
\begin{align}
\partial_x^2H
&+\frac{AN}{r^2}
\left[
(m+4\pi r^3p)(1-c_s^{-2})+2(r-2m)
\right]\partial_xH
\nonumber\\
&+\left\{
\frac{\omega^2}{c_s^2}
-\frac{2N^2}{r^2}
\left[
\frac L2-2\pi r^2w(3+c_s^{-2})
\right]
\right\}H
\nonumber\\
={}&
\frac{AN(m+4\pi r^3p)(1-c_s^{-2})}{2r^3}
\Biggl[
\frac{8\pi N^2wr}{D}
\left(
\omega^2+\frac{N^2}{r^3}(3m+4\pi r^3p)
\right)\partial_xH
\nonumber\\
&\hspace{4em}
+\frac{8\pi N^3w}{AD}
\left\{
\left[1+\frac{A^2(m+4\pi r^3p)}r\right]\omega^2
+\frac{N^2}{r^3}
\left[
4\pi r^3w
-2A^2(m+4\pi r^3p)
\left(\frac L2-8\pi r^2w\right)
\right]
\right\}H
\Biggr]
\nonumber\\
&+\frac{N^2}{r^2}
\left[
\frac{2(m+4\pi r^3p)^2}{r^2(r-2m)c_s^2}
-\frac{(m+4\pi r^3p)(1-c_s^{-2})}{2r^2}
-4\pi r(3p+\varepsilon)
\right]
\nonumber\\
&\hspace{2em}\times
\left[
-\frac{8\pi AN^3w(m+4\pi r^3p)}{Dr}\partial_xH
+\frac{8\pi N^2wr}{D}
\left(\omega^2+4\pi N^2w\right)H
\right]\,.
\label{eq:appendix-decoupled-H-x}
\end{align}
The gravitational coefficients are
\begin{align}
U_g
={}&-\frac{4N^4}{r^5D}
\left[
\frac{(m+4\pi r^3p)^2}{r-2m}
+4\pi\varepsilon r^3-3m
\right]
\nonumber\\
&\times
\left\{
\frac{r\omega^2}{AN}
+\frac{N}{Ar^2}
\left[
4\pi r^3w
-\frac L2 A^2(m+4\pi r^3p)
+4\pi r^2A^2(m+4\pi r^3p)w
\right]
\right\}\,,
\\
V_g
={}&
\frac{2N^2}{r^3}
\left[
\frac L2r-2\pi r^3(\varepsilon+3p)-m
\right]
+\frac{4N^4}{r^5D}
\left[
\frac{(m+4\pi r^3p)^2}{r-2m}
+4\pi\varepsilon r^3-3m
\right]
\nonumber\\
&\times
\left\{
\left(\frac L2+\frac mr+4\pi r^2p\right)\omega^2
+\frac{N^2}{r^3}
\left[
-\frac L2(m-4\pi r^3\varepsilon)
+4\pi r^2A^2(m+4\pi r^3p)
(1+8\pi r^2p)w
\right]
\right\}\,,
\label{eq:appendix-gravity-diagonal-coefficients}
\end{align}
where
\begin{align}
D(r,\omega)
={}&\omega^4
+\frac{\omega^2N^2}{r^4}
\left[
A^2(m+4\pi r^3p)^2
+4mr+4\pi r^4(3p+\varepsilon)
\right]
\nonumber\\
&+\frac{2N^4}{r^6}
\left[
8\pi r^3(m+2\pi r^3p)w
-A^2(m+4\pi r^3p)^2
\left(\frac L2-8\pi r^2w\right)
\right]\,.
\label{eq:appendix-aaks-elimination-denominator}
\end{align}
\end{widetext}
These expressions apply where $D(r,\omega)\ne0$~\footnote{For the stellar model considered in this work and $\ell=2$, $D(R,\omega)=0$ at $M\omega\simeq\pm0.0590234$ and $M\omega\simeq\pm0.115258\,i$.
Since the channel-decoupling approximation is motivated by the high-frequency limit, its numerical use should be restricted to sufficiently large $|\omega|$.
Accordingly, in contour integrations involving $G^{\rm pr}$, the circle around the origin is chosen large enough to exclude the low-frequency region, including these zeros, from the integration path.}.

We introduce the inward gravitational and acoustic optical
depths,
\begin{equation}
\xi=\int_r^R\frac{A}{N}\,\mathrm dr'\,,
\qquad
s=\int_r^R\frac{A}{Nc_s}\,\mathrm dr'\,,
\label{eq:appendix-surface-optical-depths}
\end{equation}
so that $\partial_x=-\partial_\xi=-c_s^{-1}\partial_s$.
The gravitational equation becomes
\begin{equation}
\partial_\xi^2S-U_g(\xi,\omega)\partial_\xi S
+\bigl[\omega^2-V_g(\xi,\omega)\bigr]S=0\,.
\label{eq:appendix-scalar-gravity-equation}
\end{equation}
We retain the full radial and frequency dependence of
$U_g$ and $V_g$ in the numerical calculation.

For the present stellar model,
$c_s^2=O(R-r)$ and $w=O(c_s^2)$ near the surface.
Provided $D(R,\omega)\ne0$, the coefficients of $H$ and
$\partial_xH$ on the right-hand side of
Eq.~\eqref{eq:appendix-decoupled-H-x} are therefore finite.
After changing to $s$ and multiplying the equation by $c_s^2$,
these terms contribute only $O(s)\partial_sH$ and $O(s^2)H$.
In contrast, the original acoustic derivative term gives
$s^{-1}\partial_sH$ at leading order. Thus the surface
equation takes the form
\begin{equation}
\partial_s^2H
+\left[\frac1s+O(s)\right]\partial_sH
+\left[\omega^2+O(s^2)\right]H=0\,.
\end{equation}
Retaining the leading surface terms yields
\begin{equation}
\partial_s^2H+\frac1s\partial_sH+\omega^2H=0\,.
\label{eq:appendix-surface-acoustic-equation}
\end{equation}

Within this decoupled approximation, the inward gravitational and acoustic channels can be written as
\begin{equation}
\boldsymbol g_{\rm in}
=
\begin{pmatrix}
S_{{\rm in},g}\\
0
\end{pmatrix}\,,
\qquad
\boldsymbol f_{\rm in}
=
\begin{pmatrix}
0\\
H_{{\rm in},f}
\end{pmatrix}\,,
\end{equation}
where, up to overall normalization,
\begin{align}
S_{{\rm in},g}(\xi,\omega)
&=
\exp\!\left[
\int_0^\xi p_g(\zeta,\omega)\,\mathrm d\zeta
\right]\,,\nonumber\\
H_{{\rm in},f}(s,\omega)
&=
H_0^{(1)}(\omega s)\,.
\label{eq:appendix-inward-channel-carriers}
\end{align}
Here $H_0^{(1)}$ is the Hankel function of the first kind,
and the gravitational logarithmic derivative $p_g=\partial_\xi S_{{\rm in},g}/S_{{\rm in},g}$ obeys
\begin{equation}
\partial_\xi p_g
=
-p_g^2+U_gp_g-\omega^2+V_g\,.
\label{eq:appendix-gravity-riccati}
\end{equation}
With the time dependence $e^{-i\omega t}$, the inward
branches are selected by $p_g\sim i\omega$ in the
high-frequency limit and $H_0^{(1)}(\omega s)\sim\sqrt{2/(\pi\omega s)}\,e^{i(\omega s-\pi/4)}$.

To determine $p_g$, we analytically continue the stellar
background and the coefficients $U_g,V_g$ along a complex
ray $\xi=e^{i\theta}\lambda$, with $\lambda\ge0$.
For the inward channel, we choose
\begin{equation}
e^{i\theta}
=-\frac{\overline{i\omega}}{|i\omega|}\,,
\end{equation}
so that the leading inward carrier $e^{i\omega\xi}$
decays along the ray. At a finite endpoint $\xi_{\rm end}=e^{i\theta}\lambda_{\rm max}$, we initialize
the logarithmic derivative with 
\begin{align}
p_g(\xi_{\rm end},\omega)
&=
\frac{U_g(\xi_{\rm end},\omega)}{2}\nonumber\\
&+i\sqrt{
\omega^2-V_g(\xi_{\rm end},\omega)
-\frac{U_g(\xi_{\rm end},\omega)^2}{4}
}\,,
\label{eq:appendix-gravity-endpoint-condition}
\end{align}
where the square-root branch is chosen to approach $\omega$
at high frequency. We then integrate Eq.~\eqref{eq:appendix-gravity-riccati} backward along the same ray to the stellar surface $\xi=0$, obtaining
\begin{equation}
p_g(0,\omega)
=
\left.
\frac{\partial_\xi S_{{\rm in},g}}
{S_{{\rm in},g}}
\right|_{\xi=0}\,.
\end{equation}
The sensitivity to the finite endpoint is monitored by
repeating the integration with increasing $\lambda_{\rm max}$,
and we have numerically verified that the resulting variations
are sufficiently small for the accuracy required here.

Finally, we form the inward combination
\begin{equation}
\boldsymbol U_{\rm I}
=b_g\boldsymbol g_{\rm in}+b_f\boldsymbol f_{\rm in}\,,
\end{equation}
with the relative amplitudes fixed by the free-surface condition $\Delta p=0$. The corresponding metric perturbations are then transformed to the exterior Zerilli variable, yielding the surface impedance
\begin{equation}
m_{\rm I}(\omega)
=
\left.
\frac{\partial_{r_*}Z_{\rm I}}{Z_{\rm I}}
\right|_{r=R}\,.
\end{equation}
Similarly, $m_{\rm reg}$ is obtained from the center-regular solution satisfying the same surface condition, using standard stellar-QNM integration methods (see, e.g., Ref.~\cite{Miao:2025utd}).
Substituting these impedances into Eq.~\eqref{eq:appendix-reflection-impedance} gives $\mathcal R_\ell^{\rm pr}$ and $\mathcal R_\ell$, and hence $\mathcal R_\ell^{\rm cent}
=\mathcal R_\ell-\mathcal R_\ell^{\rm pr}$.

\section{Time-domain simulations}
\label{sec:time-domain-validation-appendix}

We present two time-domain simulations in the main text: a localized impulse experiment to validate the Green function, and an evolution driven by the incident chirp prescribed at $\mathscr I^-$.
Both calculations use the AAKS formulation described in Appendix~\ref{sec:aaks-appendix}, with the stellar background specified in Sec.~\ref{sec:green-numerical-validation}.
We use the optical coordinate $x$ defined in
Eq.~\eqref{eq:numerical-optical-coordinate}, for which the stellar surface is located at $x=x_R$ and $\partial_x=\partial_{r_*}$ in the exterior.
The metric fields are evolved throughout the interior and exterior, whereas the fluid enthalpy perturbation is evolved only inside the star.
At the center, we impose regularity conditions on the evolved fields. At the surface, we use the regular $c_s^2\to0$ limits of the AAKS equations and enforce matching to the vacuum exterior.
Below, we provide implementation details for the two simulations.

\subsection{Green-function validation in Sec.~\ref{sec:green-numerical-validation}}
\label{sec:green-time-domain-implementation}

For the Green-function test, we evolve $(F,\partial_t F,S,\partial_t S,H,\partial_t H)$ from vanishing initial data, which satisfy the Hamiltonian constraint.
The variable $F$ is updated using its evolution equation, Eq.~\eqref{eq:appendix-aaks-F-evolution}, while the Hamiltonian constraint is monitored during the evolution as a consistency check.
The localized source is the normalized spacetime Gaussian specified in Eq.~\eqref{eq:green-validation-source}, centered at $r'=2R$ and $t'=5\sigma_t$. Its widths are $\sigma_t=\sigma_{r_*}=0.1\,M_\odot\simeq0.0715\,M$, and the signal is extracted at $r=3R$.

To implement this source in the $F$ and $S$ equations, we use the exterior Zerilli reconstruction,
\begin{equation}
Z=
\frac{4r^3}{L[(L-2)r+6M]}
\left[
-\frac{\partial_xF}{r}
+\frac{(L+2)r-2M}{2r^3}F+S
\right]\,.
\label{eq:validation-zerilli-reconstruction}
\end{equation}
Combining this relation with the vacuum exterior limit of
the Hamiltonian constraint, Eq.~\eqref{eq:appendix-aaks-time-domain-constraint}, gives
\begin{align}
F={}&r\partial_xZ
+\frac{L(L-2)r^2+6(L-2)Mr+24M^2}
{2r[(L-2)r+6M]}Z\,,
\label{eq:validation-F-reconstruction}\\
S={}&\frac{L[(L-2)r+6M]}{4r^3}Z
+\frac{\partial_xF}{r}
-\frac{(L+2)r-2M}{2r^3}F\,.
\label{eq:validation-S-reconstruction}
\end{align}

Denoting the source specified in Eq.~\eqref{eq:green-validation-source} by $\mathcal S_Z$, we therefore obtain the corresponding $F$ and $S$ sources,
\begin{align}
\mathcal S_F
={}&r\partial_x\mathcal S_Z
+\frac{L(L-2)r^2+6(L-2)Mr+24M^2}
{2r[(L-2)r+6M]}\mathcal S_Z\,,
\label{eq:validation-F-source}\\
\mathcal S_S
={}&\frac{L[(L-2)r+6M]}{4r^3}\mathcal S_Z
+\frac{\partial_x\mathcal S_F}{r}
-\frac{(L+2)r-2M}{2r^3}\mathcal S_F\,.
\label{eq:validation-S-source}
\end{align}
We then add $\mathcal S_F$ and $\mathcal S_S$ to the $F$ and $S$ evolution equations, Eqs.~\eqref{eq:appendix-aaks-F-evolution} and~\eqref{eq:appendix-aaks-time-domain-equations}, respectively. No source is added directly to the $H$ equation.

The computational domain is $x\in[0,104.9]\,M$, corresponding to $r\in[0,95.6]\,M$.
The outer boundary is placed sufficiently far from the observer to prevent reflected signals from returning during the recorded
interval, which extends to $\tau=150\,M$.
We discretize the domain using a uniform optical grid of $58\,665$ points with spacing
$\Delta x\simeq1.789\times10^{-3}\,M$, including $5\,659$ points in the stellar interior.
Spatial derivatives are evaluated using second-order centered differences, and time integration uses the fourth-order Runge--Kutta method with $\Delta t\simeq1.431\times10^{-3}\,M$.

\subsection{Evolution of the incident chirp in Sec.~\ref{sec:waveform-numerical-implementation}}
\label{sec:chirp-time-domain-implementation}

For the chirp experiment, the physical input is the incident
waveform $F(v)$ prescribed at $\mathscr I^-$ in
Sec.~\ref{sec:chirping-source}.
We first generate Cauchy initial data by an independent
characteristic evolution of the vacuum Zerilli equation.
Writing $y=r_*$, $u=t-y$, and $v=t+y$, this equation is
\begin{equation}
4\partial_u\partial_v Z
+V^Z\!\left(r\!\left(\frac{v-u}{2}\right)\right)Z=0\,.
\label{eq:simulation-characteristic-zerilli}
\end{equation}
We impose the characteristic data
\begin{equation}
Z(u,v_{\rm on})=0\,,
\qquad
\lim_{u\to-\infty}Z(u,v)=F(v)\,,
\end{equation}
corresponding to an initially unperturbed exterior and
the prescribed incoming radiation at $\mathscr I^-$.

To deal with $u=-\infty$, we introduce the compactified coordinate
\begin{equation}
q=-\frac{v_{\rm on}-2t_0}{u}\,,
\qquad 0\leq q\leq1\,,
\end{equation}
so that $\mathscr I^-$ lies at $q=0$, while the intersection of the initial slice $t=t_0$ with the incident wavefront $v=v_{\rm on}$ lies at $q=1$.
Defining $Y(q,v)=Z(q,v)-F(v)$ and integrating Eq.~\eqref{eq:simulation-characteristic-zerilli} from $q=0$ gives
\begin{equation}
\partial_vY(q,v)
=
-\int_0^q\mathrm d\bar q\,
\frac{(v_{\rm on}-2t_0)V_\ell^Z(r(\bar q,v))}
{4\bar q^2}
\left[F(v)+Y(\bar q,v)\right]\,,
\end{equation}
where
\begin{equation}
r_*(r(q,v))
=\frac12\left(v+\frac{v_{\rm on}-2t_0}{q}\right).
\end{equation}
We discretize $q\in[0,1]$ using the Chebyshev--Lobatto points
\begin{equation}
q_j=\frac{1-\cos(j\pi/N)}{2}\,,
\qquad j=0,\ldots,N\,,
\end{equation}
with $N=128$.
Starting from $Y(q,v_{\rm on})=0$, we evaluate the integral over $q$ using a Chebyshev integration matrix and advance in $v$ using the fourth-order Runge--Kutta method, with $\Delta v=0.179\,M$.

This characteristic evolution determines $Z(u,v)$ in the exterior region needed to construct the initial data.
We interpolate the solution onto the initial Cauchy slice $t=t_0\simeq-3935.5\,M$, obtaining
\begin{align}
Z(t_0,y)
&=Z(u,v)\big|_{u=t_0-y,\;v=t_0+y}\,,\\
\partial_tZ(t_0,y)
&=(\partial_u+\partial_v)Z(u,v)
\big|_{u=t_0-y,\;v=t_0+y}\,.
\end{align}
This slice precedes the arrival of the incident wavefront at the stellar surface, $t=v_{\rm on}-R_*\simeq-2723.71\,M$,
so the stellar interior is initially unperturbed.
We transform $Z$ and $\partial_tZ$ into the AAKS variables using Eqs.~\eqref{eq:validation-F-reconstruction}
and~\eqref{eq:validation-S-reconstruction},
and determine the initial $F$ and $\partial_tF$ from the Hamiltonian constraint and its time derivative.
We subsequently evolve $S$ and $H$, determining $F$ from the Hamiltonian constraint at every time substep.
The same characteristic solution supplies consistent time-dependent outer-boundary data throughout the evolution.

The computational domain is $x\in[0,21466.5]\,M$.
Inside the star, we use a uniform grid with $\Delta x_{\rm in}\simeq0.0143\,M$.
In the exterior, successive points are generated according to
\begin{align}
x_{i+1}-x_i
&=\Delta x_{\rm in}
+(\Delta x_{\rm out}-\Delta x_{\rm in})
\left(3\eta_i^2-2\eta_i^3\right)\,,\\
\eta_i
&=\min\left\{\frac{x_i-x_R}{L_{\rm tr}},\,1\right\}\,,
\qquad x_i\geq x_R\,,
\end{align}
where $\Delta x_{\rm out}\simeq0.716\,M$ and
$L_{\rm tr}\simeq42.933\,M$.
Thus the spacing increases smoothly across the exterior transition region and becomes uniform farther out.
Spatial derivatives are evaluated using five-point finite-difference stencils on this nonuniform grid, while time integration uses a fourth-order symmetric composition of implicit Newmark steps, with $\Delta t\simeq0.358\,M$. 
The evolution covers a duration of $\Delta T\simeq1.288\times10^4\,M$.

\section{Extension to Schwarzschild black holes}
\label{sec:black-hole-bilinear-appendix}
In this appendix, we develop an analogous framework for the
response of a Schwarzschild black hole to a prescribed incident
gravitational field. Unlike a fluid star, whose perturbations
involve coupled metric and fluid degrees of freedom and require
matching across the stellar surface, the even-parity perturbations
of a Schwarzschild black hole are described entirely by the
exterior Zerilli master field, i.e., $\boldsymbol{\Psi}=(Z, \Pi_Z)^T$. The stellar interior and surface matching conditions are replaced by a purely ingoing boundary condition at the event horizon, allowing radiation to be absorbed by the black hole.

We again start from the frequency-domain Green's function of
the Zerilli equation, which, for an observer outside the source, reads
\begin{equation}
\widetilde G_\ell(\omega;y,y')
=
\frac{
Z_\omega^{\rm H}(y')Z_\omega^{\rm out}(y)
}{
W_\ell(\omega)
}\,,
\qquad y>y'\,.
\label{eq:bh-frequency-green-function}
\end{equation}
Here
\begin{equation}
y:=r_*
=
r+2M\ln\!\left(\frac{r}{2M}-1\right)
\end{equation}
is the Schwarzschild tortoise coordinate, and
$Z_\omega^{\rm H}$ is the horizon-ingoing solution, normalized by
\begin{equation}
Z_\omega^{\rm H}(y)\sim e^{-i\omega y}\,,
\qquad y\to-\infty\,.
\end{equation}
At spatial infinity, it has the asymptotic form
\begin{equation}
Z_\omega^{\rm H}(y)
\sim
A_\ell^{\rm in}(\omega)e^{-i\omega y}
+
A_\ell^{\rm out}(\omega)e^{i\omega y}\,,
\qquad y\to+\infty\,.
\end{equation}
Introducing the exterior basis solutions normalized by
\begin{equation}
Z_\omega^{\rm in}(y)\sim e^{-i\omega y}\,,
\qquad
Z_\omega^{\rm out}(y)\sim e^{i\omega y}\,,
\qquad y\to+\infty\,,
\end{equation}
we may write the exact decomposition
\begin{equation}
Z_\omega^{\rm H}
=
A_\ell^{\rm in}(\omega)Z_\omega^{\rm in}
+
A_\ell^{\rm out}(\omega)Z_\omega^{\rm out}\,.
\label{eq:bh-horizon-solution-decomposition}
\end{equation}
The corresponding Wronskian is independent of $y$ and is given by
\begin{align}
W_\ell(\omega)
&:=
Z_\omega^{\rm H}\partial_yZ_\omega^{\rm out}
-
(\partial_yZ_\omega^{\rm H})Z_\omega^{\rm out}
\nonumber\\
&=
2i\omega A_\ell^{\rm in}(\omega)\,.
\label{eq:bh-wronskian}
\end{align}
With this notations, we may split the Green's function as
\begin{equation}
\widetilde G=\widetilde G^-+\widetilde G^+\,,
\end{equation}
where
\begin{align}
\widetilde G^-(\omega;y,y')
&=
\frac{
Z_\omega^{\rm out}(y)Z_\omega^{\rm in}(y')
}{
2i\omega
},\\
\widetilde G^+(\omega;y,y')
&=
\mathcal R_\ell(\omega)\frac{Z_\omega^{\rm out}(y)Z_\omega^{\rm out}(y')}{2i\omega}\,,
\end{align}
and
\begin{equation}
\mathcal R_\ell(\omega)
:=
\frac{A_\ell^{\rm out}(\omega)}
{A_\ell^{\rm in}(\omega)}\,.
\end{equation}

The retarded time-domain Green's function is obtained by the inverse Fourier transform,
\begin{equation}
G_\ell(\tau;y,y')
=
\frac{1}{2\pi}
\int_{-\infty+ic}^{\infty+ic}\mathrm d\omega\,
e^{-i\omega\tau}
\widetilde G_\ell(\omega;y,y')\,,
\quad c>0\,,
\end{equation}
where $\tau:=t-t'$.
The response to the effective source $S^{Z,{\rm d}}$ is then
\begin{equation}
Z^{\rm r}(t,y)
=
\int_{-\infty}^{t}\mathrm dt'
\int_{-\infty}^{\infty}\mathrm dy'\,
G_\ell(t-t';y,y')S^{Z,{\rm d}}(t',y')\,.
\end{equation}

Similar to the stellar case discussed in Sec.~\ref{sec:stellar-green-function} (see also Ref.~\cite{Su:2026fvj}), we partition the source events relative to a fixed observer into three regions, with the corresponding windows
\begin{align}
\chi_{\rm I}(\tau;y,y')
&=\Theta(\tau-\tau_{\rm c})\,,\\
\chi_{\rm II}(\tau;y,y')
&=\Theta(\tau-\tau_{\rm f})\Theta(\tau_{\rm c}-\tau)\,,\\
\chi_{\rm III}(\tau;y,y')
&=\Theta(\tau_{\rm f}-\tau)\,,
\end{align}
where we define
\begin{equation}
\tau_{\rm f}:=y-y'\,,
\qquad
\tau_{\rm c}:=y+|y'|\,.
\end{equation}
Following the terminology of Ref.~\cite{Su:2026fvj}, Region I admits the QNM-plus-tail representation, Region II contains the branch-cut direct contribution, and Region III lies outside the causal support of the Green's function. For $y'<0$, we have $\tau_{\rm c}=\tau_{\rm f}$, so Region II is absent.

The Green's function therefore takes the form
\begin{equation}
    G = G_{\rm dir}+G_{\rm QNM}+G_{\rm tail}\,,
\end{equation}
where $G_{\rm dir}$ is supported in Region II and the QNM and tail kernels are supported in Region I, with the switching boundaries understood distributionally.
The waveform at future null infinity is obtained at fixed retarded time $u=t-y$:
\begin{equation}
G^{\mathscr I^+}(u;y') := \lim_{y\to\infty}G(u+y;y,y')\,.
\label{eq::Inf_GF_BH_Zerilli}
\end{equation}
Consequently,
\begin{equation}
Z^{\mathscr I^+}(u)
=
\int_{-\infty}^{\infty}\mathrm dt'
\int_{-\infty}^{\infty}\mathrm dy'\,
G^{\mathscr I^+}(u-t';y')
S^{Z,{\rm d}}(t',y')\,,
\label{eq:green-covolution-black-hole}
\end{equation}
and
\begin{equation}
Z^{\mathscr I^+}
=
Z_{\rm dir}^{\mathscr I^+}
+
Z_{\rm QNM}^{\mathscr I^+}
+
Z_{\rm tail}^{\mathscr I^+}\,.
\end{equation}
Each component is obtained by replacing $G^{\mathscr I^+}$ in Eq.~\eqref{eq:green-covolution-black-hole} with the corresponding component kernel.

Analogously to the stellar case, we define the complex symmetric bilinear form for the black-hole problem as
\begin{align}
\left\langle\boldsymbol\psi_1,\boldsymbol\psi_2\right\rangle
:={}&
\int_{\mathcal C}\mathrm dy\,
\left(
Z_1\Pi_{Z,2}+\Pi_{Z,1}Z_2
\right)
\nonumber\\ 
&+\frac{C_\ell}{i(\omega_1-\omega_2)}
\left[
Z_2\partial_yZ_1-Z_1\partial_yZ_2
\right]_{\partial\mathcal C}\,,
\end{align}
where both $\boldsymbol{\psi}_{1}$ and $\boldsymbol{\psi}_{2}$ are homogeneous solutions with frequencies $\omega_1$ and $\omega_2$, respectively.
For QNMs, the radial functions grow exponentially toward
both ends of the real tortoise-coordinate axis.
The bilinear form must therefore be understood through
an appropriate analytic continuation and contour regularization, rather than by separately evaluating divergent integrals and boundary terms at real infinity; see, e.g.,
Ref.~\cite{Ma:2024qcv,Minucci:2026dgo,Green:2022htq}.

Using this bilinear form, we can express the black-hole response as source projections onto its QNM and continuum contributions,
following the same procedure as for the stellar problem
in the main text.

\textbf{Region I: QNM contribution.} 
Denoting the QNM wavefunction as $Z_n(y) := Z^{\rm out}_{\omega_n}(y)$ and $\boldsymbol{\psi}_n= (Z_n, -i\omega_nC_\ell Z_n)^T$, we can write the QNM response as
\begin{align}
Z_{\rm QNM}^{\rm r}(t,y)
=
\sum_n a_{n,{\rm I}}(t,y)Z_n(y)\,,
\end{align}
with
\begin{align}
a_{n,{\rm I}}(t,y)
=
\frac{1}{\mathcal N_n}
\int_{-\infty}^{t}\mathrm dt'\,
e^{-i\omega_n(t-t')}
\left\langle
\boldsymbol\psi_n,
\boldsymbol S_{\rm I}^{\rm d}(t';t,y)
\right\rangle\,.
\label{eq:black-hole-QNM-amplitude}
\end{align}
Here $\mathcal N_n :=\left\langle\boldsymbol\psi_n, \boldsymbol\psi_n \right\rangle$.
Differentiating Eq.~\eqref{eq:black-hole-QNM-amplitude} gives the evolution equation
\begin{align}
(\partial_t+i\omega_n)a_{n,{\rm I}}(t,y)
=
\frac{1}{\mathcal N_n}
\left\langle
\boldsymbol\psi_n,
e^{-i\omega_n\tau_{\rm c}}
\boldsymbol S^{\rm d}(t-\tau_{\rm c},y')
\right\rangle\,.
\end{align}

\textbf{Region I: tail contribution.}
Similar to the QNM contribution, the tail part can be organized as
\begin{align}
Z_{\rm tail}^{\rm r}(t,y)
=
\int_{0}^{-i\infty}\mathrm d\sigma\,
a_{\sigma,{\rm I}}^{\rm tail}(t,y)
Z_{\sigma}^{\rm H}(y)\,,
\end{align}
with 
\begin{align}
a_{\sigma,{\rm I}}^{\rm tail}(t,y)
&=
\frac{\sigma\,q_\ell^Z(i\sigma)}
{\pi C_\ell|W_\ell(\sigma+0)|^2}\nonumber\\
&\times
\int_{-\infty}^{t}\mathrm dt'\,
e^{-i\sigma(t-t')}
\left\langle
\boldsymbol\psi_\sigma^{\rm H},
\boldsymbol S_{\rm I}^{\rm d}(t';t,y)
\right\rangle\,.
\end{align}
The amplitude satisfies 
\begin{align}
(\partial_t+i\sigma)a_{\sigma,{\rm I}}^{\rm tail}(t,y)
=&{}
\frac{\sigma\,q_\ell^Z(i\sigma)}
{\pi C_\ell|W_\ell(\sigma+0)|^2}\nonumber\\
&\times \left\langle
\boldsymbol\psi_\sigma^{\rm H},
e^{-i\sigma\tau_{\rm c}}
\boldsymbol S^{\rm d}_{\rm I}(t-\tau_{\rm c},y')
\right\rangle\, .
\end{align}

\textbf{Region II: direct contribution.} 
Introduce the auxiliary phase-space profiles
\begin{align}
\boldsymbol\psi_\sigma^-
=
\begin{pmatrix}
Z_{\sigma}^{\rm in}\\
-i\sigma C_\ell Z_{\sigma}^{\rm in}
\end{pmatrix}\,,
\qquad
\boldsymbol\psi_\sigma^+
=
\begin{pmatrix}
Z_{\sigma}^{\rm out}\\
-i\sigma C_\ell Z_{\sigma}^{\rm out}
\end{pmatrix}\,,
\label{eq:direct_part_state_vec}
\end{align}
the direct part can be written as
\begin{align}
Z_{\rm II}^{\rm r}(t,y)
={}&
\int_{\mathcal{C}_{\rm NIA}}\mathrm d\sigma\,
a_{\sigma,{\rm II}}^{\rm dir,-}(t,y)Z_\sigma^{\rm out}(y)
\nonumber\\
&+
\int_{\mathcal{C}_{\rm PIA}}\mathrm d\sigma\,
a_{\sigma,{\rm II}}^{{\rm dir,+}}(t,y)Z_\sigma^{\rm out}(y),
\end{align}
where
\begin{align}
a_{\sigma,{\rm II}}^{\rm dir,-}(t,y)
&={}
\frac{1}{4\pi i C_\ell\sigma}\nonumber\\
&\times\int_{-\infty}^{t}\mathrm dt'\,
e^{-i\sigma(t-t')}
\left\langle
\boldsymbol\psi_\sigma^{-},
\boldsymbol S_{\rm II}^{\rm d}(t';t,y)
\right\rangle\,,\\
a_{\sigma,{\rm II}}^{\rm dir,+}(t,y)
&={}
\frac{\mathcal{R}_\ell(\sigma)}{4\pi i C_\ell\sigma}\nonumber\\
&\times\int_{-\infty}^{t}\mathrm dt'\,
e^{-i\sigma(t-t')}
\left\langle
\boldsymbol\psi_\sigma^{+},
\boldsymbol S_{\rm II}^{\rm d}(t';t,y)
\right\rangle\,.
\end{align}
The contours $\mathcal{C}_{\rm NIA}$ and $\mathcal{C}_{\rm PIA}$ are the same as those used in the main text, see also in Fig.~\ref{fig:nia-pia-contours}.
These amplitudes satisfy
\begin{align}
&(\partial_t+i\sigma)a_{\sigma,{\rm II}}^{\rm dir,-}(t,y)
={}\frac{1}{4\pi i C_\ell\sigma}
\nonumber\\
&\hspace{1.2cm}\times\Bigg[
\left\langle
\boldsymbol\psi_\sigma^{-},
e^{-i\sigma\tau_{\rm f}}
\boldsymbol S^{\rm d}(t-\tau_{\rm f},y')
\right\rangle\nonumber\\
&\hspace{1.2cm}-\left\langle
\boldsymbol\psi_\sigma^{-},
e^{-i\sigma\tau_{\rm c}}
\boldsymbol S^{\rm d}(t-\tau_{\rm c},y')
\right\rangle
\Bigg].\\
&(\partial_t+i\sigma)a_{\sigma,{\rm II}}^{\rm dir,+}(t,y)
={}\frac{\mathcal{R}_\ell(\sigma)}{4\pi i C_\ell\sigma}
\nonumber\\
&\hspace{1.2cm}\times\Bigg[
\left\langle
\boldsymbol\psi_\sigma^{+},
e^{-i\sigma\tau_{\rm f}}
\boldsymbol S^{\rm d}(t-\tau_{\rm f},y')
\right\rangle\nonumber\\
&\hspace{1.2cm}-\left\langle
\boldsymbol\psi_\sigma^{+},
e^{-i\sigma\tau_{\rm c}}
\boldsymbol S^{\rm d}(t-\tau_{\rm c},y')
\right\rangle
\Bigg].
\end{align}

\end{document}